\documentclass[11pt]{article}
\usepackage[dvipdfmx]{graphicx}
\usepackage{amsfonts,amsmath,amssymb,epsf}
\usepackage{amsmath,braket}
\usepackage{color}
\usepackage{hyperref}
\usepackage[toc,page]{appendix}
\usepackage[normalem]{ulem}
\usepackage{url}
\usepackage{cancel}
\usepackage{tikz}
\usepackage{microtype}
\usepackage{soul}
\usepackage{physics}
\hypersetup{
    bookmarks=true,         
    unicode=false,          
    pdftoolbar=true,        
    pdfmenubar=true,        
    pdffitwindow=false,     
    pdfstartview={FitH},    
    pdfnewwindow=true,      
    colorlinks=true,       
    linkcolor=violet,          
    citecolor=teal,        
    filecolor=blue,      
    urlcolor=blue           
}
\usepackage[T1]{fontenc}
\usepackage[latin9]{inputenc}

\usepackage{braket}
\usepackage{amsmath}
\usepackage{amssymb}
\usepackage{color}

\makeatletter
\numberwithin{figure}{section}
\numberwithin{equation}{section}
\numberwithin{table}{section}

\newcommand{\be}{\begin{equation}}
\newcommand{\ba}{\begin{eqnarray}}
\newcommand{\ea}{\end{eqnarray}}
\newcommand{\ee}{\end{equation}}

\newcommand{\bea}{\begin{eqnarray}}
\newcommand{\eea}{\end{eqnarray}}
\newcommand{\bes}{\begin{equation*}}
\newcommand{\beas}{\begin{eqnarray*}}
\newcommand{\eeas}{\end{eqnarray*}}
\newcommand{\bas}{\begin{array*}}
\newcommand{\eas}{\end{array*}}
\newcommand{\ees}{\end{equation*}}
\newcommand{\nn}{\nonumber}

\makeatother
\usepackage{comment}
\usepackage{babel}

\definecolor{LightGray}{rgb}{0.8, 0.8, 0.8}
\allowdisplaybreaks

\begin{document}

    \begin{titlepage}
    \thispagestyle{empty}
    
    \begin{flushright}
    YITP-26-24
    \\
    KUNS-3092
    \end{flushright}
    
    \bigskip
    
    \begin{center}
    \noindent{\Large \bf More on Bulk Local State Reconstruction in Flat/Carr CFT}\\
    \vspace{1cm}
   Peng-Xiang Hao$^{a,c}$,
Kotaro Shinmyo$^a$,
Yu-ki Suzuki$^a$,
Shunta Takahashi$^{b,d}$
     \vspace{1cm}\\
    \textit{{}$^a$ Center for Gravitational Physics and Quantum Information,
    Yukawa Institute for Theoretical Physics, Kyoto University, Kyoto 606-8502, Japan} \\
\vspace{0.2cm}    
    \textit{{}$^b$ Department of Physics, Kyoto University, Kyoto 606-8502, Japan} \\
\vspace{0.2cm}  
    \textit{{}$^c$ Yau Mathematical Sciences Center, Tsinghua University, Haidian District, Beijing 100084, China} \\
\vspace{0.2cm}  
    \textit{{}$^d$ Perimeter Institute for Theoretical Physics, Waterloo, Ontario N2L 2Y5, Canada} \\
    \vspace{0.5cm}
\noindent {\small Emails: \texttt{pxhao@yukawa.kyoto-u.ac.jp}, \texttt{kotaro.shinmyo@yukawa.kyoto-u.ac.jp}, \texttt{yu-ki.suzuki@yukawa.kyoto-u.ac.jp}, \texttt{shunta@gauge.scphys.kyoto-u.ac.jp}}

    \end{center}

\begin{abstract}
We revisit and extend the construction of bulk local states in flat holography, focusing on the induced representation obtained from the flat limit of the AdS highest-weight conditions. In three dimensions we clarify the scaling mismatch between bra and ket states in the flat basis and resolve it by introducing a dual basis, which yields a smooth flat limit and reproduces the correct Green's function. For higher dimensions we construct bulk local states explicitly, both in the momentum basis and in an alternative tilde basis. The flat limit of the AdS$_{d+1}$ construction is shown to be non-uniform in the descendant level and the Riemann-sum treatment over the scaling window $n\sim l$ converts the discrete descendant expansion into the continuum momentum representation, recovering the massive propagator. The tilde basis generalizes seamlessly to any dimension and is related to the three-dimensional flat basis by a sign factor. These results establish the induced representation as the correct algebraic foundation for bulk reconstruction in flat holography and provide a unified framework valid for arbitrary dimension.
\end{abstract}
    
    \end{titlepage}
    
    \newpage
    
    \tableofcontents

    \section{Introduction}
        The holographic principle \cite{tHooft:1993dmi,Susskind:1994vu} breaks an impasse on the complete understanding of quantum gravity, though much of the progress has concentrated on AdS/CFT \cite{Maldacena:1997re,Gubser:1998bc,Witten:1998qj}.
        Whether this narrative extends to more realistic models like flat holography or dS/CFT \cite{Strominger:2001pn,Maldacena:2002vr,Hikida:2021ese,Doi:2024nty} remains elusive.
        Towards the full picture of bulk-boundary correspondence in 4d flat holography which is highly relevant to our 4d universe, this paper aims to clarify the correct representation of the algebra that the boundary theory is supposed to have and to reconstruct the bulk state from that representation, extending the result in 3d in Ref.\,\cite{Hao:2025btl}.

        Before proceeding to our contribution, here are the contexts in flat holography.
        The \textit{Flat/Carrollian CFT correspondence} \cite{Barnich:2010eb, Bagchi:2010zz, Fareghbal:2013ifa} posits that the the quantum gravity on asymptotically flat space is equivalent to the Carrollian CFT (CCFT), or sometimes called the BMS (Bondi-Metzner-Sachs) field theory \cite{Bondi:1962px,Sachs:1962wk,Bagchi:2009ca,Barnich:2014kra,Barnich:2015uva}, on the codimension one asymptotic null boundary with time direction.
        It is shown that the Bekenstein-Hawking entropy of flat cosmological horizons coincides with the Cardy formula in CCFTs \cite{Barnich:2012xq, Bagchi:2012xr}, and the partition functions are discussed in Refs.\,\cite{Barnich:2012rz, Barnich2015one}, where the one-loop determinant corresponds to the CCFT vacuum character.
        The bulk reconstruction in 3d bulk are provided for different representations on the boundary \cite{Chen:2023naw, Hao:2025btl}.
        The CCFT entanglement entropy is calculated for various geometries \cite{Bagchi:2014iea, Hao:2025naz} and it can be compared to the swing surface proposal \cite{Jiang:2017ecm}, the holographic entanglement entropy formula in flat holography.
        Parallel to holographic applications, the intrinsic properties of CCFTs as novel type of quantum field theories have garnered significant interest \cite{Hao:2021urq, Chen:2021xkw, Yu:2022bcp, Hao:2022xhq, Banerjee:2022ocj, deBoer:2023fnj, Hao:2025hfa}, with the developments on the correlation functions \cite{Bagchi:2009ca}, bootstrap method \cite{Bagchi:2016geg, Bagchi:2017cpu, Chen:2020vvn, Chen:2022jhx}, among others.
        Overall reviews are available in Refs.\,\cite{Bagchi:2025vri, Nguyen:2025zhg}.

        A distinct but complementary framework is provided by celestial holography \cite{Pasterski:2016qvg, Pasterski:2017kqt}, where CFT exists on a codimension-\textit{two} boundary without time direction.
        By expressing gravitational $S$-matrix elements within a boost eigenstate basis, this approach reveals their striking resemblance to conformal correlation functions, making it a powerful tool for probing gravitational scattering.
        While the precise mapping between the Celestial and Carrollian (Flat/CCFT) perspectives is an area of active refinement \cite{Donnay:2022aba, Donnay:2022wvx, Bagchi:2022emh, Bagchi:2023fbj, Kulp:2024scx}, we restrict our present attention to the latter.
        However, we restrict our attention to Flat/CCFT and hereby present our new results on bulk reconstruction in 4d Flat/CCFT correspondence. \\

        \noindent\colorbox{LightGray}{Main results} \\
     Assuming the AdS$_4$/CFT$_3$ correspondence, we argue that the physically sensible representation on the boundary side of Flat$_4$/CCFT$_3$ is the \textbf{induced representation} obtained by taking the flat limit $l\to\infty$ of the AdS highest-weight conditions. For a scalar field this representation is characterized by
\begin{align}
P_0|\xi\rangle = \xi|\xi\rangle,\qquad P_a|\xi\rangle = 0\;\;(a=1,2,3), \qquad J_{ab}|\xi\rangle = 0, \label{eq:main_induced}
\end{align}
where $P_\mu$ are translation generators of the Poincar\'e algebra and $J_{ab}$ are rotations.  This is the natural analog of the three-dimensional induced representation studied in Refs.\,\cite{Barnich:2014kra,Barnich:2015uva} and similar to those appeared in the string theory literature \cite{Ruzziconi:2026isv, Bekaert:2024uuy, Bekaert:2025kjb}.  It differs from the highest weight representation of the full BMS$_4$ algebra \cite{Bagchi:2016bcd}, but captures the global Poincar\'e subgroup which is minimal yet sufficient for reconstructing scalar bulk local states. We would also like to point out that the supergravity amplitudes in 4d flat space are analyzed in Ref.\,\cite{Lipstein:2025jfj} by taking the flat limit of M-theory/ABJM theory correpsondence on AdS$_4\times S^7$. However, our discussion involves the reconstruction of more fundamental states from pure representation theoretic perspective, which is also without supersymmetry. 

We substantiate this claim by explicitly constructing the bulk local state in four-dimensional flat space, both in the momentum basis and in an alternative ``tilde basis''
.\footnote{In the Introduction we present the final formulas in Lorentzian signature. In the detailed derivations, some descendant sums and overlaps are most conveniently regulated by a temporary Wick rotation $t=-i \tau$ with $\tau>0$.
Equivalently, the corresponding Lorentzian expression is understood with the standard $i \epsilon$ prescription at least in our free theory case.
Moreover, as discussed in section \ref{sec:conclusion}, it follows that the robustness is not spoiled even when interactions are introduced, as long as the structure of spacetime is preserved.}  
In the momentum basis the state reads
\begin{equation}
|\phi_{\text{flat}}(t,\mathbf{0})\rangle = \int\frac{d^3p}{(2\pi)^3\,2E}\,e^{-i Et}\,|\mathbf{p}\rangle,
\end{equation}
and its inner product reproduces the positive-frequency massive Wightman function
\begin{align}
    G_{\text{flat}}(t)=\int \frac{d^3 p}{(2 \pi)^3 2E}e^{-iEt}.
\end{align}
On the AdS$_4$ side, Refs.\,\cite{Miyaji:2015fia, Nakayama:2015mva} construct the corresponding bulk local state from rotationally invariant descendants $(P^2)^n|\Delta\rangle$, whose flat limit $\Delta=\xi l+\mathcal{O}(1)$ is non-uniform in the descendant level $n$ and requires a Riemann sum over the scaling window $n\sim l$.  This yields precisely the flat-space momentum representation, confirming that the flat result is firmly anchored in the AdS construction.

The tilde basis, defined by $P_a|\tilde n\rangle = 2in\xi K_a|\widetilde{n-1}\rangle$, provides a complementary description that simplifies the stabilizer condition.  In this basis the flat bulk local state takes the universal form
\begin{equation}
|\phi(t,\mathbf{0})\rangle = \sum_{n=0}^{\infty}\frac{1}{n!\,(2i\xi t)^n}\,|\tilde n\rangle. \label{eq:flat solution intro}
\end{equation}
The same basis can be defined in AdS$_4$ (with a dimension-dependent modification of the eigenvalue $\lambda_n(l)$) and its flat limit smoothly recovers the flat-space expression \eqref{eq:flat solution intro}.

All these constructions are then generalized to arbitrary spacetime dimension $d+1$. 
The induced representation conditions remain the same, and the momentum-basis bulk local state is
\begin{equation}
|\phi_{\text{flat}}(t,\mathbf{0})\rangle = \int\frac{d^dp}{(2\pi)^d\,2E}\,e^{-iEt}\,|\mathbf{p}\rangle,
\qquad
G_{\text{flat}}(t)=\int \frac{d^d p}{(2 \pi)^d 2E}e^{-iEt}.
\end{equation}
The AdS$_{d+1}$ descendant basis, with norms $G_n = 16^n n! (\frac d2)_n (\Delta)_n (\Delta-\frac d2+1)_n G_0$, leads to a bulk local state whose flat limit again reorganizes into a continuum spectral integral.
The tilde basis generalizes straightforwardly, with the flat space action $P_a|\tilde n\rangle = 2in\xi K_a|\widetilde{n-1}\rangle$ and the same universal expansion \eqref{eq:general_bulk_state_tilde_flat}.
In three dimensions ($d=2$) the coefficient $a_n(\Delta)=1$ makes the flat limit uniform, so the limit commutes with the sum. 
For $d\ge 3$ the growth of $a_n(\Delta)$ forces the scaling window $n\sim l$ and the Riemann-sum treatment is essential. 
The three-dimensional flat basis is recovered from the general tilde basis by the simple sign relation $|\tilde n\rangle = (-1)^n |n\rangle_{\text{Flat}_3}$.

These results establish the induced representation as the correct starting point for bulk reconstruction in flat holography, provide explicit state realizations in both momentum and tilde bases, and demonstrate their seamless connection through the flat limit of AdS.\\

\noindent\colorbox{LightGray}{The plan of the paper} \\
This paper is organized as follows. In Section~\ref{sec:bulk_local_3d}, we revisit the three-dimensional construction, reviewing the induced representation, the explicit form of the bulk local state in both AdS$_3$ and flat space, the scaling mismatch of the bra state, and its resolution via a dual basis. We also compute the Green's function and verify its flat limit.

Section~\ref{sec:bulk_local_4d} is devoted to the higher-dimensional generalization. We begin in Subsection~\ref{sec:algebra_representation} with the algebraic foundations for four dimensions from the isometries of AdS$_4$ and Flat$_4$ to their flat limit, and the resulting induced representation of the Poincar\'e group. We then translate the bulk primary conditions into constraints on boundary states in the dual Carrollian CFT.

In Subsection~\ref{subsec:bulk_local_4d_momentum} we construct the bulk local state in the momentum basis, first directly in Flat$_4$ and then as the flat limit of the corresponding AdS$_4$ state. We show that the flat limit is non-uniform in the descendant level and requires a Riemann sum over the scaling window $n\sim l$, which reproduces the expected flat-space Green's function and momentum-space wavefunction.

Subsection~\ref{subsec:bulk_local_4d_tilde} introduces an alternative description in the tilde basis which simplifies the action of the translation generators and makes contact with the flat basis used in three dimensions. We construct this basis in flat space, extend it to AdS$_4$, and show that its flat limit again yields the same bulk local state.

Finally, Subsection~\ref{subsec:arbitrary_dim} generalizes the entire discussion to arbitrary spacetime dimension $d+1$. We present the algebraic setup, the momentum-basis construction, and the tilde-basis construction in full generality.

We conclude in Section~\ref{sec:conclusion} with a summary of our results and an outlook on future directions. Several appendices collect explicit expressions for symmetry generators, conventions for the Osterwalder-Schrader reflection, derivations of descendant norms, and an alternative treatment of the flat limit via Legendre functions. \\

\noindent\colorbox{LightGray}{Notations} \\
We use the calligraphy letters $\mathcal{H}$, $\mathcal{P}_a$, ... for AdS conformal algebra, while the normal letters $P_a$, $K_a$, ... for Poincar\'{e} (flat) generators.

\section{Bulk local state in 3d revisited} \label{sec:bulk_local_3d}

In this section, we revisit the construction of bulk local states in three-dimensional flat spacetime and their relation to the AdS$_3$ counterparts explored in Ref.\,\cite{Hao:2025btl}. After summarizing the necessary background on the Flat$_3$/CCFT$_2$ correspondence and the associated algebraic structures, we explicitly construct the bulk local states in both AdS$_3$ and flat space, paying special attention to their behavior under the flat limit. We then turn to the definition of the corresponding bra states and examine why a naive flat limit leads to a scaling mismatch. This puzzle is resolved by introducing a dual basis, which naturally accounts for the divergent inner product in the limit. Finally, we compute the Green's function from the inner product of these states and verify that the result matches the expected flat space propagator.

\subsection{Background and bulk local states in AdS$_3$ and Flat$_3$}
\label{subsec:3d_background_states}

\subsubsection{Brief review on Flat$_3$/CCFT$_2$ correspondence}
\label{section: background}

To set the stage, we briefly recall the essential features of the Flat$_3$/CCFT$_2$ correspondence. The asymptotic symmetry algebra of three-dimensional asymptotically flat spacetimes in Einstein gravity is the BMS$_3$ algebra. For the specific case of the Flat$_3$/CCFT$_2$ correspondence, this algebra takes the form
\begin{align}
    \begin{aligned}
        \left[L_{n},\,L_{m}\right] &= (n-m)L_{m+n}+\frac{c_L}{12} n(n^2-1)\delta_{m+n,0}, \\
	    \left[L_{n},\,M_{m}\right] &= (n-m)M_{m+n}+\frac{c_M}{12}  n(n^2-1)\delta_{m+n,0}, \\
	    \left[M_{n},\,M_{m}\right] &= 0.
    \end{aligned}
    \label{eq: bms_3 algebra}
\end{align}
The central charges in this algebra are given by
\begin{equation}
c_L=0,\ \ c_M=\frac{3}{G},  \label{eq: BMS central charge}
\end{equation}
where $G$ denotes the three-dimensional Newton constant. The corresponding gravitational solutions can be written in the Bondi gauge as
\begin{equation}
ds^2=\Theta(\phi)du^2-2dudr+2\bigg(\Xi(\phi)+\frac{u\partial_\phi \Theta(\phi)}{2}\bigg)dud\phi+r^2d\phi^2,\ \ \phi\sim\phi+2\pi,
\end{equation}
with $u$ being the retarded time. The constant mode solution, characterized by constant functions $\Theta(\phi)=M$ and $\Xi(\phi)=J$, is given by
\begin{equation}\label{eq: const mode solution}
ds^2=Mdu^2-2dudr+2Jdud\phi+r^2d\phi^2,\ \ \phi\sim\phi+2\pi.
\end{equation}

These flat space solutions can be systematically obtained by taking the large AdS radius limit $l\rightarrow \infty$. In this limit, the boundary cylinder recedes to infinity and the bulk region near the center of AdS becomes flat. For $ M > 0$, the solution possesses a cosmological horizon located at $r_c = {|J|}/{\sqrt{M}}$. For $ M < 0$, the solution describes a spinning particle, which introduces a conical defect and a twist in the time identification. Although locally flat, such configurations have a delta-functional source. The case $M = -1$ corresponds precisely to global Minkowski space.

According to the Flat/CCFT correspondence, the dual field theory is a two-dimensional BMS$_3$-invariant theory, also known as BMSFT$_2$ or a 2D Carrollian conformal field theory (CCFT$_2$). This is a two-dimensional quantum field theory invariant under the transformations
\begin{equation}
\phi\to f(\phi), \quad u \to f^{\prime}(\phi)u + g(\phi),
\end{equation}
where $\phi$ is a spatial coordinate and $u$ is a null time coordinate. The infinitesimal transformations are generated by the Fourier modes
\begin{align}
L_{n} &= ie^{in\phi }\partial_{\phi} -n e^{in\phi}u\partial_{u},\\
M_{n} &= ie^{in\phi }\partial_{u}, \label{lmncylinder}
\end{align}
which realize the BMS$_3$ algebra \eqref{eq: bms_3 algebra} without central extensions. The algebra admits two quadratic Casimir operators,
\begin{equation}
C_1=M_0^2-M_{-1}M_1,\ \ C_2=2L_0M_0-\frac{1}{2}(L_{-1}M_1+L_1M_{-1}+M_1L_{-1}+M_{-1}L_1).   \label{eq: quadratic casimir in flat}
\end{equation}

In the literature two types of representations are commonly studied, namely the highest weight representation \cite{Bagchi:2009ca} and the induced representation \cite{Barnich:2014kra,Barnich:2015uva}. The induced representation is unitary and can be obtained from an flat (ultra-relativistic) limit of the highest weight representations of the CFT$_2$ dual to AdS$_3$. The highest weight representation of the BMS algebra, while analogous to that of the Virasoro algebra, is non-unitary.

In this work, we focus on the induced representation as it gives a physically sensible results in the subsequent discussion. In this framework, the states are labeled by the action under $L_0$ and $M_0$ with the primary condition,
\begin{align}
    \begin{gathered}
L_0|\Delta,\xi\rangle=\Delta|\Delta,\xi\rangle,\ \ M_0|\Delta,\xi\rangle=\xi|\Delta,\xi\rangle, \\
        M_n|\Delta,\xi\rangle=0,\quad  ( n\neq0).
    \end{gathered}
    \label{eq: bms induced rep}
\end{align}
Here, $\Delta$ and $\xi$ are referred to as the conformal weight and the boost charge of the state, respectively. Descendant states are generated by applying the $L_{n\neq0}$ operators on the primary state. The eigenvalues of the Casimir operators \eqref{eq: quadratic casimir in flat} on such a state are
\begin{equation}\label{eq: induced casimir}
C_1|\Delta,\xi\rangle=\xi^2|\Delta,\xi\rangle,\ \ C_2|\Delta,\xi\rangle=2\Delta \xi|\Delta,\xi\rangle.
\end{equation}
Since the Casimir operators commute with all global generators, these eigenvalues are shared by all global descendant states in the module. In particular, the induced vacuum state $|0_{\rm I}\rangle$ must satisfy
\begin{equation}\label{i72v}
L_0|0_{\rm I}\rangle=M_n|0_{\rm I}\rangle=0,\ \ \quad\quad \forall n\in \mathbb{Z}.
\end{equation}
In free field realizations \cite{Hao:2021urq,Hao:2022xhq}, the induced vacuum behaves as a direct product state, leading to ultra-local correlation functions.

For completeness, we also recall the highest weight representation of the BMS algebra, which directly generalizes the Virasoro case. To ensure that $L_0$ is bounded from below, the highest weight conditions are imposed as
\begin{align}
    \begin{gathered}
        L_0|\Delta,\xi\rangle=\Delta|\Delta,\xi\rangle,\ \ M_0|\Delta,\xi\rangle=\xi|\Delta,\xi\rangle, \\
        L_n|\Delta,\xi\rangle=M_n|\Delta,\xi\rangle=0,\quad (n>0)
    \end{gathered}
    \label{eq: bms highest weight rep}
\end{align}
defining a highest weight primary state $|\Delta,\xi\rangle$. Descendant states are obtained by applying $L_{-n}$ and $M_{-n}$ for $n>0$. The resulting module has Casimir eigenvalues
\begin{equation}
C_1|\Delta,\xi\rangle=\xi^2|\Delta,\xi\rangle,\ \ C_2|\Delta,\xi\rangle=2(\Delta-1) \xi|\Delta,\xi\rangle.
\end{equation}
We will not explore further details of the highest weight representation as we do not use it in the following analysis. \\

\noindent\textbf{Flat limit from Virasoro algebra} \\
For comparison, in the AdS$_3$ context, the dual CFT$_2$ possesses left and right Virasoro algebras $\text{Vir}\times\overline{\text{Vir}}$ with central charges $c_{\rm AdS}=\overline{c_{\rm AdS}}=\frac{3l}{2G}$. The commutation relations are
\begin{align}
    [\mathcal{L}_m,\mathcal{L}_n]&=(m-n)\mathcal{L}_{m+n}+\frac{c_{\rm AdS}}{12}m(m^2-1)\delta_{m+n},\nn\\
    [\bar{\mathcal{L}}_m,\bar{\mathcal{L}}_n]&=(m-n)\bar{\mathcal{L}}_{m+n}+\frac{\bar{c}_{\rm AdS}}{12}m(m^2-1)\delta_{m+n},\quad (m,n\in \mathbf{Z})\nn\\
    [\mathcal{L}_m,\bar{\mathcal{L}}_n]&=0.
\end{align}
Physical states are described by highest weight representations. Primary states $\ket*{h,\bar{h}}$ satisfy
\begin{align}
\mathcal{L}_0\ket*{h,\bar{h}}&=\bar{h}\ket*{h,\bar{h}},\quad \bar{\mathcal{L}}_0\ket*{h,\bar{h}}=h\ket*{h,\bar{h}},\nn\\
\mathcal{L}_n\ket*{h,\bar{h}}&= \bar{\mathcal{L}}_n\ket*{h,\bar{h}}=0. \quad (n>0) \label{eq: highest weight condition}
\end{align}
The Hilbert space is spanned by states of the form
\be
\mathcal{L}_{-n_1}\cdots\mathcal{L}_{-n_k}\bar{\mathcal{L}}_{-\bar{n}_1}\cdots\bar{\mathcal{L}}_{-\bar{n}_k}\ket*{h,\bar{h}},\quad (n_1\geq\cdots n_k\geq0, \,\,\bar{n}_1\geq\cdots\bar{n}_k\geq0).
\ee
Unitarity of the representation imposes the conjugation rules
\be
\mathcal{L}_n^\dagger=\mathcal{L}_{-n},\quad \bar{\mathcal{L}}_n^\dagger=\bar{\mathcal{L}}_{-n}. \label{eq:ads conjugation}
\ee

To study the flat limit, it is convenient to define a new set of generators that depend on the AdS radius $l$,
\be
L_n(l)=\mathcal{L}_n-(-1)^n\bar{\mathcal{L}}_{-n},\quad M_n(l)=\frac{1}{l}(\mathcal{L}_n+(-1)^n\bar{\mathcal{L}}_{-n}), \label{eq: generator relation between flat and AdS}
\ee
whose commutation relations are
 \begin{align}
                \begin{aligned}
                    [L_m(l),L_n(l)] & =(m-n)L_{m+n}(l)+\frac{c_L}{12}(m^3-m)\delta_{m+n,0}, \\
                    [L_m(l),M_n(l)] & =(m-n)M_{m+n}(l)+\frac{c_M}{12}(m^3-m)\delta_{m+n,0}, \\
                    [M_m(l),M_n(l)] & =\frac{1}{l^2}\Big((m-n)L_{m+n}(l)+\frac{c_L}{12}(m^3-m)\delta_{m+n,0}\Big).
                \end{aligned} \label{eq: BMS-like commutators before the flat limit}
            \end{align}
The parameters $c_L$ and $c_M$ are defined as
            \begin{align}
                    c_L:=c_{\rm AdS}-\overline{c}_{\rm AdS}=0,\quad c_M:=\frac{c_{\rm AdS }+\overline{c}_{\rm AdS}}{l}=\frac{3}{G},  \label{eq: BMS-like central charge}
            \end{align}
which coincide exactly with the BMS central charges \eqref{eq: BMS central charge}. From the AdS conjugation rules \eqref{eq:ads conjugation}, we obtain the following relations for the new generators,
            \begin{align}
                L_n(l)^{\dagger}=L_{-n}(l),\;\; M_n(l)^{\dagger}=M_{-n}(l). \label{eq: flat conjugation before flat limit}
            \end{align}
In the limit $l\to\infty$, the commutators \eqref{eq: BMS-like commutators before the flat limit} reduce precisely to the BMS$_3$ algebra \eqref{eq: bms_3 algebra} with the relation $\lim_{l\to\infty}L_n(l)=L_n$ and $\lim_{l\to\infty}M_n(l)=M_n$, recovering the flat space generators.

We now verify that the highest weight representation of the Virasoro algebra flows to the induced representation of the BMS algebra in the flat limit.
Rewriting the highest weight conditions \eqref{eq: highest weight condition} in terms of the generators $L_n(l)$ and $M_n(l)$ yields
            \begin{gather}
                L_0(l)|h,\overline{h}\rangle=(h-\overline{h})|h,\overline{h}\rangle,\:\:M_0(l)|h,\overline{h}\rangle=\frac{h+\overline{h}}{l}|h,\overline{h}\rangle, \\
                \Big(M_n(l)+\frac{1}{l}L_n(l)\Big)|h,\overline{h}\rangle=0,\:\:\Big(M_{-n}(l)-\frac{1}{l}L_{-n}(l)\Big)|h,\overline{h}\rangle=0\quad(n>0). \label{eq: highest weight condition in AdS represented in flat generators}
            \end{gather}
Introducing the finite limit variables $\displaystyle\xi:=\lim_{l\to\infty}\frac{1}{l}(h+\overline{h})$ and $\displaystyle\Delta:=\lim_{l\to\infty}(h-\overline{h})$, these conditions become, in the $l\to\infty$ limit,
            \begin{align}
                M_0|\Delta,\xi\rangle=\xi|\Delta,\xi\rangle,\:M_n|\Delta,\xi\rangle=0\,(n\neq 0),\:L_0|\Delta,\xi\rangle=\Delta|\Delta,\xi\rangle,  \label{eq: BMS induced from AdS highest}
            \end{align}
where $\displaystyle L_n:=\lim_{l\to\infty}L_n(l),\;M_n:=\lim_{l\to\infty}M_n(l),\;|\Delta,\xi\rangle:=\lim_{l\to\infty}|h,\overline{h}\rangle$.
In this paper, we restrict our attention to scalar fields, where $h=\overline{h}=\frac{1}{2}+\frac{1}{2}\sqrt{m^2l^2+1}$. In this case, the limit values $\xi=m$ and $\Delta=0$ are finite, and eq.\,\eqref{eq: BMS induced from AdS highest} agrees completely with the scalar BMS$_3$ induced condition \eqref{eq: bms induced rep}.

\subsubsection{Bulk local state in AdS$_3$ and Flat$_3$}

With the algebraic preliminaries in place, we now construct explicit bulk local states in both flat and AdS spacetimes. We begin by considering the three-dimensional flat Minkowski space with the metric
\be
ds^2=-dt^2+dx^2+dy^2.
\ee
By symmetry, a bulk local state of the scalar excitation inserted at the origin should satisfy the condition (See Ref.\,\cite[eq.\,(2.29)]{Hao:2025btl})
\begin{equation}\label{eq: origin equation}
L_{0,\pm1}|\phi(0,0,0)\rangle=0.
\end{equation}
More generally, for a state inserted at a point $(t,0,0)$ on the $x$-axis, the symmetry constraints are modified to
\begin{equation}
(L_1-itM_1)|\phi(t,0,0)\rangle=(L_{-1}+itM_{-1})|\phi(t,0,0)\rangle=L_0|\phi(t,0,0)\rangle=0.
\end{equation}
A solution to these conditions within the induced representation was found in \cite{Hao:2025btl} and takes the form
\begin{equation}\label{eq: t solution}
    |\phi(t,0,0)\rangle=e^{i \xi 
   (t-1)}\sum_k\frac{2^{-k} \xi ^{-k} \left(-\frac{i}{t}\right)^{k+1} }{k!}|k\rangle,
\end{equation}
where
\be
|k\rangle:= L_{-1}^kL_1^k|\xi\rangle.
\ee

Next, we construct the analogous bulk local state in AdS$_3$ and examine its flat limit. We work in global AdS$_3$ coordinates,
\be
 ds^2=-\Big(1+\frac{r^2}{l^2}\Big)dt^2+\frac{1}{1+\frac{r^2}{l^2}}dr^2+r^2d\phi^2.   \label{eq: global coordinate of AdS}
\ee
In the large radius limit $l\to \infty$, this metric reduces to the flat Minkowski metric, with the point $(t,r,\phi)=(t,0,0)$ mapping to $(t,x,y)=(t,0,0)$. The bulk local state at $(t,0,0)$ in AdS$_3$ satisfies
  \be
                 L_0\ket*{\phi(t,0,0)}=0,\quad (L_{\pm 1}\mp it M_{\pm 1})\ket*{\phi(t,0,0)}=0.\label{eq:UR_limit_of_constraint_from_AdS}
            \ee
The explicit answer is given by
\be
  \ket*{\Psi_\text{AdS}(t,0,0)}=\sum_{k=0}^\infty \frac{e^{ik(\pi+\frac{2t}{l})}}{\Gamma (k+1) (2h)_k}\mathcal{L}_{-1}^k\overline{\mathcal{L}}_{-1}^k|h,h\rangle.
\ee
To facilitate the flat limit, we re-express the AdS$_3$/CFT$_2$ descendants in terms of ``flat descendants'' generated by $L_{-1}(l)$ and $L_{1}(l)$ acting on the primary,
            \begin{align} 
                \mathcal{L}_{-1}^k\overline{\mathcal{L}}_{-1}^k|h,h\rangle=\sum_{a=0}^k f_{k,a}\big(L_{-1}(l)\big)^a\big(L_{1}(l)\big)^a|h,h\rangle, \label{eq: AdS descendants and fla descendants}
            \end{align}
where the coefficients are
            \begin{align}   \label{eq: AdS flat descendant coefficient}
               f_{k,a}= \frac{(-1)^{k-a} \Gamma (k+1)^2 \Gamma (k+2h)}{\Gamma (a+1)^2 \Gamma (-a+k+1) \Gamma (a+2h)}.
            \end{align}
Using this expansion, the AdS bulk local state becomes
            \begin{align}
                 \ket*{\Psi_\text{AdS}(t,0,0)}=\sum_{a=0}^\infty\underbrace{\sum_{k=0}^\infty\frac{e^{ik(\pi+\frac{2t}{l})}}{\Gamma (k+1) (2h)_k}f_{k,a}}_{=:\,f_a}\big(L_{-1}(l)\big)^{a}\big(L_{1}(l)\big)^{a}|h,h\rangle.
            \end{align}
The coefficient $f_a$ can be summed in closed form,
            \begin{equation}
                f_a=\sum_{k=0}^\infty \frac{e^{ik(\pi+2\frac{t}{l})}}{\Gamma (k+1) (2h)_k}f_{k,a}=\frac{(-1)^{-a} \, _2\tilde{F}_1\big(1,1;1-a;e^{\frac{2it}{l}}\big)}{\Gamma (a+1)^2 (2h)_a},
            \end{equation}
where
            \begin{equation}
                _2\tilde{F}_1\big(p,q;r;z\big):=\frac{_2F_1\big(p,q;r;z\big)}{\Gamma(r)}=\sum_{n=0}^{\infty}\frac{(p)_n(q)_n}{\Gamma(r+n)}\frac{z^n}{n!} \label{eq: regularized hypergeometric function}
            \end{equation}
is the regularized hypergeometric function.
Taking the $l\to\infty$ limit, one finds
            \begin{equation}
                \lim_{l\to\infty}\frac{f_a}{l}=-\frac{2^{-a-1}m^{-a}(-\frac{i}{t})^{a+1}}{\Gamma (a+1)},\;\;\lim_{l\to\infty}\big(L_{-1}(l)\big)^{a}\big(L_{1}(l)\big)^{a}|h,h\rangle=L_{-1}^{a}L_{1}^{a}|0,\xi\rangle,
            \end{equation}
which leads to
            \begin{equation}
            -\lim_{l\to\infty}\frac{2}{l}\ket*{\Psi_\text{AdS}(t,0,0)}=\sum_{a=0}^\infty\frac{2^{-a}m^{-a}(-\frac{i}{t})^{a+1}}{\Gamma (a+1)}L_{-1}^{a}L_{1}^{a}|0,\xi\rangle=\ket*{\Psi_\text{Flat}(t,0,0)}.
            \end{equation}
Thus, up to an overall factor, the bulk local state in flat space is recovered from the AdS one in the large $l$ limit.

\subsection{Bra state and its flat limit from AdS$_3$ in the flat basis}
\label{subsec:3d_bra_flatlimit}

We now turn to the construction of the corresponding bra states. As noted in Ref.\,\cite{Hao:2025btl}, the bra state obtained by simply taking the flat limit from AdS$_3$ does not scale in the same way as the ket state. Specifically, the ket state rescales as
\be
\ket*{\Psi_{\rm flat}}=\lim_{l \to \infty}-\frac{2}{l}\ket*{\Psi_{\rm AdS}},
\ee
where $\ket*{\Psi_{\rm flat}}$ and $\ket*{\Psi_{\rm AdS}}$ are the states constructed from isometry considerations. In contrast, the Green's function rescales as
\be
G_{\rm flat}=\lim_{l \to \infty} \frac{1}{l}G_{\rm AdS}.
\ee
If we assume $G_{\rm flat}=\braket*{\Psi_{\rm flat}}$ and $G_{\rm AdS}=\braket*{\Psi_{\rm AdS}}$, consistency would require the bra state to scale as
\be
\bra*{\Psi_{\rm flat}}\sim\lim_{l \to \infty}\bra*{\Psi_{\rm AdS}},
\ee
up to an $\mathcal{O}(1)$ factor. This is incompatible with the standard Hermitian conjugate relation, which demands that bra and ket states scale identically with $l$.

In this subsection, we resolve this apparent contradiction. The key observation is that the inner product itself diverges in the flat limit, and an appropriate rescaling must be applied. We first illustrate this phenomenon in AdS$_3$.

In AdS$_3$, the bra state is defined by the usual Hermitian conjugate,
\be
\bra*{\Psi_{\rm AdS}}:= (\ket*{\Psi_{\rm AdS}})^\dagger.
\ee
One can verify that the inner product of such states correctly reproduces the AdS$_3$ Green's function \cite{Miyaji:2015fia},
\be
\langle\Psi_{\rm AdS}(x)|\Psi_{\rm AdS}(x')\rangle=G_{\rm AdS}(x,x').
\ee
Naively taking the flat limit would suggest
\be
G_{\rm flat}(x,x')=\langle\Psi_{\rm flat}(x)|\Psi_{\rm flat}(x')\rangle\overset{?}{=}\lim_{l\to\infty}\frac{4}{l^2}\langle\Psi_{\rm AdS}(x)|\Psi_{\rm AdS}(x')\rangle=\lim_{l\to\infty}\frac{4}{l^2}G_{\rm AdS}(x,x').
\ee
However, we established in Ref.\,\cite{Hao:2025btl} that
\begin{equation}
    G_{\text{flat}}(x,x')=\lim_{l\to\infty}\frac{1}{l}G_{\text{AdS}}(x,x').
\end{equation}
The mismatch in scaling indicates that the bra state should be treated as an $\mathcal{O}(l^0)$ quantity in the flat limit. This motivates the introduction of a dual basis. As already proposed in Ref.\,\cite{Hao:2025btl}, we consider the same dual basis here. Our new insight is that the non-commutativity of the flat limit and the Hermitian conjugate operation arises because the inner product diverges.

It should be noted that the overall normalization of Green's functions is convention-dependent. We follow the conventions of Ref.\,\cite{Berenstein:1998ij}, which slightly differ from those used above. This does not affect the physical conclusions.

\subsubsection{Why does the bra state scale as $\mathcal{O}(l^0)$?} \label{subsubsec:scaling}
\subsubsection*{Toy model: Divergence of the Gram matrix}
 
To clarify the issue, let us first consider a simple qubit toy model to build intuition before proceeding to the case for AdS$_3$. Define the basis states
 \be\ket*{0}=
\begin{pmatrix}
 1\\0
\end{pmatrix},\quad \ket*{1}=\begin{pmatrix}
 0\\1
\end{pmatrix}.
 \ee
The Gram matrix is chosen to be
 \be
 G=
\begin{pmatrix}
  1 & 0 \\
  0 & l^2
\end{pmatrix},
 \ee
so that $\langle 0|0\rangle=1$ and $\langle 1|1\rangle=l^2$. The raising operator is
 \be
 L=
\begin{pmatrix}
  0 & 0 \\
  1 & 0
\end{pmatrix},
 \ee
with action $L\ket*{0}=\ket*{1}$ and $L\ket*{1}=0$.
The bra state is defined as $\bra*{v}=v^\dagger G(l)$, giving
\be
\bra*{0}=\begin{pmatrix}
  1 & 0 
\end{pmatrix},\quad \bra*{1}=\begin{pmatrix}
  0 & l^2
\end{pmatrix}.
\ee
Note that $\bra*{1}$ diverges as $l\rightarrow\infty$.

 \subsubsection*{Dual basis}
A dual basis ${}^\vee\!\bra*{i}$ can be introduced such that ${}^\vee\!\langle i|j\rangle=\delta_{ij}$ \cite[subsection 2.3]{Hao:2025btl}. Equivalently,
\be
{}^\vee\!\bra*{i}:=\sum_k (G^{-1})_{ik}\bra*{k}.
\ee
In our toy model, this yields
 \be
{}^\vee\!\bra*{0}=\begin{pmatrix}
  1 & 0 
\end{pmatrix},\quad {}^\vee\!\bra*{1}=\begin{pmatrix}
  0 & \frac{1}{l^2} 
\end{pmatrix}.
 \ee
The conjugate of an operator $A$ with respect to this dual basis is defined as
 \be
A^{\dagger(l)}=G^{-1}A^\dagger G.
 \ee
For the raising operator $L$, we obtain
 \be
L^{\dagger(l)}=\begin{pmatrix}
  0 & l^2 \\
  0 & 0
\end{pmatrix}.
 \ee
Rescaling this operator as $M\equiv\frac{1}{l^2}L^{\dagger(l)}=\begin{pmatrix}
  0 & 1 \\
  0 & 0
\end{pmatrix}$ yields a well-behaved raising operator in the dual basis,
\be
{}^\vee\!\bra*{0}M={}^\vee\!\bra*{1}.
\ee
Thus, in the large $l$ limit, the original operator $L$ is effectively replaced by $M$ when acting on the dual basis. This structure mirrors what we will encounter in the AdS context and its flat limit.

\subsubsection{AdS$_3$ case}
Let us apply the above ideas to AdS$_3$. For simplicity, we consider states up to level two first. Define the ket states
\be
\ket*{e_0}:=\ket*{h,h},\quad \ket*{e_1}:=\mathcal{L}_{-1}\bar{\mathcal{L}}_{-1}\ket*{h,h},\quad \ket*{e_2}:=\mathcal{L}_{-1}^2\bar{\mathcal{L}}_{-1}^2\ket*{h,h}.
\ee
Their inner products are
\be
\langle h|\mathcal{L}_{1}^n\mathcal{L}_{-1}^n|h\rangle=(2h)_nn!,\quad \langle e_m|e_n\rangle=\delta_{mn}G_n,\quad G_n=((2h)_nn!)^2,
\ee
where $G_n$ includes contributions from both holomorphic and anti-holomorphic sectors.
These AdS basis states can be expanded in terms of a ``flat'' basis defined using the generators $L_n(l)$ and $M_n(l)$,
\be
\ket*{f_a}:=L_{-1}^a(l)L_1^a(l)\ket*{h,h},
\ee
noting that $L_n(l)$ and $M_n(l)$ are $l$-dependent and reduce to the flat generators $L_n$ and $M_n$ only in the $l\to\infty$ limit, as noted earlier.
The expansion coefficients \eqref{eq: AdS flat descendant coefficient} are
\be
\ket*{e_k}=\sum_{a=0}^kf_{k,a}\ket*{f_a},\quad f_{k,a}=\frac{(-1)^{k-a}\Gamma(k+1)^2\Gamma(k+2h)}{\Gamma(a+1)^2\Gamma(k-a+1)\Gamma(a+2h)}.
\ee
Up to level two, they can be explicitly written down as
\be
\begin{pmatrix}
 \ket*{e_0}\\ \ket*{e_1}\\ \ket*{e_2}
\end{pmatrix}=F\ \begin{pmatrix}
 \ket*{f_0}\\ \ket*{f_1}\\ \ket*{f_2}
\end{pmatrix},\quad F=\begin{pmatrix}
  1 & 0 &0\\
  -2h & 1 &0\\
  4h(2h+1)&-4(2h+1)&1
\end{pmatrix}.
\ee
We now introduce a dual basis for the flat states, defined by the relation
\be
{}^\vee\!\langle f_a|f_b\rangle=\delta_{ab}.
\ee
Given that $\langle e_m|e_n\rangle=\delta_{mn}G_n$, we find
\be
{}^\vee\!\bra*{f_n}=\frac{1}{G_n}\bra*{e_n}F.
\ee
The explicit expressions for the dual basis states are
\begin{align}
    {}^\vee\!\bra*{f_0}&=\bra*{e_0}-\frac{1}{2h}\bra*{e_1}+\frac{1}{4h(2h+1)}\bra*{e_2},\\
     {}^\vee\!\bra*{f_1}&=\frac{1}{4h^2}\bra*{e_1}-\frac{1}{4h^2(2h+1)}\bra*{e_2},\\
      {}^\vee\!\bra*{f_2}&=\frac{1}{16h^2(2h+1)^2}\bra*{e_2}.
\end{align}
Crucially, these dual basis states scale as $\mathcal{O}(l^0)$. For example,
\begin{align}
\bra*{e_1}&=\bra*{h,h}\mathcal{L}_{1}\bar{\mathcal{L}}_1\nn\\
&=\bra*{h,h}\frac{(L_1(l)+lM_1(l))(L_{-1}(l)-lM_{-1}(l))}{4}\nn\\
&\sim -\frac{l^2}{4}\bra*{h,h}M_1(l)M_{-1}(l)+\mathcal{O}(l),  
\end{align}
leading to
\be
{}^\vee\!\bra*{f_1}\sim\frac{1}{4h^2}\bra*{e_1}\sim-\frac{1}{4\xi^2}\bra*{h,h}M_1(l)M_{-1}(l),
\ee
which is indeed $\mathcal{O}(1)$ and matches the result in Ref.\,\cite[eq.\,(2.55)]{Hao:2025btl}. We emphasize that in the $l\to\infty$ limit, $L_n(l)\to L_n$ and $M_n(l)\to M_n$.

To summarize, the Gram matrix (inner product) diverges in the flat limit. Introducing a dual basis, which involves multiplying by the inverse of the inner product, removes this divergence at each level. The transformation of states under the large $l$ limit can be tracked, and the resulting basis coincides with the flat basis in the limit. In the following, we will compute Green's functions using this dual basis. It is important to note that the dual basis is simply a choice of basis; we are \textit{not} altering the definition of Hermitian conjugation.

\subsubsection{Bra state of the bulk local state}

We now analyze the behavior of the bra of the bulk local state in the dual basis. With the dual basis defined by $\,{}^\vee\!\bra*{f_a(l)}f_b(l)\rangle=\delta_{ab}$, the completeness relation reads
\begin{equation}
\mathbf 1=\sum_{a=0}^{\infty} \ket*{f_a(l)}\,{}^\vee\!\bra*{f_a(l)}.
\label{eq:identity_f_dual}
\end{equation}
Inserting this identity, the Hermitian bra of the bulk local state can be expanded as
\begin{equation}
\langle\Psi_{\mathrm{AdS}_3}^E(\tau)|
=\sum_{a=0}^\infty \psi_a(l,\tau)\;{}^\vee\!\bra*{f_a(l)},
\qquad
\psi_a(l,\tau)=\langle\Psi_{\mathrm{AdS}_3}^E(\tau)|f_a(l)\rangle.
\end{equation}
Below, we use the Euclidean signature to converge the series since the factor regarding the time evolution $e^{-\tau/l}$ leads to the convergence. The computation in this paper can be justified for $\tau\in \mathbf{C}$ with $\Re[\tau]>0$. By setting $\tau=\epsilon+i t$, where $\epsilon>0$ and $t$ are regulator and Lorentzian time, respectively, we can recover the bulk local state in the Lorentzian signature. This is nothing but the usual $i\epsilon$ prescription. This is true in our free theory case. Notice that however, if we include interactions and quantum $1/N$ corrections, the flat limit and Wick rotation does not commute since the pole and branch cut structure is modified in the flat limit \cite{Fitzpatrick:2011hu}.

Using the inverse form, which can be verified via the binomial inverse theorem,
\begin{equation}
|f_a(l)\rangle=\sum_{n=0}^{a} g_{a,n}\,|e_n\rangle,
\qquad
g_{a,n}
=\frac{\Gamma(a+1)^2\Gamma(a+2h)}
{\Gamma(n+1)^2\Gamma(a-n+1)\Gamma(n+2h)},
\label{eq:AdS3_Finv_matrix}
\end{equation}
we compute the coefficients
\begin{align}
\psi_a(l,\tau)
&
=e^{-2h\tau/l}\sum_{n=0}^{a}\frac{(-q)^n}{n!(2h)_n}\,g_{a,n}\,G_n \nn\\
&=G_0\,e^{-2h\tau/l}\;a!\,(2h)_a\,(1-q)^a,
\label{eq:AdS3_bra_coeff_dual_f}
\end{align}
where $q=e^{-2\tau/l}$.
In the flat limit $l\to\infty$, with $2h=\xi l+\mathcal O(1)$ and $1-q=1-e^{-2\tau/l}\sim 2\tau/l$, we obtain a finite result
\begin{equation}
\psi_a(l,\tau)\sim G_0\,e^{-\xi\tau}\;a!\,(2\xi\tau)^a
=\mathcal O(l^0)\qquad(l\to\infty,\ a\ \text{fixed}).
\end{equation}
Thus, the bra of the bulk local state has a smooth flat limit without any additional $l$-dependent rescaling.
The flat space bra state is therefore given by
\be
\bra*{\Psi^E_{\rm flat}(\tau)}=\sum_{a=0}^\infty G_0e^{-\xi\tau}a!(2\xi\tau)^a \,\,{}^\vee\!\bra*{f_a}.
\label{flatdual}
\ee
In Ref.\,\cite[eq.\,(2.63)]{Hao:2025btl}, based on the condition ${}^\vee\!\bra*{0}L_{\pm1,0}=0$, we anticipated
\be
\bra*{\Psi_{\rm flat}(0,0,0)}\propto {}^\vee\!\bra*{0}.
\ee
Taking $\tau\to 0$ in eq.\,\eqref{flatdual}, the only term that survives is $a=0$, confirming this expectation.

The time evolution in eq.\,\eqref{flatdual} can also be derived purely from the flat space algebra. In Ref.\,\cite{Hao:2025btl}, the inner product
\be
f_{0,k}(a):={}^\vee\!\langle0|\;e^{aM_0}\;|k\rangle
\ee
was computed using recursion relations, yielding
\be
f_{0,k}(a)=\Gamma(k+1)\,(-2a\xi)^k\,e^{a\xi}.
\ee
After Wick-rotating to Euclidean time ($a=-\tau$), the flat bra can be defined as
\be
\bra*{\Psi^E_{\rm flat}(\tau)}:=
G_0\;{}^\vee\!\bra*{0}\;e^{-\tau M_0}.
\ee
Expanding in the dual basis by inserting the complete set $\sum_k \ket*{k}\,{}^\vee\bra*{k}$, we obtain
\begin{align}
    \bra*{\Psi^E_{\rm flat}(\tau)}&=\sum_{k=0}^\infty G_0\;{}^\vee\!\bra*{0}\;e^{-\tau M_0}\ket*{k}\,{}^\vee\bra*{k}\nn\\
    &=\sum_{k=0}^\infty G_0 e^{-\xi\tau}k!(2\xi\tau)^k\,{}^\vee\bra*{k},
\end{align}
which coincides with eq.\,\eqref{flatdual}) after identifying ${}^\vee\!\bra*{f_k}={}^\vee\bra*{k}$.

\subsubsection{Bulk local state in ket}
We now examine the behavior of the bulk local state in the ket sector when expanded in the $\ket*{f_n(l)}$ basis.
A direct computation gives
\begin{equation}
|\Psi_{\rm AdS_3}^E(\tau)\rangle=e^{-\frac{2h\tau}{l}}\sum_{a=0}^{\infty}\psi_a(l,\tau)\,|f_a(l)\rangle,
\qquad
\psi_a(l,\tau)\equiv \sum_{k=a}^{\infty}
\frac{(-)^ke^{-\frac{2\tau k}{l}}}{\Gamma(k+1)\,(2h)_k}\;f_{k,a}.
\label{eq:AdS3_bulk_local_fa_expand_def}
\end{equation}
The coefficient can be simplified to the closed form
\begin{equation}
\psi_a(l,\tau)
=
\frac{(-1)^a}{a!\,(2h)_a}\;
\frac{q^a}{(1-q)^{a+1}},
\qquad
q=e^{-\frac{2\tau}{l}}.
\label{eq:AdS3_psi_closed}
\end{equation}
Now consider the flat limit.
Using the asymptotic behavior
\begin{equation}
2h=1+\sqrt{1+\xi^2l^2}=\xi l+\mathcal{O}(1),
\qquad
(2h)_a=(\xi l)^a\bigl(1+\mathcal{O}(l^{-1})\bigr),
\label{eq:2h_asympt}
\end{equation}
we find that the coefficients grow linearly with $l$,
\begin{equation}
\psi_a(l,\tau)
=
\frac{l}{2^{a+1}}\;
\frac{(-1)^a\xi^{-a}}{\Gamma(a+1)}\;
\left(\frac{1}{\tau}\right)^{a+1}
\Bigl(1+\mathcal{O}(l^{-1})\Bigr),
\quad (l\to\infty,\ a\ \text{fixed}).
\label{eq:AdS3_psi_large_l}
\end{equation}
Hence, for fixed level $a$, each coefficient is of order $l$. This leads to the rescaled ket state
\begin{equation}
-\frac{2}{l}\,|\Psi^E_{\rm AdS}(\tau)\rangle
\ \xrightarrow[l\to\infty]{}\
\ket*{\Psi^E_{\rm flat}(\tau)}\equiv e^{-\xi\tau}\sum_{a=0}^{\infty}
\frac{2^{-a}\xi^{-a}}{\Gamma(a+1)}
\left(-\frac{1}{\tau}\right)^{a+1}
\,|f_a\rangle,
\label{eq:AdS3_to_flat_bulk_local_coeff}
 \end{equation}
which exactly matches the flat space bulk local state expansion in the induced representation. The linear growth in $l$ arises from rewriting the AdS basis in terms of the flat basis.

 Notice that the AdS$_3$ case is special since eq.\,\eqref{eq:AdS3_bulk_local_fa_expand_def} reduces to the geometric series. The flat limit and the summation over all the descendant modes commute in this case, leading to the complete matching of the bulk local state at each level. This is not true in higher dimensions since the coefficient in eq.\,\eqref{eq:AdS3_bulk_local_fa_expand_def} grows as we increase the level. We will see this problem in the later sections.

\subsubsection{Flat Green's function}
\label{subsec:3d_green}

Having constructed both the bra and ket states in the flat limit, we now compute the Green's function from their inner product. To define the inner product appropriately, we employ the Osterwalder-Schrader (OS) reflection, which involves a time reflection combined with Hermitian conjugation,
\be
{}_{\rm OS}\bra*{\phi(\tau)}\equiv (\ket*{\phi(-\tau)})^\dagger.
\ee

Taking the inner product of the flat bra state \eqref{flatdual} and the flat ket state \eqref{eq:AdS3_to_flat_bulk_local_coeff}, we obtain
\begin{align}
{}_{\rm OS}\langle\Psi^E_{\rm flat}(\tau_1)|\Psi^E_{\rm flat}(\tau_2)\rangle&=\sum_{a=0}^\infty\sum_{b=0}^{\infty} G_0 e^{-\xi(-\tau_1+\tau_2)}a!(-2\xi\tau_1)^a \,\,
\frac{2^{-b}\xi^{-b}}{\Gamma(b+1)}
\left(-\frac{1}{\tau_2}\right)^{b+1}
{}^\vee\!\langle f_a|f_b\rangle\nn\\
&=-\frac{G_0 e^{-\xi(-\tau_1+\tau_2)}}{\tau_2}\sum_{a=0}^\infty \left(\frac{\tau_1}{\tau_2}\right)^a\nn\\
&=-\frac{G_0e^{-\xi(\tau_2-\tau_1)}}{\tau_2-\tau_1},\quad (|\tau_2|>|\tau_1|)
\end{align}
which correctly reproduces the flat space Green's function up to an overall normalization. Although this procedure yields the correct Green's function, the ket and bra used here are not related by standard Hermitian conjugation because the flat limit involves different rescaling behaviors for the bra and ket sectors. This is not a contradiction, as bulk local states are non-normalizable, reflecting their truly localized nature.

\subsubsection{Green's function in AdS$_3$ and flat limit}

We now demonstrate that the same result can be obtained by starting from the AdS$_3$ bulk local states and taking the flat limit, confirming the consistency of the dual basis approach.
\subsubsection*{Direct computation in the descendant basis}
To ensure convergence of the series, we Wick-rotate to Euclidean signature via $t=i\tau$. The bulk local state at $(\tau,0,0)$ then reads
\begin{equation}
\ket*{\Psi_{\rm AdS_3}^E(\tau)}
=
e^{-\frac{2h\tau}{l}}
\sum_{n=0}^\infty
\frac{(-1)^ne^{-\frac{2n\tau}{l}}}{n!\,(2h)_n}\,\ket*{e_n},\quad \ket*{e_n}:= \mathcal{L}_{-1}^n\bar{\mathcal{L}}_{-1}^n\ket*{h,h}.
\label{eq:AdS3_bulk_ket}
\end{equation}
The Green's function is computed as the OS-reflected inner product,
\begin{align}
G_{\rm AdS_3}^E(\tau_2-\tau_1)
&\equiv \langle \Psi_{\rm AdS_3}^E(\tau_1)|\Psi_{\rm AdS_3}^E(\tau_2)\rangle\nn\\
&=e^{-\frac{2h(\tau_2-\tau_1)}{l}}\sum_{n\ge0}
\frac{e^{-\frac{2n(\tau_2-\tau_1)}{l}}}{n!(2h)_n}\frac{1}{n!(2h)_n}\,G_n\nn\\
&=G_0\,e^{-\frac{2h(\tau_2-\tau_1)}{l}}\sum_{n\ge0}e^{-\frac{2n(\tau_2-\tau_1)}{l}}\nn\\
&=G_0\,e^{-\frac{2h(\tau_2-\tau_1)}{l}}\,\frac{1}{1-e^{-2(\tau_2-\tau_1)/l}}.
\label{eq:AdS3_overlap_geometric}
\end{align}
With $2h=\xi l+\mathcal O(1)$ and taking $l\to\infty$, using $1-e^{-2\tau/l}\sim 2\tau/l$, we obtain
\begin{equation}
G_{\rm flat}\equiv\lim_{l\to\infty}-\frac{2}{l}G_{\rm AdS_3}^E(\tau)=-G_0\,\frac{e^{-\xi(\tau_2-\tau_1)}}{\tau_2-\tau_1},
\end{equation}
where the prefactor $-\frac{2}{l}$ accounts for the scaling of the ket state. Since the inner product itself diverges, the operations of Hermitian conjugation and taking the flat limit do not commute; no additional rescaling of the bra state is required. The flat limit result is consistent with the one obtained above.

\subsubsection*{Inserting the dual basis}
We present an alternative derivation of the inner product using the dual basis.

Inserting the complete set of dual basis states, the AdS Green's function can be written as
\be
G_{\rm AdS_3}^E(\tau)
=
\sum_{a\ge0}{}_{\rm OS}\langle\Psi_{\rm AdS_3}^E(0)|\,f_a(l)\rangle\;{}^\vee\!\bra*{f_a(l)}\Psi_{\rm AdS_3}^E(\tau)\rangle.
\ee
The bulk local state at $\tau=0$ satisfies the conditions
\begin{equation}
L_{0,\pm1}(l)\ket*{\Psi_{\rm AdS_3}^E(0)}=0,
\end{equation}
and consequently $\bra*{\Psi_{\rm AdS_3}^E(0)}L_{0,\pm1}(l)=0$. Therefore, only the $a=0$ term survives in the sum,
\begin{align}
{}_{\rm OS}\langle\Psi_{\rm AdS_3}^E(0)|\,f_a(l)\rangle&={}_{\rm OS}\langle\Psi_{\rm AdS_3}^E(0)|\,L_{-1}^a(l)L_{1}^a(l)\ket*{h,h}\nn\\
&={}_{\rm OS}\langle\Psi_{\rm AdS_3}^E(0)|\,h,h\rangle\,\delta_{a,0}.
\end{align}
The Green's function thus reduces to
\begin{align}
    G_{\rm AdS_3}^E(\tau)
&=
\langle\Psi_{\rm AdS_3}^E(0)|f_0(l)\rangle\;{}^\vee\!{}_{\rm OS}\langle f_0(l)|\,\Psi_{\rm AdS_3}^E(\tau)\rangle\nn\\
&=G_0 e^{-\frac{2h\tau}{l}}\sum_{k\geq0}e^{-\frac{2k\tau}{l}}\nn\\
&=G_0\frac{e^{-\frac{2h\tau}{l}}}{1-e^{-\frac{2\tau}{l}}},
\end{align}
reproducing eq.\,\eqref{eq:AdS3_overlap_geometric}. This computation confirms that the dual basis constructed from the flat generators $L_n(l)$ provides a complete and convenient basis for analyzing the flat limit.


    \section{Higher dimensional generalization} 
    \label{sec:bulk_local_4d}
Previously we have focused on the three-dimensional case to illustrate the construction of bulk local states and their flat limit. We now extend this analysis to general spacetime dimension $d+1$. For simplicity and clarity, we will first present the construction in four spacetime dimensions ($d=3$) as a concrete example. In Subsection \ref{subsec:arbitrary_dim}, we provide the results for arbitrary dimension, emphasizing that the main structure remains essentially unchanged, with only some dimension-dependent coefficients appearing, where the complete expressions are summarized.
\subsection{Algebra and representation} \label{sec:algebra_representation}

In this subsection, we lay the algebraic foundation for bulk reconstruction in flat holography. We begin by reviewing the isometries of the four-dimensional bulk spacetimes, both Minkowski$_4$ and AdS$_4$, and establish the precise connection between their symmetry algebras via the flat limit. This leads to a representation of the Poincar\'e group that is induced from a one-dimensional little group, suitable for describing bulk states. We then introduce the dual boundary theory, a three-dimensional Carrollian conformal field theory, and detail its symmetry algebra, namely the BMS$_4$ algebra, along with its finite transformations. Finally, we translate the bulk representation conditions into constraints on boundary primary states, setting the stage for the construction of bulk operators.

\subsubsection{Bulk isometries: From AdS$_4$ to Flat$_4$}
\label{subsec:bulk_isometries}

The isometries of four-dimensional Minkowski space (Flat$_4$) are generated by the Poincar\'e algebra $\mathfrak{iso}(1,3) \cong \mathfrak{so}(1,3) \ltimes \mathbb{R}^4$.  In the standard Cartesian coordinates $(x^0, x^1, x^2, x^3)$ with metric $\eta_{\mu\nu}= \mathrm{diag}(-1,1,1,1)$, the generators are translations $P_\mu$ and Lorentz rotations $J_{\mu\nu}$, which satisfy the commutation relations:
\begin{align}
    [J_{\mu\nu},J_{\rho\sigma}] &= i(\eta_{\nu\rho}J_{\mu\sigma}-\eta_{\mu\rho}J_{\nu\sigma}-\eta_{\nu\sigma}J_{\mu\rho}+\eta_{\mu\sigma}J_{\nu\rho}), \label{eq:poincare_lorentz}\\
    [J_{\mu\nu},P_{\rho}] &= i(\eta_{\mu\rho}P_{\nu}-\eta_{\nu\rho}P_{\mu}), \label{eq:poincare_mix}\\
    [P_{\mu},P_{\nu}] &= 0. \label{eq:poincare_trans}
\end{align}
For later convenience, it is useful to split the Lorentz generators into rotations $J_a$ and boosts $K_a$ ($a=1,2,3$) by defining
\begin{equation}
    K_a := J_{0a}, \qquad J_a := -\frac{1}{2}\epsilon_{abc}J_{bc}.
\end{equation}
In this basis, the algebra \eqref{eq:poincare_lorentz} -- \eqref{eq:poincare_trans} take the form
\begin{align}
    [J_a,J_b] &= i\epsilon_{abc}J_c, \quad [K_a,K_b] = -i\epsilon_{abc}J_c, \quad [J_a,K_b] = i\epsilon_{abc}K_c, \label{eq:sl2c_subalgebra}\\
    [P_0, K_a] &= iP_a, \quad [P_a, K_b] = i\delta_{ab} P_0, \quad [J_a, P_b] = i\epsilon_{abc}P_c. \label{eq:poincare_mix_explicit}
\end{align}
The subalgebra \eqref{eq:sl2c_subalgebra} is isomorphic to $\mathfrak{sl}(2,\mathbb{C})$.

We now consider AdS$_4$ with radius $l$, which can be embedded as a hyperboloid $-x_{-1}^2-x_0^2+x_1^2+x_2^2+x_3^3=-l^2$ in $\mathbb{R}^{2,3}$. Its isometry algebra is $\mathfrak{so}(2,3)$, which is isomorphic to the conformal algebra in three dimensions, $\mathfrak{conf}(1,2)$.  In a basis adapted to the AdS/CFT correspondence, the generators are the dilatation $\mathcal{H}$, translations $\mathcal{P}_a$, special conformal transformations $\mathcal{K}_a$, and rotations $\mathcal{M}_{ab}$, satisfying
\begin{align}
    [\mathcal{H},\mathcal{P}_a] &= \mathcal{P}_a, \quad [\mathcal{H},\mathcal{K}_a] = -\mathcal{K}_a, \quad [\mathcal{K}_a,\mathcal{P}_b] = 2(\delta_{ab}\mathcal{H}-\mathcal{M}_{ab}), \label{eq:conf_algebra_1}\\
    [\mathcal{K}_a,\mathcal{M}_{bc}] &= \delta_{ab}\mathcal{K}_c-\delta_{ac}\mathcal{K}_b, \quad [\mathcal{P}_a,\mathcal{M}_{bc}] = \delta_{ab}\mathcal{P}_c-\delta_{ac}\mathcal{P}_b, \label{eq:conf_algebra_2}\\
    [\mathcal{M}_{ab},\mathcal{M}_{cd}] &= \delta_{bc}\mathcal{M}_{ad}+\delta_{ad}\mathcal{M}_{bc}-\delta_{ac}\mathcal{M}_{bd}-\delta_{bd}\mathcal{M}_{ac}. \label{eq:conf_algebra_3}
\end{align}

The connection between the AdS$_4$ and Flat$_4$ isometries is established via the flat limit, $l \to \infty$. By considering appropriate linear combinations of the AdS generators that are rescaled with powers of $l$, we can retrieve the Poincar\'e algebra in this limit.  We define the following $l$-dependent generators
\begin{equation}
    P_0(l) := \frac{1}{l}\mathcal{H}, \qquad P_a(l) := -\frac{1}{2l}(\mathcal{P}_a+\mathcal{K}_a), \qquad K_a(l) := -\frac{i}{2}(\mathcal{P}_a-\mathcal{K}_a), \qquad J_{ab}(l) := i\mathcal{M}_{ab}. \label{eq:flat_limit_generators}
\end{equation}
In the limit $l \to \infty$, one can verify that these generators satisfy the Poincar\'e algebra \eqref{eq:poincare_lorentz} -- \eqref{eq:poincare_trans} by comparing their explicit differential operator realization provided in Appendix \ref{app:explicit_generators}. For instance, $P_\mu := \lim_{l\to\infty} P_\mu(l)$ become the translation generators, and $J_{ab} := \lim_{l\to\infty} J_{ab}(l)$ become the rotation generators, with the boosts recovered as $K_a := \lim_{l\to\infty} K_a(l)$.

This limiting procedure also dictates how the representations of the two algebras are related.  In AdS$_4$/CFT$_3$, a scalar primary state $|\Delta\rangle$ is defined by the highest weight conditions
\begin{equation}
    \mathcal{K}_a|\Delta\rangle = 0, \qquad \mathcal{H}|\Delta\rangle = \Delta|\Delta\rangle. \label{eq:ads_highest_weight}
\end{equation}
Using the definitions \eqref{eq:flat_limit_generators}, these conditions become
\begin{equation}
    P_a(l)|\Delta\rangle = -\frac{i}{l}K_a(l)|\Delta\rangle, \qquad P_0(l)|\Delta\rangle = \frac{\Delta}{l}|\Delta\rangle. \label{eq:intermediate_rep}
\end{equation}
Taking the flat limit $l\to\infty$, and assuming the existence of
\begin{equation}
    |\xi\rangle := \lim_{l\to\infty} |\Delta\rangle, \qquad \xi := \lim_{l\to\infty} \frac{\Delta}{l},
\end{equation}
we arrive at the defining conditions for a state $|\xi\rangle$ in the flat space bulk
\begin{align}
    P_0 |\xi\rangle &= \xi |\xi\rangle, \label{eq:flat_primary_energy}\\
    P_a |\xi\rangle &= 0, \label{eq:flat_primary_momentum}\\
    J_a |\xi\rangle &= 0. \label{eq:flat_primary_spin}
\end{align}
These equations characterize a massive scalar particle at rest, with $P_\mu$ being the 4-momentum operator and $J_a$ the rotation generators. The state $|\xi\rangle$ is thus a primary state of an ``induced" representation of the Poincar\'e group, where all other momentum eigenstates are generated by acting the boost generators $K_a$. This representation will be the fundamental building block for constructing bulk fields.

\subsubsection{Boundary Carrollian conformal algebra}
\label{subsec:boundary_algebra}

The putative dual field theory for asymptotically flat spacetimes in four dimensions is a three-dimensional Carrollian conformal field theory (also referred to as a BMS field theory, BMSFT$_3$). Its symmetries are generated by the BMS$_4$ algebra, which is the asymptotic symmetry algebra of Flat$_4$ at null infinity.  The global part of this algebra, which is relevant for our discussion, consists of the generators $\{L_{-1},L_0,L_1, \bar L_{-1},\bar L_0,\bar L_1, M_{0,0}, M_{1,0}, M_{0,1}, M_{1,1}\}$. In coordinates $(z,\bar z, y)$, where $z$ and $\bar z$ are complex coordinates on the celestial sphere and $y$ is a coordinate along the null generators, these generators are realized as differential operators
\begin{align}
    L_n &= -z^{n+1}\partial_z - \frac{1}{2}(n+1)z^n y \partial_y, \quad n = -1,0,1,\\
    \bar L_n &= -\bar z^{n+1}\partial_{\bar z} - \frac{1}{2}(n+1)\bar z^n y \partial_y, \quad n = -1,0,1,\\
    M_{r,s} &= -z^r \bar z^s \partial_y, \quad r,s = 0,1.
\end{align}
They obey the global BMS$_4$ algebra:
\begin{align}
    [L_m, L_n] &= (m-n)L_{m+n}, \quad [\bar L_m, \bar L_n] = (m-n)\bar L_{m+n}, \label{eq:bms_global_virasoro}\\
    [L_m, M_{r,s}] &= \left(\frac{m+1}{2} - r\right) M_{r+m, s}, \quad [\bar L_m, M_{r,s}] = \left(\frac{m+1}{2} - s\right) M_{r, s+m}, \label{eq:bms_global_mix}\\
    [M_{r,s}, M_{p,q}] &= 0. \label{eq:bms_global_trans}
\end{align}
Here, $L_n$ and $\bar L_n$ generate superrotations (conformal transformations on the sphere), and $M_{r,s}$ generate supertranslations. The finite transformations of the coordinates under these generators can be derived by solving the first-order differential equations $\frac{d}{d\lambda}(z,\bar z, y) = X(z,\bar z, y)$ for a generator $X$ with parameter $\lambda$. The results for the global generators are summarized in Table \ref{tab:finite_transformations}.

\begin{table}[h!]
\centering
\begin{tabular}{c|c}
\hline
Generator & Finite Transformation $(z(\lambda), \bar z(\lambda), y(\lambda))$ \\
\hline
$L_{-1} = -\partial_z$ & $(z_0 - \lambda, \bar z_0, y_0)$ \\
$L_{0} = -z\partial_z - \frac{1}{2}y\partial_y$ & $(z_0 e^{-\lambda}, \bar z_0, y_0 e^{-\lambda/2})$ \\
$L_{1} = -z^2\partial_z - z y\partial_y$ & $\left(\frac{z_0}{1+\lambda z_0}, \bar z_0, \frac{y_0}{1+\lambda z_0}\right)$ \\
$\bar L_{-1} = -\partial_{\bar z}$ & $(z_0, \bar z_0 - \lambda, y_0)$ \\
$\bar L_{0} = -\bar z\partial_{\bar z} - \frac{1}{2}y\partial_y$ & $(z_0, \bar z_0 e^{-\lambda}, y_0 e^{-\lambda/2})$ \\
$\bar L_{1} = -\bar z^2\partial_{\bar z} - \bar z y\partial_y$ & $\left(z_0, \frac{\bar z_0}{1+\lambda \bar z_0}, \frac{y_0}{1+\lambda \bar z_0}\right)$ \\
$M_{0,0} = -\partial_y$ & $(z_0, \bar z_0, y_0 - \lambda)$ \\
$M_{1,0} = -z\partial_y$ & $(z_0, \bar z_0, y_0 - z_0 \lambda)$ \\
$M_{0,1} = -\bar z\partial_y$ & $(z_0, \bar z_0, y_0 - \bar z_0 \lambda)$ \\
$M_{1,1} = -z\bar z\partial_y$ & $(z_0, \bar z_0, y_0 - z_0\bar z_0 \lambda)$ \\
\hline
\end{tabular}
\caption{Finite transformations for the global BMS$_4$ generators. The initial condition is $(z(0),\bar z(0),y(0)) = (z_0,\bar z_0, y_0)$.}
\label{tab:finite_transformations}
\end{table}

The BMS$_4$ generators are linearly related to the Poincar\'e generators, which is given by
\begin{align}
    M_{0,0} &= -P_0 + P_3, \quad M_{1,1} = -P_0 - P_3,\quad M_{1,0} = P_1 + iP_2,\quad M_{0,1} = P_1 - iP_2, \\
    L_{-1} &= \tfrac{1}{2}(iK_1+K_2+J_1-iJ_2),\quad L_1 = \tfrac{1}{2}(-iK_1+K_2-J_1-iJ_2),\\
    L_0 &= \tfrac{1}{2}(-J_3-iK_3),\quad \bar L_0 = \tfrac{1}{2}(J_3-K_3),\\
    \bar L_{-1} &= \tfrac{1}{2}(iK_1-K_2-J_1-iJ_2),\quad \bar L_1 = \tfrac{1}{2}(-iK_1-K_2+J_1-iJ_2).
\end{align}
This map can be inverted to yield the expressions in Table \ref{tab:poincare_finite_transformations}.

\begin{table}[h!]
\centering
\begin{tabular}{c|c}
\hline
Generator & Finite Transformation $(z(\lambda), \bar z(\lambda), y(\lambda))$ \\
\hline
$P_0 = \frac{1+z\bar z}{2}\partial_y$ & $(z_0,\bar z_0, y_0 + \frac{1+z_0\bar z_0}{2}\lambda)$ \\
$P_1 = -\frac{z+\bar z}{2}\partial_y$ & $(z_0,\bar z_0, y_0 - \frac{z_0+\bar z_0}{2}\lambda)$ \\
$P_2 = -\frac{z-\bar z}{2i}\partial_y$ & $(z_0,\bar z_0, y_0 - \frac{z_0-\bar z_0}{2i}\lambda)$ \\
$P_3 = \frac{z\bar z-1}{2}\partial_y$ & $(z_0,\bar z_0, y_0 + \frac{z_0\bar z_0-1}{2}\lambda)$ \\
\hline
\end{tabular}
\caption{Finite transformations for translation generators, derived from the map to BMS$_4$ generators.}
\label{tab:poincare_finite_transformations}
\end{table}

\subsubsection{Bulk primaries in Carrollian CFTs}
\label{subsec:boundary_conditions}

We now translate the bulk primary conditions \eqref{eq:flat_primary_energy} -- \eqref{eq:flat_primary_spin} for the state $|\xi\rangle$ into constraints involving the boundary BMS$_4$ generators.  Using the linear map between the two sets of generators, the conditions $P_a |\xi\rangle = 0$ directly imply
\begin{equation}
    M_{1,0}|\xi\rangle = (P_1+iP_2)|\xi\rangle = 0, \qquad M_{0,1}|\xi\rangle = (P_1-iP_2)|\xi\rangle = 0. \label{eq:boundary_condition_transverse}
\end{equation}
The energy condition $P_0 |\xi\rangle = \xi |\xi\rangle$ and the condition from $P_3$ (which is not independent but follows from the algebra) combine to give
\begin{equation}
    M_{0,0}|\xi\rangle = (-P_0+P_3)|\xi\rangle = -\xi |\xi\rangle, \qquad M_{1,1}|\xi\rangle = (-P_0-P_3)|\xi\rangle = -\xi |\xi\rangle. \label{eq:boundary_condition_energy}
\end{equation}
The state is thus a simultaneous eigenstate of the two translation generators $M_{0,0}$ and $M_{1,1}$ with equal eigenvalues $-\xi$.

Finally, the condition that the state is a Lorentz scalar, $J_a |\xi\rangle = 0$, translates into a set of relations between the $L_n$ and $\bar L_n$ generators.  From the definitions of $J_a$ in terms of $L_n$ and $\bar L_n$, we obtain
\begin{align}
    (L_{-1} + \bar L_1)|\xi\rangle &= 0, \label{eq:boundary_condition_scalar_1}\\
    (L_1 + \bar L_{-1})|\xi\rangle &= 0, \label{eq:boundary_condition_scalar_2}\\
    (\bar L_0 + i L_0)|\xi\rangle &= 0. \label{eq:boundary_condition_scalar_3}
\end{align}
These conditions are the imprint of the bulk Lorentz invariance on the boundary primary state. Together with eqs.\,\eqref{eq:boundary_condition_transverse} and \eqref{eq:boundary_condition_energy}, they provide a complete characterization of the state $|\xi\rangle$ within the boundary CCFT, which serves as the dual description of the bulk scalar primary. All other states in its representation, corresponding to moving particles with non-zero momentum, are generated by acting with the boost generators $K_a$, or equivalently with the appropriate combinations of $L_n$ and $\bar L_n$. This algebraic dictionary is the cornerstone for the construction of local bulk operators, which we will develop in the subsequent sections.

\subsection{Bulk local states in the AdS$_4$ descendant basis and the flat limit in the momentum basis }
\label{subsec:bulk_local_4d_momentum}
In this subsection, we consider the bulk local state in AdS$_4$ and its flat limit in the momentum basis. The construction relies entirely on Poincar\'{e} symmetry and representation theory, without reference to a specific field-theoretic realization. We derive the state by imposing invariance conditions \eqref{eq:flat_primary_energy} -- \eqref{eq:flat_primary_spin} that localize an operator at a bulk point. In the AdS$_3$ case, we just take the flat limit at each descendant level and find the complete agreement with the flat result. In AdS$_4$, that naive expectation does not work. To see that we first take the flat limit of the Green's function.
\subsubsection{Bulk local state in Flat$_4$ and Green's function}

Let us consider the bulk local state in the momentum basis in Flat$_4$ first.

\subsubsection*{Bulk local state in the flat space}

Let $\ket*{\xi}$ be the static scalar primary, which satisfies
\begin{equation}
P_a\ket*{\xi}=0,\quad P_0\ket*{\xi}=\xi\ket*{\xi},\quad J_a\ket*{\xi}=0.
\end{equation}
Define the momentum basis by boosting $\ket*{\xi}$
\begin{equation}
\label{eq:flat_momket_def}
\ket*{\mathbf p}:=e^{\eta\,p_a K^a}\ket*{\xi},
\quad \tanh\eta=\frac{p}{E},\quad E=\sqrt{\xi^2+p^2},
\end{equation}
where
\begin{equation}
P_0\ket*{\mathbf p}=E\ket*{\mathbf p},\quad P_a\ket*{\mathbf p}=p_a\ket*{\mathbf p},\quad J_a\ket*{\mathbf p}=0.
\end{equation}
The normalization of the momentum basis reads
\begin{equation}
\label{eq:flat_mom_norm}
\braket*{\mathbf p}{\mathbf p'}=(2\pi)^3\,2E\;\delta^{(3)}(\mathbf p-\mathbf p').
\end{equation}

 Next, let us construct the bulk local state in the momentum basis.
At $(t,\mathbf 0)$ the stabilizer condition is (See Appendix \ref{app:poincare_spherical})
\begin{equation}
\label{eq:flat_stabilizer}
(K_a-tP_a)\ket*{\phi(t,\mathbf 0)}=0,\quad J_a \ket*{\phi(t,\mathbf 0)}=0.
\end{equation}
We can expand a general rotation-invariant state as
\begin{equation}
\ket*{\phi(t,\mathbf 0)}=\int \frac{d^3p}{(2\pi)^3\,2E}\;\psi(\mathbf p)\ket*{\mathbf p}.
\end{equation}
Using eq.\,\eqref{eq:flat_stabilizer}, we obtain a first-order equation for $\psi(\mathbf p)$
\begin{equation}
\Big(iE\,\partial_{p_a}-t\,p_a\Big)\psi(\mathbf p)=0,
\end{equation}
whose solution is given by $\psi(\mathbf p)=c\,e^{-itE}$.
Therefore, the solution to eq.\,\eqref{eq:flat_stabilizer} is given by
\begin{equation}
\ket*{\phi(t,\mathbf 0)}=
c\int \frac{d^3p}{(2\pi)^3\,2E}\;e^{-itE}\ket*{\mathbf p}.
\end{equation}
In Euclidean time ($t=-i\tau$), it becomes
\begin{equation}
\label{eq:flat_bulk_ket_E_mom}
\ket*{\phi_E(\tau,\mathbf 0)}=
c\int \frac{d^3p}{(2\pi)^3\,2E}\;e^{-\tau E}\ket*{\mathbf p},
\qquad (\tau>0).
\end{equation}
This indicates that the bulk local state diverges at $\tau=0$, while for $\tau>0$ highly excited states are exponentially suppressed. This is also true for the bulk local state at the center of AdS. It appears to converge at the center, but this is not true since the norm of descendant states diverges as the level increases. 

Since the momentum basis is  orthonormal, the Hermitian conjugate bra is 
\begin{equation}
\bra*{\phi_E(\tau,\mathbf 0)}=(\ket*{\phi_E(\tau,\mathbf 0)})^\dagger
=c^*\int \frac{d^3p}{(2\pi)^3\,2E}\;e^{-\tau E}\bra*{\mathbf p}.
\end{equation}

Let us reproduce the Green's function, which is defined by
\begin{equation}
\label{eq:flat_green_def}
G^E_{\rm flat}(\tau_2-\tau_1):=
{}_{\rm OS}\bra*{\phi_E(\tau_1,\mathbf 0)}\phi_E(\tau_2,\mathbf 0)\rangle.
\quad (\tau_2>\tau_1)
\end{equation}
Using eq.\,\eqref{eq:flat_bulk_ket_E_mom}, we find
\begin{align}
G^E_{\rm flat}(\tau)
&=|c|^2\int \frac{d^3p}{(2\pi)^3\,2E}\;e^{-\tau E}
\nonumber\\
&=\frac{|c|^2}{4\pi^2}\int_0^\infty dp\;\frac{p^2}{E}\,e^{-\tau E},
\qquad E=\sqrt{\xi^2+p^2}.
\label{eq:flat_green_mom_int}
\end{align}
The integral is standard and yields
\begin{equation}
\label{eq:flat_green_Bessel}
G^E_{\rm flat}(\tau)=|c|^2\,\frac{\xi}{4\pi^2\,\tau}\,K_1(\xi\tau).
\end{equation}
Its short distance behavior is given by $K_1(z)\sim 1/z + O(z\log z)$, leading to $G^E_{\rm flat}(\tau)\sim (4\pi^2\tau^2)^{-1}+\cdots$.

\subsubsection{Bulk local state in AdS$_4$ and Green's function}

Let us consider the bulk local state in AdS$_4$. We define a primary state as follows
\be
\mathcal{K}_a\ket*{\Delta}=0,\qquad \mathcal{H}\ket*{\Delta}=\Delta\ket*{\Delta},\qquad \mathcal{M}_{ab}\ket*{\Delta}=0.
\ee
The descendant states are obtained by applying $\mathcal{P}^2$
\be
\ket*{e_n}:= (\mathcal{P}^2)^n\ket*{\Delta},\qquad \mathcal{P}^2:= \mathcal{P}_a\mathcal{P}_a,\qquad n=0,1,2,\dots
\label{eq:e_n_def}
\ee
and their conjugate bras are defined as
\be
\bra*{e_n}:= \bra*{\Delta}(\mathcal{K}^2)^n,\qquad \mathcal{K}^2:= \mathcal{K}_a\mathcal{K}_a.
\ee
Then, we have orthogonal relation
\be
\langle e_m|e_n\rangle=\delta_{mn}\,G_n,\quad G_n=16^n\,n!\,\Bigl(\frac{3}{2}\Bigr)_n\,(\Delta)_n\Bigl(\Delta-\frac12\Bigr)_n\,G_0.
\label{eq:Gram_diag}
\ee
See Appendix \ref{subsubsec:derive_Gn} for detailed derivation of this equality. 

A bulk local state at  $(t,0,0, 0)$ is invariant under the isometries
leaving the point fixed. We Wick-rotate the time to Euclidean time  via $t=-i\tau$ ($\tau>0$) and the 
 stabilizer conditions can be written as
\be
\Bigl(e^{-\tau/l}\mathcal{P}_a - e^{+\tau/l}\mathcal{K}_a\Bigr)\ket*{\Psi_{\rm AdS}^E(\tau,0,0, 0)}=0,
\qquad
\mathcal{M}_{ab}\ket*{\Psi_{\rm AdS}^E(\tau,0,0, 0)}=0.
\label{eq:AdS4_stabilizer_E}
\ee
At $\tau=0$ this reduces to
\be
(\mathcal{P}_a-\mathcal{K}_a)\ket*{\Psi_{\rm AdS}(0,0,0, 0)}=0,
\qquad
\mathcal{M}_{ab}\ket*{\Psi_{\rm AdS}(0,0,0,0)}=0.
\label{eq:AdS4_stabilizer_tau0}
\ee
Taking the conjugation gives the corresponding constraints on the  bra state
\be
\bra*{\Psi_{\rm AdS}(0,0,0, 0)}(\mathcal{K}_a-\mathcal{P}_a)=0,
\qquad
\bra*{\Psi_{\rm AdS}(0,0,0,0)}\mathcal{M}_{ab}=0.
\label{eq:AdS4_stabilizer_bra_tau0}
\ee
A solution to eq.\,\eqref{eq:AdS4_stabilizer_E}  can be expanded as
\begin{equation}\label{eq:ads4_solution_des}
|\Psi^E_{\rm AdS_4}(\tau)\rangle
=
e^{-\Delta \tau/l}
\sum_{k=0}^\infty
\frac{q^k}{4^k\,k!\,(\Delta-\tfrac12)_k}\;
|e_k\rangle,\quad q:=e^{-2\tau/l}.\quad (|q|<1)
\end{equation}
The derivation is summarized in the Appendix \ref{subsubsec:derive_bulk_local_center_PminusK}.

Let us compute the Green's function from the bulk local state. 
AdS$_4$ Euclidean Green's function with different time reads
\begin{equation}
G^E_{\rm AdS_4}(\tau)
\equiv
{}_{\rm OS}\langle \Psi^E_{\rm AdS}(-\tau,{\bf 0})|\,\Psi^E_{\rm AdS}(0,{\bf 0})\rangle=\langle \Psi^E_{\rm AdS}(\tau,{\bf 0})|\,\Psi^E_{\rm AdS}(0,{\bf 0})\rangle.
\quad (\tau>0)
\label{eq:AdS_G_def}
\end{equation}
By direct computation, we obtain
\begin{align}
G^{E}_{\rm AdS_4}(\tau)
&=
G_0\,e^{-\Delta\tau/l}\sum_{n=0}^{\infty} a_n(\Delta)\,q^n,
\qquad
a_n(\Delta):=\frac{\left(\frac32\right)_n(\Delta)_n}{n!\,(\Delta-\frac12)_n},\nn\\
&=G_0\,e^{-\Delta\tau/l}{}_2F_1\left[\Delta,\frac{3}{2},\Delta-\frac{1}{2},e^{-\frac{2\tau}{l}}\right],
\label{app:eq:ads4_exact_series}
\end{align}
which reproduces the result in Ref.\,\cite{Berenstein:1998ij} with the normalization 
\begin{equation}
G_0=\frac{\Gamma(\Delta)}{2\pi^{3/2}\Gamma(\Delta-\frac12)\,l^2}.
\label{app:eq:G0_ads4}
\end{equation}

\subsubsection{Flat limit of Green's function}
In this section, we explain why the flat limit $l\to\infty$ of the Euclidean
two-point function (or equivalently, the overlap of bulk-local states) is \textit{non-uniform} in the
descendant level $n$  in AdS$_4$. Concretely, the limit does not commute with the infinite sum
\begin{equation}
\lim_{l\to\infty}\sum_{n\ge0}G_0\,e^{-\Delta\tau/l}a_n(\Delta)\,q^n\ \neq\ \sum_{n\ge0}\lim_{l\to\infty} G_0\,e^{-\Delta\tau/l}a_n(\Delta)\,q^n.
\end{equation}
In particular, the dominant contribution in the flat limit comes from the scaling window $n\sim l$. We will demonstrate this first in the computation of Green's function and later do it at the state level in the next section.
We will now turn to a detailed discussion of the problem, and readers uninterested in technical details may skip to Subsubsection \ref{subsubsec: flat limit}.

\subsubsection*{Naive sum at each level}
Let us consider  the flat limit
\begin{equation}
\Delta=\xi\,l+\mathcal O(1),\quad \xi>0\,\,\,\, \text{fixed},\quad l\to\infty.
\label{app:eq:flat_limit_scaling}
\end{equation}
Using $\Gamma(\Delta)/\Gamma(\Delta-\frac12)\sim \Delta^{1/2}$, we have
\begin{equation}
G_0\sim \frac{\sqrt{\Delta}}{2\pi^{3/2}\,l^2}
\sim
\frac{\sqrt{\xi}}{2\pi^{3/2}}\,l^{-3/2},\quad (l\to\infty).
\label{app:eq:G0_asympt}
\end{equation}
Let us fix $n$ and take $l\to\infty$. Then, $q^n\to 1$ and $e^{-\Delta\tau/l}\to e^{-\xi\tau}$.
Moreover,
\begin{equation}
\frac{(\Delta)_n}{(\Delta-\frac12)_n}
=
\prod_{k=0}^{n-1}\frac{\Delta+k}{\Delta+k-\frac12}
\xrightarrow[l\to\infty]{} 1,
\qquad (n\ \text{fixed}).
\end{equation}
Hence
\begin{equation}
a_n(\Delta)\xrightarrow[l\to\infty]{}\frac{(\frac32)_n}{n!}\qquad (n\ \text{fixed}),
\end{equation}
and the termwise asymptotics of each summand
\begin{equation}
G_0\,e^{-\Delta\tau/l}a_n(\Delta)\,q^n\to
G_0\,e^{-\xi\tau}\frac{(\frac32)_n}{n!}
\sim
\mathcal O(l^{-3/2})
\qquad (l\to\infty,\ n\ \text{fixed}).
\label{app:eq:termwise_vanish}
\end{equation}
Therefore, for any \textit{fixed} cutoff $N$,
\begin{equation}
\sum_{n=0}^{N}s_n(l)\xrightarrow[l\to\infty]{}0.
\end{equation}
If we  interchange the limit and the infinite sum, we would mistakenly conclude
$
\lim_{l\to\infty}G^{E}_{\rm AdS_4}(\tau)=0
$,
recalling that the correct flat limit is a Bessel function \eqref{eq:flat_green_Bessel}.
This shows that the flat limit is non-uniform in $n$.

The above argument extends: any sublinear window $n<\mathcal{O}(l)$ does not contribute in the limit.
For large $n$, we have the standard asymptotic
\begin{equation}
\frac{(\frac32)_n}{n!}=\frac{\Gamma(n+\frac32)}{\Gamma(\frac32)\Gamma(n+1)}
\sim \frac{1}{\Gamma(\frac32)}\,n^{1/2}.
\label{app:eq:poch_asympt}
\end{equation}
Moreover, in the regime $n\le l^\alpha$ with any fixed $\alpha<1$, we have 
\be
n\ll \Delta\sim l,
\ee
which leads to
\be
\frac{(\Delta)_n}{(\Delta-\frac12)_n}=1+\mathcal O(n/\Delta)=1+\mathcal{O}(1).
\ee
Therefore, the summand can be approximated as
\begin{equation}
a_n(\Delta)\lesssim C\,n^{1/2},\quad (n\le l^\alpha,\ l\to\infty)
\end{equation}
for some constant $C$ independent of $l$.
Substituting these results, we obtain the bound
\begin{equation}
\sum_{n=0}^{\lfloor l^\alpha\rfloor}G_0\,e^{-\Delta\tau/l}a_n(\Delta)\,q^n
\ \lesssim\
l^{-3/2}\sum_{n\le l^\alpha}n^{1/2}
\ \sim\
l^{-3/2}\cdot l^{3\alpha/2}
=
l^{\frac{3}{2}(\alpha-1)}
\ \xrightarrow[l\to\infty]{}\ 0. \quad(\alpha<1)
\label{app:eq:sublinear_vanish}
\end{equation}
Thus, the dominant contribution in the sum must come from
\begin{equation}
n\sim l.
\label{app:eq:scaling_window}
\end{equation}

\subsubsection*{Correct way: Riemann sum}
    Taking the above evaluation into account, we need a new approach to sum over highly excited modes.
Let us define the scaling variable
\begin{equation}
x\equiv\frac{n}{l}\ge0,\qquad n=\lfloor lx\rfloor.
\end{equation}
Rewrite $a_n(\Delta)$ using Gamma functions as
\begin{equation}
a_n(\Delta)=
\frac{\Gamma(n+\frac32)}{\Gamma(\frac32)\Gamma(n+1)}\cdot
\frac{\Gamma(\Delta+n)}{\Gamma(\Delta)}\cdot
\frac{\Gamma(\Delta-\frac12)}{\Gamma(\Delta+n-\frac12)}.
\end{equation}
In the large $l$ limit with $\Delta=\xi l+\mathcal O(1)$ and $n=lx$ (for  $x$ fixed), the standard ratio
asymptotics give
\begin{align}
\frac{\Gamma(n+\frac32)}{\Gamma(n+1)} &\sim n^{1/2}\sim \sqrt{lx},\\
\frac{\Gamma(\Delta+n)}{\Gamma(\Delta+n-\frac12)} &\sim (\Delta+n)^{1/2}\sim \sqrt{l(\xi+x)},\\
\frac{\Gamma(\Delta-\frac12)}{\Gamma(\Delta)} &\sim \Delta^{-1/2}\sim \frac{1}{\sqrt{l\xi}}.
\end{align}
Also using $\Gamma(\frac32)=\sqrt{\pi}/2$, we obtain
\begin{equation}
a_{\lfloor lx\rfloor}(\Delta)
    \ \xrightarrow[l\to\infty]{}\
\frac{2}{\sqrt{\pi}}\sqrt{l}\sqrt{\frac{x(\xi+x)}{\xi}}
\Bigl(1+\mathcal O(l^{-1})\Bigr).
\label{app:eq:an_scaling}
\end{equation}
Multiplying by $G_0$ in eq.\,\eqref{app:eq:G0_asympt}, the powers of $l$ cancel to give the crucial scaling
\begin{equation}
G_0\,a_{\lfloor lx\rfloor}(\Delta)
\ \xrightarrow[l\to\infty]{}\
\frac{1}{\pi^2}\frac{1}{l}\sqrt{x(\xi+x)}
\Bigl(1+\mathcal O(l^{-1})\Bigr).
\label{app:eq:G0an_scaling}
\end{equation}
The exponential factors become
\begin{equation}
e^{-\Delta\tau/l}q^{\lfloor lx\rfloor}
=
e^{-\Delta\tau/l}e^{-2\tau\lfloor lx\rfloor/l}
\ \xrightarrow[l\to\infty]{}\
e^{-(\xi+2x)\tau}.
\label{app:eq:exp_scaling}
\end{equation}
Substituting eqs.\,\eqref{app:eq:G0an_scaling} and \eqref{app:eq:exp_scaling} into the exact sum
\eqref{app:eq:ads4_exact_series}, we obtain a Riemann sum with a step size $1/l$
\begin{equation}
G^{E}_{\rm AdS_4}(\tau)
\xrightarrow[l\to\infty]{}
\frac{1}{\pi^2}\int_0^\infty dx\,\sqrt{x(\xi+x)}\,e^{-(\xi+2x)\tau}.
\label{app:eq:riemann_integral}
\end{equation}
This already exhibits that the dominant contribution comes from $n=lx$ with $x=\mathcal O(1)$,
i.e.\ the non-uniform scaling window \eqref{app:eq:scaling_window}.

To match the standard flat Green's function, we change variables as
\begin{equation}
E\equiv\xi+2x.
\end{equation}
Then eq.\,\eqref{app:eq:riemann_integral} becomes
\begin{equation}
\lim_{l\to\infty}G^{E}_{\rm AdS_4}(\tau)
=
\frac{1}{4\pi^2}\int_{\xi}^{\infty} dE\,\sqrt{E^2-\xi^2}\,e^{-E\tau}.
\label{app:eq:energy_integral}
\end{equation}
The integral is evaluated as
\begin{equation}
\int_{\xi}^{\infty} dE\,\sqrt{E^2-\xi^2}\,e^{-E\tau}
=
\frac{\xi}{\tau}K_1(\xi\tau),
\end{equation}
leading to
\begin{equation}
\lim_{l\to\infty}G^{E}_{\rm AdS_4}(\tau)
=
\frac{\xi}{4\pi^2\,\tau}\,K_1(\xi\tau)
\equiv G^{E}_{\rm flat}(\tau).
\label{app:eq:flat_green_final}
\end{equation}

We also present another way to take the flat limit of Green's function in the Appendix \ref{legendre} just for completeness.

\subsubsection{Flat limit of bulk local state} \label{subsubsec: flat limit}
 Applying the above strategy, we can also reproduce the bulk local state by the flat limit as we see below.

The bulk local state at $(\tau,0,0,0)$ in AdS$_4$ reads (below we omit the spatial coordinates for notational simplicity)
\begin{equation}
\label{eq:AdS_bulk_ket_en}
\ket*{\Psi^E_{\rm AdS_4}(\tau)}=
e^{-\Delta \tau/l}\sum_{k=0}^\infty
\frac{q^k}{4^k\,k!\,(\Delta-\tfrac12)_k}\;\ket*{e_k},
\qquad
q=e^{-2\tau/l},
\end{equation}
with $\ket*{e_k}:=(\mathcal P^2)^k\ket*{\Delta}$.
The basis is orthogonal
\begin{equation}
\label{eq:AdS_orth_en}
\braket*{e_m}{e_n}=\delta_{mn}\,G_n,\quad G_n=G_0\,16^n\,n!\,(\Delta-\tfrac12)_n\,(\tfrac32)_n\,(\Delta)_n.
\end{equation}
We define the orthonormal basis as
\begin{equation}
\label{eq:AdS_k_basis}
\ket*{k}:=\frac{1}{\sqrt{G_k}}\ket*{e_k},
\qquad
\braket*{k}{k'}=\delta_{kk'},
\end{equation}
which plays the role of a momentum-like basis at the AdS center; in the flat limit $k\sim l$
this transforms into the flat momentum basis.

Using this basis, the bulk local state can be represented as
\begin{equation}
\label{eq:AdS_k_expansion}
\ket*{\Psi^E_{\rm AdS_4}(\tau)}=\sum_{k=0}^\infty F_k(l,\tau)\ket*{k},
\quad
F_k(l,\tau)=\sqrt{G_0}\;e^{-\Delta\tau/l}\;q^k\;\sqrt{a_k(\Delta)},
\end{equation}
where  
\be
a_k(\Delta):=\frac{(\tfrac32)_k(\Delta)_k}{k!\,(\Delta-\tfrac12)_k}.
\ee
Because the basis is orthonormal, the bra is simply given by
\begin{equation}
\label{eq:AdS_bra_k}
\bra*{\Psi^E_{\rm AdS_4}(\tau)}=\sum_{k=0}^\infty F_k(l,\tau)^*\bra*{k},
\quad
{}_{\rm OS}\bra*{\Psi^E_{\rm AdS_4}(\tau_1)}
=\sum_{k=0}^\infty F_k(l,-\tau_1)^*\bra*{k}.
\end{equation}
Let us take a flat limit with
\begin{equation}
\label{eq:flatlimit_Delta}
\Delta=\xi l+O(1),\qquad l\to\infty.
\end{equation}
The dominant contribution comes from $k\sim l$.
Now, introduce the continuum variable $x:=\frac{k}{l}\ge 0,$ and the normalized basis so that
\begin{equation}
\label{eq:x_basis_def}
\ket*{x}:=\sqrt{l}\,\,\ket*{k},
\quad
\langle x|x'\rangle=\delta(x-x').
\end{equation}
We evaluate the large $l$ asymptotics for each term. First, we can rewrite the $a_k(\Delta)$ using Gamma functions
\begin{equation}
a_k(\Delta)=
\frac{\Gamma(k+\frac32)}{\Gamma(\frac32)\Gamma(k+1)}\cdot
\frac{\Gamma(\Delta+k)}{\Gamma(\Delta)}\cdot
\frac{\Gamma(\Delta-\frac12)}{\Gamma(\Delta+k-\frac12)}.
\end{equation}

Using the formula
\be
\Gamma(x)\sim\sqrt{\frac{2\pi}{x}}\left(\frac{x}{e}\right)^x,\quad (x\to\infty) 
\ee
we find
\begin{align}
    G_0&=\frac{\Gamma(\Delta)}{2\pi^{\frac{3}{2}}\Gamma\left(\Delta-\frac{1}{2}\right)l^2}\sim \frac{\sqrt{\xi}}{2\pi^{\frac{3}{2}} l^{\frac{3}{2}}},\nn\\
    a_k(\Delta)&\sim \frac{2}{\sqrt{\pi}}\sqrt{l}\sqrt{\frac{x(x+\xi)}{\xi}},\nn\\
    q^ke^{-\frac{\Delta\tau}{l}}&\sim e^{-(\xi+2x)\tau}.
\end{align}
Substituting these, we evaluate the contribution $k\sim \mathcal{O}(l)$ and obtain
\begin{align}
    \lim_{l\to\infty}\ket*{\Psi^E_{\rm AdS_4}(\tau)}&= \lim_{l\to\infty}\frac{1}{l}\sum_{k=0}^{\infty} \frac{\sqrt{l}(x(x+\xi))^\frac{1}{4}}{\pi}e^{-(\xi+2x)\tau}\ket*{k}\nn\\
    &\sim\int_0^\infty dx\;\phi_l(x,\tau)\ket*{x},
\end{align}
where 
\be
\phi_l(x,\tau):=\sqrt{l}\,\,F_{\lfloor lx\rfloor}(l,\tau).
\ee
Notice that in the exact large $l$ limit, it reduces to
\be
\phi_l(x,\tau)\to\phi_{\rm flat}(x,\tau)=\frac{(x(x+\xi))^\frac{1}{4}}{\pi}e^{-(\xi+2x)\tau}.
\ee
Since $\ket*{x}$ is orthonormal, the OS-bra is again obtained by conjugation
\begin{equation}
\label{eq:flat_bra_x}
{}_{\rm OS}\bra*{\Psi^E_{\rm AdS_4}(\tau)}
\simeq \int_0^\infty dx\;\phi_l(x,\tau)^*\bra*{x}
\ \xrightarrow[l\to\infty]{}\
{}_{\rm OS}\bra*{\Psi^E_{\rm flat}(\tau)}
:=\int_0^\infty dx\;\phi_{\rm flat}(x,\tau)^*\bra*{x}.
\end{equation}

    Finally, to match the flat expression, we rescale the $\ket*{x}$ basis after the limit $l\to\infty$ as follows
\be
\ket*{x}:=\frac{(x(x+\xi))^\frac{1}{4}}{4\pi^2}\int d\Omega \ket*{\mathbf p}.
\ee
Notice that this normalization leads to the usual normalization for $\ket*{\mathbf p}$
\be
\langle x|x'\rangle=\delta(x-x')\rightarrow
\langle \mathbf p|\mathbf p'\rangle=(2\pi)^3\,2E\,\delta^{(3)}(\mathbf p-\mathbf p').
\ee
Identifying
\begin{equation}
\label{eq:x_to_p}
E(x)=\xi+2x,\quad
p=2\sqrt{x(\xi+x)},
\end{equation}
we find $E=\sqrt{\xi^2+p^2}$.
The Jacobian gives
\begin{equation}
\sqrt{x(\xi+x)}\,dx=\frac{p^2}{4E}\,dp.
\end{equation}
Substituting these, we obtain
\begin{align}
     \lim_{l\to\infty}\ket*{\Psi^E_{\rm AdS_4}(\tau)}&=\int_0^\infty dx \frac{\sqrt{x(x+\xi)}}{4\pi^3}e^{-E\tau}\int d\Omega\, \ket*{\mathbf p}\nn\\
     &=\int \frac{d^3p}{(2\pi)^3\,2E}e^{-E\tau}\ket*{\mathbf p},
\end{align}
which reproduces the flat bulk local state (\ref{eq:flat_bulk_ket_E_mom}).

\subsection{Bulk local state in the AdS$_4$ tilde basis and the flat limit }
\label{subsec:bulk_local_4d_tilde}

In the previous subsection we constructed the bulk local state in the momentum basis, which is the most natural choice from the perspective of particle physics and directly yields the familiar plane-wave expansion. However, the representation of the Poincar\' e group admits many equivalent bases, and different choices can highlight different algebraic features. Here we introduce an alternative basis, the \textit{tilde basis}, which is tailored to make the action of the translation generators particularly simple. This basis is constructed by a triangular similarity transformation from the descendant basis ${|n\rangle}$ and satisfies $P_a|\tilde n\rangle = 2i n\xi K_a|\widetilde{n-1}\rangle$, a form that greatly simplifies the imposition of the bulk stabilizer condition. Moreover, the tilde basis is directly related to the flat basis employed in the three-dimensional analysis of Flat$_3$/CCFT$_2$. Its construction in four dimensions serves as a concrete warm-up for the general-dimensional case discussed later.
\subsubsection{Bulk local state in Flat$_4$ in the tilde basis}
We begin with the algebraic setup. The isometry algebra of Flat$_4$ is the Poincar\'{e} algebra $\mathfrak{iso}(1,3)$ in eqs.\,\eqref{eq:poincare_mix} and \eqref{eq:poincare_trans}, generated by translations $P_\mu$ ($\mu=0,1,2,3$) and Lorentz generators $J_{\mu\nu}$. It is convenient to decompose the Lorentz generators into rotations $J_a$ and boosts $K_a$ ($a=1,2,3$), which satisfy the commutation relations \eqref{eq:sl2c_subalgebra}, together with
the commutators involving translations \eqref{eq:poincare_mix_explicit}.

Again, let us consider a scalar primary state $|\xi\rangle$ specified by the conditions \eqref{eq:flat_primary_energy} -- \eqref{eq:flat_primary_spin}
\begin{equation}
P_a|\xi\rangle=0,\qquad
P_0|\xi\rangle=\xi|\xi\rangle,\qquad
J_a|\xi\rangle=0 .
\end{equation}
Applying the boost generators $K_a$ on the primary $\ket*{\xi}$ yields descendant states. A rotation-invariant family is obtained by repeated application of the quadratic Casimir operator $C_K = K_a K^a$ of the boost subalgebra. The descendants
\begin{equation}
|n\rangle := C_K^{\,n} |\xi\rangle ,\qquad n=0,1,2,\dots
\label{eq:flat_descendant}
\end{equation}
satisfy $J_a|n\rangle=0$ and form a convenient basis for constructing rotation-invariant bulk operators.

To construct a bulk local state, we need the action of $P_a$ and $P_0$ on the descendant basis $\{|n\rangle\}$. Using the commutation relations
\begin{equation}
[P_a,C_K]=2iK_aP_0-P_a,\qquad
[P_0,C_K]=2iK_aP^a-3P_0,
\end{equation}
we derive the recurrence relations
\begin{align}
P_a|n\rangle &= K_a|\bar{n}\rangle, \label{eq:Pa_on_n} \\
|\bar{n}\rangle &= (C_K-1)|\overline{n-1}\rangle + 2iP_0 C_K^{\,n-1}|\xi\rangle, \label{eq:bar_n_rec1} \\
P_0 C_K^{\,n}|\xi\rangle &= (C_K+3)P_0 C_K^{\,n-1}|\xi\rangle + 2iC_K|\overline{n-1}\rangle. \label{eq:P0CK_rec}
\end{align}
Solving these with $|\bar{0}\rangle=0$ yields the closed-form expression
\begin{equation}
|\bar{n}\rangle = i\xi\,
\frac{\bigl(C_K+2\sqrt{C_K+1}+1\bigr)^{n}
-\bigl(C_K-2\sqrt{C_K+1}+1\bigr)^{n}}
{2\sqrt{C_K+1}}\,|\xi\rangle . \label{eq:bar_n_explicit}
\end{equation}
In particular, the coefficient of the highest term $K_a|n-1\rangle$ in the expansion $P_a|n\rangle=\sum_{i=0}^{n-1}A_{ni}K_a|i\rangle$ is
\begin{equation}
A_{n,n-1}=2in\xi . \label{eq:Ann1}
\end{equation}
An equivalent expansion in terms of a hypergeometric function is
\begin{equation}
P_a|n\rangle = \sum_{i=0}^{n-1} \underbrace{2in\xi\binom{n-1}{i}\,
{}_3F_2\!\left(\begin{array}{c}
i-n+1,\;\tfrac12-\tfrac n2,\;1-\tfrac n2\\[2pt]
\tfrac32,\;1-n
\end{array};4\right)}_{A_{ni}} K_a|i\rangle . \label{eq:Pa_on_n_hypergeom}
\end{equation}

The action of $P_0$ on $|n\rangle$ can also be computed explicitly
\begin{multline}
P_0|n\rangle = \xi\,
\frac{\sqrt{C_K+1}\bigl[(C_K-2\sqrt{C_K+1}+1)^n-(C_K+2\sqrt{C_K+1}+1)^n\bigr]}
{2\sqrt{C_K+1}}\,|\xi\rangle \\
+\xi\,
\frac{\sqrt{C_K+1}\bigl[(C_K+2\sqrt{C_K+1}+1)^n+(C_K-2\sqrt{C_K+1}+1)^n\bigr]}
{2\sqrt{C_K+1}}\,|\xi\rangle .
\end{multline}
An alternative expression in summation form is
\begin{equation}
P_0|n\rangle = \sum_i\sum_{t=0}^{n}
\binom{n}{t}2^{t}\binom{n-\lceil t/2\rceil}{i}\,|i\rangle . \label{eq:P0_on_n}
\end{equation}
Both $P_a$ and $P_0$ act triangularly on the basis $\{|n\rangle\}$, involving only states with index $i\le n$.
Due to this fact, we can perform a lower-triangular similarity transformation to a new basis $\{|\tilde n\rangle\}$ in which $P_a$ takes a simpler form. Specifically, we seek a basis satisfying
\begin{equation}
P_a |\tilde n\rangle = \lambda_n K_a |\widetilde{n-1}\rangle,\qquad |\widetilde{-1}\rangle\equiv0, \label{eq:new_basis_condition}
\end{equation}
where $\lambda_n$ are coefficients to be determined. 

To determine $\lambda_n$, let the new basis be expressed as $|\tilde n\rangle = \sum_{k=0}^{n} S_{nk}|k\rangle$ with $S_{nn}=1$. By substituting into eq.\,\eqref{eq:new_basis_condition} and using $P_a|k\rangle=\sum_{i=0}^{k-1}A_{ki}K_a|i\rangle$,
\begin{equation}
\sum_{k=i+1}^{n} S_{nk}A_{ki} = \lambda_n S_{n-1,i}. \label{eq:coeff_relation}
\end{equation}
For $i=n-1$, the left-hand side reduces to $S_{nn}A_{n,n-1}=A_{n,n-1}$ because $S_{nn}=1$. The right-hand side is $\lambda_n S_{n-1,n-1}$. Requiring all diagonal elements $S_{kk}=1$ (i.e., the highest-weight term in $|\tilde k\rangle$ is $|k\rangle$ with coefficient one) gives $S_{n-1,n-1}=1$, so
\begin{equation}
A_{n,n-1} = \lambda_n .
\end{equation}
From eq.\,\eqref{eq:Ann1} we have $A_{n,n-1}=2in\xi$, hence
\begin{equation}
\lambda_n = 2in\xi . \label{eq:lambda_n}
\end{equation}
Thus, under the natural normalization convention, the coefficient is uniquely fixed to $\lambda_n=2in\xi$.

Eq.\,\eqref{eq:coeff_relation} with $\lambda_n=2in\xi$ provides a recursive definition of the transformation matrix $S_{nk}$. Starting from $|\tilde0\rangle=|0\rangle$ (i.e., $S_{00}=1$), we give the example by solving it order by order. For $n=1$
\begin{equation}
S_{10}A_{10}=2\xi S_{00}\;\Longrightarrow\; S_{10}\cdot2\xi=2\xi\cdot1\;\Longrightarrow\; S_{10}=0,
\end{equation}
so $|\tilde1\rangle=|1\rangle$. For $n=2$, using $A_{10}=2\xi$, $A_{20}=4\xi$, $A_{21}=4\xi$
\begin{align}
S_{20}\cdot2\xi + S_{21}\cdot4\xi = 0,\qquad S_{21}\cdot4\xi = 4\xi\cdot1,
\end{align}
giving $S_{21}=1$, $S_{20}=-2$, and $|\tilde2\rangle=|2\rangle-2|1\rangle$. Proceeding similarly yields
\begin{align}
|\tilde3\rangle &= |3\rangle - 8|2\rangle + 9|1\rangle, \\
|\tilde4\rangle &= |4\rangle - 20|3\rangle + 96|2\rangle - 72|1\rangle .
\end{align}
These satisfy $P_a|\tilde n\rangle=2in\xi K_a|\widetilde{n-1}\rangle$, confirming the consistency of the construction.

A bulk local state located at $(t,0,0,0)$ must be invariant under the combination of a boost and a translation that leaves the point fixed (See eq.\,\eqref{eq:flat_stabilizer})
\begin{equation}
\bigl(K_a - t P_a\bigr)|\phi(t,0,0,0)\rangle = 0. \label{eq:bulk_condition}
\end{equation}
The state must also be rotation-invariant, $J_a|\phi(t,0,0,0)\rangle = 0$, which is automatically satisfied by an expansion in the basis $\{|\tilde n\rangle\}$.

Expanding the bulk local state as $|\phi(t,0,0,0)\rangle = \sum_{n=0}^{\infty} f_n(t)\,|\tilde n\rangle$ and inserting into \eqref{eq:bulk_condition}, it follows
\begin{equation}
f_n = \frac{1}{n!\,(2i\xi t)^n},
\end{equation}
after choosing $f_0=1$ (up to an overall normalization).
The solution to eq.\,\eqref{eq:bulk_condition} is thus
\begin{equation}
|\phi(t,0,0,0)\rangle = \sum_{n=0}^{\infty} \frac{1}{n!\,(2i\xi t)^n}\;|\tilde n\rangle . \label{eq:bulk_state_series}
\end{equation}

\subsubsection{Extension of the tilde basis in AdS$_4$}
We begin with the scalar primary state $|\Delta\rangle$ in AdS$_4$ satisfying (See eq.\,\eqref{eq:intermediate_rep})
\begin{equation}
K_a(l)|\Delta\rangle = 0,\qquad 
P_a(l)\ket*{\Delta}=-\frac{i}{l}K_a(l)\ket*{\Delta},\qquad 
J_{ab}(l)|\Delta\rangle = 0,
\label{eq:AdS_primary_conditions}
\end{equation}
where $\xi = \Delta/l$ and $l$ is the AdS radius. The algebra is then
\begin{align}
[P_a(l), K_b(l)] &= i\,\delta_{ab}\,P_0(l) - \frac{1}{l}\, J_{ab}(l), \label{eq:AdS_comm_KP} \\
[ P_0(l), P_a(l)] &= \frac{i}{l^2}\, K_a(l), \label{eq:AdS_comm_P0P} \\
[ P_0(l), K_a(l)] &= i\,P_a(l), \label{eq:AdS_comm_P0K}
\end{align}
together with the standard rotation algebra $[ J_{ab}(l), J_{cd}(l)]$. In the flat limit $l\to\infty$ these relations contract to the Poincare algebra \eqref{eq:poincare_mix_explicit}.

The bulk local state at the point $(t,0,0,0)$ must be invariant under the combination of a boost and a translation that leaves the point fixed. This condition can be rewritten as 
\begin{equation}
\Bigl(e^{-i\frac{t}{l}}P_a(l) - e^{i\frac{t}{l}}K_a(l)\Bigr)|\Psi_{\mathrm{AdS}}(t,0,0,0)\rangle = 0,\qquad 
J_{ab}(l)\,|\Psi_{\mathrm{AdS}}(t,0,0,0)\rangle = 0 .
\label{eq:AdS_bulk_condition_original}
\end{equation}
Equivalently, the first condition can be recast into the form
\begin{equation}
\Bigl(l\sin\frac{t}{l}\,P_a(l) + \cos\frac{t}{l}\,K_a(l)\Bigr)|\Psi_{\mathrm{AdS}}(t,0,0,0)\rangle = 0 .
\label{eq:AdS_bulk_condition_simplified}
\end{equation}

To solve these equations we construct a convenient basis of rotation-invariant descendants. Define the quadratic Casimir of the boost subalgebra
\begin{equation}
C_K(l) := K_a(l)K^a(l).
\label{eq:AdS_boost_casimir}
\end{equation}
Starting from the primary we define the states 
\begin{equation}
|n\rangle_l := \bigl(C_K(l)\bigr)^n |\Delta\rangle ,\qquad n=0,1,2,\dots
\label{eq:AdS_descendant_basis_CK}
\end{equation}
which are manifestly invariant under rotations, $J_{ab}(l)|n\rangle_l=0$.

The action of $P_a(l)$ on this basis is not diagonal. From the commutation relations one obtains the recursive relations
\begin{align}
P_a(l) |n\rangle_l &\equiv K_a(l) |A_n\rangle, \label{eq:Pa_on_nl} \\
|A_n\rangle &= -\bigl(C_K(l)+1\bigr)|A_{n-1}\rangle - 2\xi\,C_K(l)\,|n-1\rangle_l, \label{eq:An_rec1} \\
P_0(l) |n\rangle_l &= \bigl(C_K(l)+3\bigr)P_0(l)|n-1\rangle_l + 2C_K(l)|A_{n-1}\rangle, \label{eq:H_on_nl}
\end{align}
with the initial condition $|A_0\rangle = -\frac{i}{l}|\Delta\rangle$. 
The solution to this equations gives the explicit coefficients for expanding $|A_n\rangle$ and $P_0(l)|n\rangle_l$ in the basis $\{|k\rangle_l\}$, which is given by
\begin{align}
|A_n\rangle &= \sum_{i=0}^n d_i \bigl(C_K(l)\bigr)^i |\Delta\rangle, \label{eq:An_expansion} \\
d_i &= (-1)^{n-i+1}\frac{i}{2l}\sum_{j=0}^n \binom{n}{j} 2^j \binom{n-\lceil j/2\rceil}{i}
        \bigl[(2-l\xi)(-1)^j + l\xi\bigr], \label{eq:di_explicit} \\
P_0(l)|n\rangle_l &= \sum_{i=0}^n c_i \bigl(C_K(l)\bigr)^i |\Delta\rangle, \label{eq:Hn_expansion} \\
c_i &= \xi\sum_{k=0}^n \binom{n}{k} 2^k \binom{n-\lceil k/2\rceil}{i} 
      - \frac{1}{l}\sum_{\substack{k=1 \\ k\ \text{odd}}}^n \binom{n}{k} 2^k \binom{n-\frac{k+1}{2}}{i-1}. \label{eq:ci_explicit}
\end{align}

Because the matrix representing $P_a(l)$ in the basis $\{|n\rangle_l\}$ is lower triangular, we can perform a similarity transformation to a new basis $\{|\tilde n\rangle_l\}$ in which $P_a(l)$ takes a simpler form. Define 
\begin{equation}
|\tilde n\rangle_l = \sum_{k\le n} S_{nk}(l)\,|k\rangle_l,\qquad S_{nn}(l)=1,
\label{eq:AdS_tilde_basis_def}
\end{equation}
where the coefficients $S_{nk}(l)$ are chosen so that 
\begin{equation}
P_a(l)\,|\tilde n\rangle_l = \lambda_n(l)\,K_a(l)\,|\widetilde{n-1}\rangle_l
- \frac{i}{l}\,K_a(l)\,|\tilde n\rangle_l,\qquad |\widetilde{-1}\rangle_l\equiv0 .
\label{eq:AdS_tilde_basis_action}
\end{equation}
The coefficient $\lambda_n(l)$ is determined by the highest weight part of the action of $\mathcal{P}_a(l)$. By direct computation, we find
\begin{equation}
    \lambda_n(l)=d_{n-1}=\frac{in}{l}(2l\xi+2n-3).
\end{equation}
It behaves as
\begin{equation}
\lambda_n(l) = i\bigl(2n\xi + \mathcal{O}(l^{-1})\bigr),\qquad 
\lim_{l\to\infty}\lambda_n(l) = i\,2n\xi .
\label{eq:lambda_n_behavior}
\end{equation}

Now expand the bulk local state in this improved basis
\begin{equation}
|\Psi_{\mathrm{AdS}}(t,0,0,0)\rangle = \sum_{n=0}^{\infty} f_n(l,t)\,|\tilde n\rangle_l .
\label{eq:AdS_bulk_expansion_tilde}
\end{equation}
Substituting this expansion into the invariance condition \eqref{eq:AdS_bulk_condition_simplified} and using eq.\,\eqref{eq:AdS_tilde_basis_action}
\begin{equation}
\sum_n f_n(l,t) \Bigl[ l\sin\frac{t}{l} \bigl( \lambda_n(l) K_a |\widetilde{n-1}\rangle_l
- \frac{i}{l} K_a |\tilde n\rangle_l \bigr) + \cos\frac{t}{l} K_a |\tilde n\rangle_l \Bigr] = 0 .
\label{eq:AdS_bulk_expansion_substituted}
\end{equation}
The solution  with the normalization $f_0(l,t)=1$ (up to an overall constant) is
\begin{equation}
f_n(l,t) = \left( \,\frac{e^{-it/l}}{l\sin(t/l)} \right)^{\!n}\;
\prod_{k=1}^{n} \frac{1}{\lambda_k(l)} .
\label{eq:AdS_f_n_solution}
\end{equation}
Finally, we can exploit the explicit relations between the descendant states \eqref{eq:arb:AdS_en_def} and the tilde basis $\{|\tilde n\rangle_l\}_n$ as follows
\begin{equation}
|\tilde 0\rangle_l = |e_0\rangle,\quad
|\tilde n\rangle_l
=
\,n!\,(\Delta-\tfrac12)_n
\sum_{k=1}^{n}(-1)^{\,n-k}
\binom{n-1}{k-1}\;
\frac{1}{4^k\,k!\,(\Delta-\tfrac12)_k}\;
|e_k\rangle
\qquad (n\ge1).
\end{equation}
The detailed derivation can be found in Appendix \ref{subsec:relation_descendant}.

\subsubsection{Flat limit}

In the flat limit $l\to\infty$ the AdS metric reduces to the Minkowski metric, and the isometry algebra contracts to the Poincar\'{e} algebra. Correspondingly, the generators scale as
\begin{equation}
P_a(l) \longrightarrow P_a,\qquad
K_a(l) \longrightarrow K_a,\qquad
P_0(l) \longrightarrow P_0,
\end{equation}
and the primary state becomes
\begin{equation}
K_a|\xi\rangle = 0,\qquad
P_0|\xi\rangle = \xi|\xi\rangle,\qquad
J_{ab}|\xi\rangle = 0,
\label{eq:flat_primary_conditions}
\end{equation}
with $\xi=m$. The quadratic operator $C_K(l)$ asymptotes to the flat space operator $C_K = K_aK^a$, and the transformed basis $|\tilde n\rangle_l$ reduces to the flat space basis $|\tilde n\rangle$ defined in Eq.~\eqref{eq:new_basis_condition}.

Taking the limit of the coefficients \eqref{eq:AdS_f_n_solution} we use $\sin(t/l)\sim t/l$, $\cos(t/l)\sim 1$, $e^{-it/l}\sim 1$ and eq.\,\eqref{eq:lambda_n_behavior} to obtain
\begin{equation}
\lim_{l\to\infty} f_n(l,t) = 
\frac{1}{n!\,(2i\xi t)^n}\; \bigl(1 + \mathcal{O}(l^{-1})\bigr).
\label{eq:flat_limit_f_n}
\end{equation}
Thus, up to an overall normalization that can be absorbed into $f_0$, the limiting coefficients coincide with the flat space coefficients found in eq.\,\eqref{eq:bulk_state_series}. Consequently,
\begin{equation}
\lim_{l\to\infty} |\Psi_{\mathrm{AdS}}(t,0,0,0)\rangle
= \sum_{n=0}^{\infty} \frac{1}{n!\,(2i\xi t)^n}\,|\tilde n\rangle
= |\phi(t,0,0,0)\rangle_{\mathrm{flat}},
\label{eq:flat_limit_bulk_state}
\end{equation}
which is precisely the bulk local state in Minkowski space. This demonstrates the consistency of the bulk local operator construction in the flat limit of AdS/CFT.

\subsection{Arbitrary dimensional case}
\label{subsec:arbitrary_dim}

We have shown that for a scalar field in four dimensions, the bulk local state can be constructed purely from the induced-like representation of the Poincar\'e algebra in both momentum basis and tilde basis. The state satisfies the correct stabilizer condition and, upon taking inner products, reproduces the bulk Green function. Moreover, it emerges as the flat-space limit of the corresponding AdS$_4$ bulk local state, confirming the consistency of our approach. We now generalize this discussion to arbitrary dimensions. The reasoning parallels the four-dimensional case; we will indicate below the specific equations that undergo generalization.

\subsubsection{Algebra and representation}
\label{subsubsec:algebra_general}

\paragraph{Bulk isometries: from $\text{AdS}_{d+1}$ to $\text{Flat}_{d+1}$.}
The isometries of $(d+1)$-dimensional Minkowski space $\text{Flat}_{d+1}$ are generated by the Poincar\'e algebra $\mathfrak{iso}(1,d)$. In Cartesian coordinates $(x^0,x^1,\dots,x^d)$ the generators are translations $P_\mu$ and Lorentz rotations $J_{\mu\nu}=-J_{\nu\mu}$ ($\mu,\nu=0,\dots,d$). They satisfy
\begin{align}
[J_{\mu\nu},J_{\rho\sigma}] &= i(\eta_{\nu\rho}J_{\mu\sigma}-\eta_{\mu\rho}J_{\nu\sigma}-\eta_{\nu\sigma}J_{\mu\rho}+\eta_{\mu\sigma}J_{\nu\rho}),\\
[J_{\mu\nu},P_\rho] &= i(\eta_{\mu\rho}P_\nu-\eta_{\nu\rho}P_\mu),\\
[P_\mu,P_\nu] &= 0,
\end{align}
where $\eta_{\mu\nu}=\mathrm{diag}(-1,1,\dots,1)$. For a scalar field it is convenient to split the Lorentz generators into rotations $J_{ab}$ ($a,b=1,\dots,d$) and boosts $K_a:= J_{0a}$. The commutation relations become
\begin{align}
[J_{ab},J_{cd}] &= i(\delta_{ac}J_{bd}-\delta_{bc}J_{ad}-\delta_{ad}J_{bc}+\delta_{bd}J_{ac}),\\
[K_a,K_b] &= -iJ_{ab},\qquad [J_{ab},K_c] = i(\delta_{ac}K_b-\delta_{bc}K_a),\\
[P_0,K_a] &= iP_a,\qquad [P_a,K_b] = i\delta_{ab}P_0,\qquad [J_{ab},P_c] = i(\delta_{ac}P_b-\delta_{bc}P_a).
\end{align}

Anti-de Sitter space $\text{AdS}_{d+1}$ of radius $l$ can be embedded as a hyperboloid in $\mathbb{R}^{2,d}$. Its isometry algebra is $\mathfrak{so}(2,d)$, which is isomorphic to the conformal algebra in $d$ dimensions, $\mathfrak{conf}(1,d-1)$. In a basis better adapted to the AdS/CFT correspondence, the generators are the dilatation $\mathcal{H}$, translations $\mathcal{P}_a$, special conformal transformations $\mathcal{K}_a$ and rotations $\mathcal{M}_{ab}$ ($a,b=1,\dots,d$). They satisfy
\begin{align}
[\mathcal{H},\mathcal{P}_a] &= \mathcal{P}_a,\quad [\mathcal{H},\mathcal{K}_a] = -\mathcal{K}_a,\quad [\mathcal{K}_a,\mathcal{P}_b] = 2(\delta_{ab}\mathcal{H}-\mathcal{M}_{ab}),\\
[\mathcal{K}_a,\mathcal{M}_{bc}] &= \delta_{ab}\mathcal{K}_c-\delta_{ac}\mathcal{K}_b,\quad [\mathcal{P}_a,\mathcal{M}_{bc}] = \delta_{ab}\mathcal{P}_c-\delta_{ac}\mathcal{P}_b,\\
[\mathcal{M}_{ab},\mathcal{M}_{cd}] &= \delta_{bc}\mathcal{M}_{ad}+\delta_{ad}\mathcal{M}_{bc}-\delta_{ac}\mathcal{M}_{bd}-\delta_{bd}\mathcal{M}_{ac}.
\end{align}

The connection between the two algebras is established via the flat limit $l\to\infty$. Define the following $l$-dependent generators
\begin{equation}
P_0(l) := \frac{1}{l}\mathcal{H},\qquad P_a(l) := -\frac{1}{2l}(\mathcal{P}_a+\mathcal{K}_a),\qquad
K_a(l) := -\frac{i}{2}(\mathcal{P}_a-\mathcal{K}_a),\qquad J_{ab}(l) := i\mathcal{M}_{ab}.
\end{equation}
In the limit $l\to\infty$ one confirms that these satisfy the Poincar\'e algebra above. For instance, the commutator
\[
[P_a(l),K_b(l)] = i\delta_{ab}P_0(l) - \frac{1}{l}J_{ab}(l)
\]
asymptotes to $[P_a,K_b]=i\delta_{ab}P_0$ as $l\to\infty$.

\paragraph{Representations and the flat limit.}
In $\text{AdS}_{d+1}$/CFT$_d$, a scalar primary state $|\Delta\rangle$ is defined by the highest-weight conditions
\begin{equation}
\mathcal{K}_a|\Delta\rangle = 0,\qquad \mathcal{H}|\Delta\rangle = \Delta|\Delta\rangle,\qquad \mathcal{M}_{ab}|\Delta\rangle = 0.
\label{eq:general_ads_highest_weight}
\end{equation}
Using the rescaled generators, these become
\begin{equation}
P_a(l)|\Delta\rangle = -\frac{i}{l}K_a(l)|\Delta\rangle,\qquad P_0(l)|\Delta\rangle = \frac{\Delta}{l}|\Delta\rangle.
\end{equation}
Taking the flat limit $l\to\infty$ and assuming the existence of finite quantities
\begin{equation}
|\xi\rangle := \lim_{l\to\infty} |\Delta\rangle,\qquad \xi := \lim_{l\to\infty} \frac{\Delta}{l},
\end{equation}
we arrive at the defining conditions for a state $|\xi\rangle$ in the flat space bulk:
\begin{align}
P_0|\xi\rangle &= \xi|\xi\rangle, \label{eq:general_flat_primary_energy}\\
P_a|\xi\rangle &= 0, \label{eq:general_flat_primary_momentum}\\
J_{ab}|\xi\rangle &= 0. \label{eq:general_flat_primary_spin}
\end{align}
These equations characterize a massive scalar particle at the rest frame, which we also call the \textit{induced representation}. Here, $P_\mu$ are the $(d+1)$-momentum operators and $J_{ab}$ are the rotation generators. The state $|\xi\rangle$ is thus a primary state of an induced representation of the Poincar\'e group, where all other momentum eigenstates are generated by applying the boost generators $K_a$. This representation is the fundamental building block for constructing bulk fields in any dimension.

\subsubsection{Bulk local state in the momentum basis}
\label{subsubsec:momentum_general}

\paragraph{Bulk local state in Flat$_{d+1}$ in the momentum basis.}
Let $|\xi\rangle$ be the static scalar primary in Flat$_{d+1}$ satisfying eqs.\,\eqref{eq:general_flat_primary_energy} -- \eqref{eq:general_flat_primary_spin}. Define the momentum basis $|\mathbf{p}\rangle$ by boosting $|\xi\rangle$
\begin{equation}
P_0|\mathbf{p}\rangle=E_{\mathbf{p}}|\mathbf{p}\rangle,\qquad P_a|\mathbf{p}\rangle=p_a|\mathbf{p}\rangle,\qquad E_{\mathbf{p}}:=\sqrt{\xi^2+\mathbf{p}^{\,2}},
\end{equation}
where we adopt the relativistic normalization
\begin{equation}
\langle\mathbf{p}|\mathbf{p}'\rangle=(2\pi)^d\,2E_{\mathbf{p}}\,\delta^{(d)}(\mathbf{p}-\mathbf{p}'),\qquad
\mathbf{1}=\int\frac{d^d p}{(2\pi)^d\,2E_{\mathbf{p}}}\,|\mathbf{p}\rangle\langle\mathbf{p}|.
\label{eq:arb:Flat_mom_norm}
\end{equation}
At $(t,\mathbf{0})$ the stabilizer condition is $(K_a-tP_a)|\phi(t,\mathbf{0})\rangle=0$. Expanding a rotation-invariant state as $|\phi(t,\mathbf{0})\rangle=\int\frac{d^dp}{(2\pi)^d2E_{\mathbf{p}}}\psi(\mathbf{p})|\mathbf{p}\rangle$ and using $[K_a,|\mathbf{p}\rangle]=iE_{\mathbf{p}}\frac{\partial}{\partial p_a}|\mathbf{p}\rangle$ yields $(iE_{\mathbf{p}}\partial_{p_a}-tp_a)\psi(\mathbf{p})=0$, whose solution is $\psi(\mathbf{p})=c e^{-itE_{\mathbf{p}}}$. After Wick rotation $t=-i\tau$ ($\tau>0$) we obtain the Euclidean bulk local state
\begin{equation}
|\phi^E_{\text{flat}}(\tau,\mathbf{0})\rangle = \int\frac{d^d p}{(2\pi)^d\,2E_{\mathbf{p}}}\,e^{-\tau E_{\mathbf{p}}}\,|\mathbf{p}\rangle,
\label{eq:arb:Flat_bulk_local_momentum}
\end{equation}
where we have set the constant $c=1$ for simplicity. The Euclidean Green's function follows as
\begin{equation}
G^E_{\text{flat}}(\tau) = \langle\phi^E_{\text{flat}}(0,\mathbf{0})|\phi^E_{\text{flat}}(\tau,\mathbf{0})\rangle
= \int\frac{d^d p}{(2\pi)^d\,2E_{\mathbf{p}}}\,e^{-\tau E_{\mathbf{p}}}
= \frac{\xi^{\frac{d-1}{2}}}{(2\pi)^{\frac{d+1}{2}}}\;\frac{K_{\frac{d-1}{2}}(\xi\tau)}{\tau^{\frac{d-1}{2}}},
\label{eq:arb:Flat_GE_Bessel}
\end{equation}
where $K_\nu$ is the modified Bessel function.

\paragraph{Bulk local state in AdS$_{d+1}$ in the momentum-descendant basis.}
In AdS$_{d+1}$ we work with Euclidean time $\tau>0$ ($t=-i\tau$). Let $|\Delta\rangle$ be a scalar primary satisfying eq.\,\eqref{eq:general_ads_highest_weight}. Define rotation-invariant descendants by repeated action of $P^2:= P_a(l)P_a(l)$
\begin{equation}
|e_n\rangle := (P^2)^n|\Delta\rangle,\qquad n=0,1,2,\dots
\label{eq:arb:AdS_en_def}
\end{equation}
These states are orthogonal,
\begin{equation}
\langle e_m|e_n\rangle = \delta_{mn}G_n,\qquad
G_n = 16^n\,n!\,\Bigl(\frac{d}{2}\Bigr)_n(\Delta)_n\Bigl(\Delta-\frac{d}{2}+1\Bigr)_n\,G_0,
\label{eq:arb:AdS_Gn}
\end{equation}
with $G_0=\langle\Delta|\Delta\rangle$. Following Ref.\,\cite{Berenstein:1998ij} we choose the normalization
\begin{equation}
G_0 = \frac{\Gamma(\Delta)}{2\pi^{d/2}\,\Gamma\!\left(\Delta-\frac{d}{2}+1\right)}\;\frac{1}{l^{\,d-1}}.
\label{eq:arb:AdS_G0}
\end{equation}
Introduce the orthonormal basis
\begin{equation}
|n\rangle_E := \frac{1}{\sqrt{G_n}}|e_n\rangle,\qquad {}_E\langle m|n\rangle_E=\delta_{mn}.
\label{eq:arb:AdS_orthonormal}
\end{equation}

The Euclidean bulk local state at the AdS center, invariant under $(e^{-\tau/l}P_a-e^{\tau/l}K_a)|\Psi^E_{\text{AdS}}(\tau)\rangle=0$, admits the expansion
\begin{equation}
|\Psi^E_{\text{AdS}}(\tau)\rangle = e^{-\Delta\tau/l}\sum_{n=0}^{\infty}
\frac{q^n}{4^n\,n!\,\bigl(\Delta-\frac{d}{2}+1\bigr)_n}\,|e_n\rangle,\qquad q=e^{-2\tau/l}.
\label{eq:arb:AdS_bulk_local_ket_en}
\end{equation}
In the orthonormal basis this becomes
\begin{equation}
|\Psi^E_{\text{AdS}}(\tau)\rangle = \sum_{n=0}^{\infty} F_n(l,\tau)\,|n\rangle_E,\quad
F_n(l,\tau)=\sqrt{G_0}\,e^{-\Delta\tau/l}\,q^n\sqrt{a_n(\Delta)},\quad
a_n(\Delta)=\frac{\bigl(\frac{d}{2}\bigr)_n(\Delta)_n}{n!\,\bigl(\Delta-\frac{d}{2}+1\bigr)_n}.
\label{eq:arb:AdS_bulk_local_orth}
\end{equation}

\paragraph{Flat limit for general $d\ge 2$.}
Take the flat limit
\begin{equation}
\Delta = \xi l + \mathcal{O}(1),\qquad l\to\infty,\qquad \xi>0\ \text{fixed}.
\label{eq:arb:flat_limit_Delta}
\end{equation}
For $d\ge 3$ the sum over $n$ is dominated by the scaling window $n\sim l$. Introduce the scaling variable $x:=n/l\ge 0$ and the continuum-normalized basis
\begin{equation}
|x\rangle := \sqrt{l}\,\,|n\rangle_E\Big|_{n=\lfloor lx\rfloor},\qquad \langle x|x'\rangle=\delta(x-x')\quad (l\to\infty).
\label{eq:arb:x_basis}
\end{equation}
Using the asymptotics
\begin{align}
G_0\,a_{\lfloor lx\rfloor}(\Delta) &\sim \frac{1}{2\pi^{d/2}\Gamma(d/2)}\;\frac{1}{l}\;\bigl[x(\xi+x)\bigr]^{\frac{d-2}{2}},\\
e^{-\Delta\tau/l}q^{\lfloor lx\rfloor} &\longrightarrow e^{-(\xi+2x)\tau},
\end{align}
we obtain
\begin{equation}
|\Psi^E_{\text{AdS}}(\tau)\rangle \;\xrightarrow[l\to\infty]{}\;
\int_0^\infty dx\; \frac{\bigl[x(\xi+x)\bigr]^{\frac{d-2}{4}}}{\sqrt{2\pi^{d/2}\Gamma(d/2)}}\;e^{-(\xi+2x)\tau}\,|x\rangle.
\label{eq:arb:state_limit_x}
\end{equation}

To connect with the flat space momentum basis, change to the energy variable $E=\xi+2x$ (so that $p^2=E^2-\xi^2$). The density of states is
\begin{equation}
\rho(E)=\int\frac{d^dp}{(2\pi)^d\,2E}\,\delta(E-\sqrt{\xi^2+p^2})
      = \frac{\Omega_{d-1}}{2(2\pi)^d}\,(E^2-\xi^2)^{\frac{d-2}{2}},\qquad
\Omega_{d-1}=\frac{2\pi^{d/2}}{\Gamma(d/2)}.
\label{eq:arb:rho_E}
\end{equation}
Define the energy-basis states
\begin{equation}
|E\rangle = \frac{1}{\sqrt{\rho(E)}}\int\frac{d^dp}{(2\pi)^d\,2E}\,\delta(E-\sqrt{\xi^2+p^2})\,|\mathbf{p}\rangle,
\label{eq:arb:E_basis}
\end{equation}
which satisfy $\langle E|E'\rangle=\delta(E-E')$ and are related to $|x\rangle$ by $|x\rangle = \sqrt{2}\,|E\rangle$. Substituting into \eqref{eq:arb:state_limit_x} yields
\begin{equation}
|\Psi^E_{\text{AdS}}(\tau)\rangle \;\xrightarrow[l\to\infty]{}\;
\int_{\xi}^{\infty} dE\,\sqrt{\rho(E)}\,e^{-E\tau}\,|E\rangle
= \int\frac{d^dp}{(2\pi)^d\,2E}\;e^{-E\tau}\,|\mathbf{p}\rangle = |\phi^E_{\text{flat}}(\tau,\mathbf{0})\rangle,
\label{eq:arb:to_flat_momentum}
\end{equation}
which is precisely the flat space bulk local state \eqref{eq:arb:Flat_bulk_local_momentum}.

\paragraph{Revisit the AdS$_3$ case in the momentum basis.}
For $d=2$ the general formulas simplify dramatically. Set $\Delta=2h$. Then
\begin{equation}
G_n = 16^n\,(n!)^2\,(\Delta)_n^2\,G_0 = 16^n\,(n!)^2\,(2h)_n^2\,G_0,
\end{equation}
and from eq.\,\eqref{eq:arb:AdS_bulk_local_orth}
\begin{equation}
a_n(\Delta)\big|_{d=2}= \frac{(1)_n(\Delta)_n}{n!(\Delta-1+1)_n}=1.
\end{equation}
Hence the bulk local state reduces to
\begin{equation}
|\Psi^{E}_{\mathrm{AdS}_3}(\tau)\rangle
= \sqrt{G_0}\,e^{-\Delta\tau/l}\sum_{n=0}^{\infty}q^n\,|n\rangle_E,\qquad q=e^{-2\tau/l}.
\label{eq:arb:AdS3_bulk_local_ket}
\end{equation}

\subparagraph{Flat limit.} Introduce $x=n/l$ and $|x\rangle=\sqrt{l}\,|n\rangle_E|_{n=\lfloor lx\rfloor}$. Using $G_0\,a_{\lfloor lx\rfloor}(\Delta)\big|_{d=2}\sim\frac{1}{2\pi}\frac{1}{l}$ we obtain
\begin{equation}
|\Psi^{E}_{\mathrm{AdS}_3}(\tau)\rangle
\xrightarrow[l\to\infty]{} \int_0^\infty dx\,
\frac{1}{\sqrt{2\pi}}\,e^{-(\xi+2x)\tau}\,|x\rangle.
\label{eq:arb:AdS3_state_limit_x}
\end{equation}
Switch to the energy basis $E=\xi+2x$, $|E\rangle=\frac{1}{\sqrt{2}}|x\rangle$ (so $\langle E|E'\rangle=\delta(E-E')$). Then
\begin{equation}
|\Psi^{E}_{\mathrm{AdS}_3}(\tau)\rangle
\xrightarrow[l\to\infty]{} \int_{\xi}^{\infty} dE\,
\sqrt{\rho_3(E)}\,e^{-E\tau}\,|E\rangle,\qquad \rho_3(E)=\frac{1}{4\pi}.
\label{eq:arb:AdS3_state_limit_E}
\end{equation}
In Flat$_3$ the momentum basis is normalized as $\langle\mathbf{p}|\mathbf{p}'\rangle=(2\pi)^2\,2E_{\mathbf{p}}\,\delta^{(2)}(\mathbf{p}-\mathbf{p}')$, and the energy basis is related by
\begin{equation}
|E\rangle = \frac{1}{\sqrt{\rho_3(E)}}\int\frac{d^2p}{(2\pi)^2\,2E_{\mathbf{p}}}\,
\delta\!\left(E-E_{\mathbf{p}}\right)|\mathbf{p}\rangle,\qquad
\rho_3(E)=\int\frac{d^2p}{(2\pi)^2\,2E_{\mathbf{p}}}\,
\delta\!\left(E-E_{\mathbf{p}}\right)=\frac{1}{4\pi}.
\label{eq:arb:Flat3_E_basis}
\end{equation}
Substituting gives the flat space bulk local state in momentum representation
\begin{equation}
|\Psi^{E}_{\mathrm{AdS}_3}(\tau)\rangle
\xrightarrow[l\to\infty]{} |\phi^{E}_{\mathrm{flat}_3}(\tau,\mathbf 0)\rangle
= \int\frac{d^2p}{(2\pi)^2\,2E_{\mathbf{p}}}\ e^{-\tau E_{\mathbf{p}}}\ |\mathbf{p}\rangle.
\label{eq:arb:AdS3_to_flat3}
\end{equation}

\subparagraph{Comment on the AdS$_3$ dual-basis issue.}
The ``bra vs. ket scaling" issue encountered in the AdS$_3$ analysis when using the BMS/flat basis (See Subsubsection \ref{subsubsec:scaling}) arises because the change of basis from AdS$_3$ descendants to the flat descendants involves an $l$-dependent Gram matrix whose entries diverge in the limit. Consequently, Hermitian conjugation and the flat limit do not commute at each level. In contrast, the momentum-like orthonormal basis $|n\rangle_E$ used here is well-behaved: its elements have finite norms and the state-level flat limit is manifest, connecting directly to the standard relativistic momentum basis of Flat$_3$.

\subparagraph{Comment on AdS$_3$ vs higher dimensions.}
For $d=2$ the coefficient $a_n(\Delta)=1$ is level independent, so the flat limit can be taken uniformly: one may exchange the limit $l\to\infty$ with the infinite sum over $n$, obtaining a geometric series that directly yields the flat space momentum representation. In higher dimensions ($d\ge 3$) the coefficient $a_n(\Delta)$ grows with $n$ as $n^{(d-2)/2}$, forcing the dominant contribution to come from the scaling window $n\sim l$. The limit and the infinite sum do not commute, and the correct procedure is to rewrite the sum as a Riemann integral over the scaling variable $x=n/l$ before taking $l\to\infty$. This distinction is captured by the dimension-dependent factor $\bigl(\frac{d}{2}\bigr)_n/(\Delta-\frac{d}{2}+1)_n$, which for $d=2$ equals $1$ and for $d\ge 3$ grows with $n$.

\subsubsection{Bulk local state in the tilde basis}
\label{subsubsec:tilde_general}

We now extend the tilde-basis construction to general dimension $d+1$. The procedure closely follows the four-dimensional case, and the final expression for the bulk local state takes the same simple form. We first review the flat space construction, then show how it emerges from the flat limit of an analogous construction in $\text{AdS}_{d+1}$.

\paragraph{Flat space construction.}
Let $|\xi\rangle$ be the scalar primary satisfying eqs.\,\eqref{eq:general_flat_primary_energy} -- \eqref{eq:general_flat_primary_spin}. Define the quadratic Casimir of the boost subalgebra
\[
C_K := K_a K^a,
\]
and introduce the rotation-invariant descendants
\[
|n\rangle := C_K^{\,n} |\xi\rangle,\qquad n=0,1,2,\dots
\]
These states are annihilated by the rotation generators $J_{ab}$ and form a basis for the space of rotation-invariant states.

The action of the spatial translation generators on this basis can be computed using the commutation relations. For general spacetime dimension $d$, one finds
\begin{equation}
\begin{aligned}
P_a |n\rangle = -i\,\frac{2^{1-n}\xi}{\sqrt{16C_K+(d+1)^2}} 
\Bigl[ &\bigl(-\sqrt{16C_K+(d+1)^2}+2C_K+d-1\bigr)^{n} \\
&\quad - \bigl(\sqrt{16C_K+(d+1)^2}+2C_K+d-1\bigr)^{n} \Bigr] K_a|\xi\rangle.
\end{aligned}
\label{eq:general_Pa_on_n}
\end{equation}
In the four-dimensional case $d=3$, this reduces to the expression given in eq.\,\eqref{eq:bar_n_explicit}. Eq.\,\eqref{eq:general_Pa_on_n} shows that $P_a|n\rangle$ is a linear combination of states $K_a|m\rangle$ with $m\le n-1$, i.e. the action is lower-triangular in the index $n$.

Because the matrix representing $P_a$ in the basis $\{|n\rangle\}$ is lower-triangular, we can perform a similarity transformation to a new basis $\{|\tilde n\rangle\}$ in which the action of $P_a$ takes a particularly simple form. Define $|\tilde n\rangle$ as a linear combination
\[
|\tilde n\rangle = \sum_{k=0}^{n} S_{nk}\,|k\rangle,\qquad S_{nn}=1,
\]
with the requirement that
\begin{equation}
P_a |\tilde n\rangle = \lambda_n K_a |\widetilde{n-1}\rangle,\qquad |\widetilde{-1}\rangle:=0.
\label{eq:general_tilde_condition_flat}
\end{equation}
The coefficient $\lambda_n$ is fixed by the highest-weight part of the action of $P_a$ on the original basis. From eq.\,\eqref{eq:general_Pa_on_n} one can extract the term proportional to $K_a|n-1\rangle$ and after some algebra we find the answer is $\lambda_n = 2i n \xi$ independent of the dimension $d$. Thus the tilde basis satisfies
\begin{equation}
P_a |\tilde n\rangle = 2i n \xi\, K_a |\widetilde{n-1}\rangle.
\label{eq:general_tilde_action_flat}
\end{equation}
Expanding the bulk local state $|\phi(t,\mathbf{0})\rangle$ in the tilde basis,
\[
|\phi(t,\mathbf{0})\rangle = \sum_{n=0}^{\infty} f_n(t)\,|\tilde n\rangle,
\]
and imposing $(K_a - tP_a)|\phi(t,\mathbf{0})\rangle = 0$ together with the condition \eqref{eq:general_tilde_action_flat}, we land on the recursive relation
\[
f_n - t\,(2i(n+1)\xi)\,f_{n+1}=0\quad\Longrightarrow\quad
f_{n+1} = \frac{1}{2i(n+1)\xi t}\,f_n.
\]
Choosing the normalization $f_0=1$, the bulk local state in Minkowski space in any dimension $d+1$ takes the universal form
\begin{equation}
|\phi(t,\mathbf{0})\rangle = \sum_{n=0}^{\infty} \frac{1}{n!\,(2i\xi t)^n}\;|\tilde n\rangle.
\label{eq:general_bulk_state_tilde_flat}
\end{equation} 

\paragraph{AdS construction and the flat limit.}
In $\text{AdS}_{d+1}$ the generators depend on the radius $l$. The stabilizer condition for a point at the center becomes \[\Bigl(e^{-i\frac{t}{l}}P_a(l) - e^{i\frac{t}{l}}K_a(l)\Bigr)|\Psi_{\mathrm{AdS}}(t,0,0,0)\rangle = 0,\qquad 
J_{ab}(l)\,|\Psi_{\mathrm{AdS}}(t,0,0,0)\rangle = 0 .\] To construct a basis well-suited to this condition, we need a modified version of the tilde basis. Define the $l$-dependent descendant basis
\[
|n\rangle_l := \bigl(C_K(l)\bigr)^n|\Delta\rangle,\qquad C_K(l):=K_a(l)K^a(l),
\]
and seek a new basis $|\tilde n\rangle_l$ such that
\begin{equation}
\Bigl(P_a(l)+\frac{i}{l}K_a(l)\Bigr)|\tilde n\rangle_l = \lambda_n(l)\,K_a(l)\,|\widetilde{n-1}\rangle_l,\qquad |\widetilde{-1}\rangle_l\equiv0.
\label{eq:modified_tilde_AdS}
\end{equation}
The combination $P_a(l)+\frac{i}{l}K_a(l)$ is proportional to the AdS lowering operator $-\frac{1}{l}\mathcal{K}_a$, which makes the computation of its action on descendant states tractable.

Using the AdS algebra, one can evaluate the highest-level term in the left-hand side of eq.\,\eqref{eq:modified_tilde_AdS}. For a state $|\tilde n\rangle_l$ whose highest-level component is proportional to $\mathcal{P}_a^{2n}|\Delta\rangle$, the action of $\mathcal{K}_a$ produces a term of level $2n-1$. The calculation proceeds exactly as in four dimensions, but with the commutator $[\mathcal{K}_a,\mathcal{P}^2]$ now yielding a dimension-dependent constant
\[
[\mathcal{K}_a,\mathcal{P}^2] = 4\mathcal{P}_a\mathcal{D} - 4\mathcal{P}^b\mathcal{M}_{ab} + (4-2d)\mathcal{P}_a.
\]
This leads to the explicit result for the highest-level coefficient
\[
\Bigl(\bigl(P_a(l)+\frac{i}{l}K_a(l)\bigr)|\tilde n\rangle_l\Bigr)_{\text{level }2n-1}
= -\frac{1}{l}\Bigl(-\frac{1}{4}\Bigr)^n \bigl(2n(2\Delta+2n-d)\bigr)\,\mathcal{P}_a^{2n-1}|\Delta\rangle.
\]
On the other hand, the right-hand side of eq.\,\eqref{eq:modified_tilde_AdS} at the same level is
\[
\bigl(\lambda_n(l)K_a(l)|\widetilde{n-1}\rangle_l\bigr)_{\text{level }2n-1}
= \lambda_n(l)\Bigl(-\frac{i}{2}\Bigr)\Bigl(-\frac{1}{4}\Bigr)^{n-1}\mathcal{P}_a^{2n-1}|\Delta\rangle.
\]
Comparing the two expressions fixes
\begin{equation}
\;\lambda_n(l) = \frac{i n}{l}\,\bigl(2\Delta+2n-d\bigr)\; .
\label{eq:lambda_n_AdS_general}
\end{equation}
In the flat limit $l\to\infty$ with $\Delta=\xi l+\mathcal{O}(1)$, this becomes $\lambda_n(l) \to 2i n\xi$, recovering the flat space value.
The proof of the existence of the tilde basis in eq.\,\eqref{eq:modified_tilde_AdS} is provided in Appendix \ref{app:ads tilde}.

\paragraph{Bulk local state in AdS and its flat limit.}
Expanding the Lorentzian bulk local state in the modified tilde basis,
\[
|\Psi_{\mathrm{AdS}}(t,\mathbf{0})\rangle = \sum_{n=0}^{\infty} f_n(l,t)\,|\tilde n\rangle_l,
\]
and imposing the stabilizer condition together with eq.\,\eqref{eq:modified_tilde_AdS} leads to a recursive relation for the coefficients. Following the same steps as in the four-dimensional case, the solution with the natural normalization $f_0=1$ is
\[
f_n(l,t) = \left(\frac{e^{-it/l}}{l\sin(t/l)}\right)^{\!n}\prod_{k=1}^{n}\frac{1}{\lambda_k(l)}.
\]
Using the coefficient \eqref{eq:lambda_n_AdS_general} and the flat-limit behavior $\sin(t/l)\sim t/l$, $e^{-it/l}\sim 1$, we obtain
\[
\lim_{l\to\infty} f_n(l,t) = \frac{1}{n!\,(2i\xi t)^n},
\]
while the basis vectors reduce to the flat space tilde basis: $|\tilde n\rangle_l \to |\tilde n\rangle$. Consequently,
\[
\lim_{l\to\infty} |\Psi_{\mathrm{AdS}}(t,\mathbf{0})\rangle = \sum_{n=0}^{\infty} \frac{1}{n!\,(2i\xi t)^n}\,|\tilde n\rangle = |\phi(t,\mathbf{0})\rangle_{\mathrm{flat}},
\]
which is precisely the bulk local state in Minkowski space constructed directly from the Poincar\'e algebra \eqref{eq:general_bulk_state_tilde_flat}. This demonstrates that the tilde-basis construction generalizes seamlessly to arbitrary dimensions and yields a smooth flat limit identical in form to the four-dimensional case.

Comparing the universal expression for the bulk local state in the tilde basis \eqref{eq:general_bulk_state_tilde_flat},
\begin{equation}
|\phi(t,\mathbf{0})\rangle = \sum_{n=0}^{\infty} \frac{1}{n!\,(2i\xi t)^n}\;|\tilde n\rangle,
\end{equation}
with its three-dimensional counterpart in the flat basis \eqref{eq: t solution}, 
\begin{equation}
|\phi(t,0,0)\rangle = e^{i\xi(t-1)}\sum_{k=0}^{\infty}\frac{2^{-k}\xi^{-k}(-i/t)^{k+1}}{k!}\,|k\rangle_{\text{Flat$_3$}},
\end{equation}
we observe that after removing the overall factor $e^{i\xi(t-1)}(-i/t)$ in eq.\,\eqref{eq: t solution}, the two series become identical up to an alternating sign.  Explicitly, identifying
\[
|\tilde n\rangle = (-1)^n |n\rangle_{\text{Flat$_3$}},
\]
makes eq.\,\eqref{eq:general_bulk_state_tilde_flat} coincide with eq.\,\eqref{eq: t solution} term-by-term. This relation can be also checked explicitly  algebraic relation as in Ref.\,\cite{Hao:2025btl}.  This relation demonstrates that the tilde basis in general dimensions is equivalent, up to a simple sign factor, to the flat basis used in the three-dimensional analysis.  It also underscores the special simplicity of the $d=2$ case while revealing the universal algebraic structure that underlies the tilde construction in any dimension. 
Finally, we can exploit the explicit relations between the descendant states \eqref{eq:arb:AdS_en_def} and the tilde basis $\{|\tilde n\rangle_l\}_n$ as follows
\begin{align}
|\tilde n\rangle_l &= n!\,\Bigl(\Delta-\frac{d}{2}+1\Bigr)_n
\sum_{k=1}^{n}\frac{(-1)^{n-k}}{4^k k!\,\bigl(\Delta-\frac{d}{2}+1\bigr)_k}\binom{n-1}{k-1}\,
\frac{1}{\sqrt{G_k}}\,|e_k\rangle \qquad (n\ge1), \label{eq:arb:tilde_to_en}\\
|\tilde0\rangle_l &= |e_0\rangle .
\end{align}
The detailed derivation can be found in Appendix \ref{subsec:relation_descendant_general}.


    \section{Conclusion and discussion} \label{sec:conclusion}

In this paper, we revisited and extended the reconstruction of bulk local states in flat holography.
Our main aim was to identify a physically sensible representation on the boundary side and to show
that it reproduces the expected bulk Green's functions, both directly in flat space and through the
flat limit of AdS bulk local states. For scalar states, the flat limit of the AdS highest-weight
conditions naturally leads to the induced representation of the Poincar\'e subgroup,
\begin{equation}
P_0|\xi\rangle=\xi|\xi\rangle,\qquad P_a|\xi\rangle=0,
\end{equation}
supplemented by rotational invariance. We argued that this representation provides the appropriate
starting point for bulk reconstruction in four-dimensional flat holography.

We first revisited the three-dimensional case. In the flat basis, the bra and ket states
appear to scale differently in the flat limit, which at first sight seems incompatible with the
standard Hermitian conjugation as claimed in Ref.\,\cite{Hao:2025btl}. We clarified that this puzzle is caused by the $l$-dependent
divergence of the Gram matrix: taking the Hermitian conjugate and taking the flat limit do not
commute level by level. Introducing the dual basis as a complete basis removes this divergence and yields a smooth flat
limit of the bra state, reproducing the expected flat Green's function. This ensures that the bulk local state of bra is described by the time evolution of ${}^\vee\bra*{0}$ and it reduces to ${}^\vee\bra*{0}$ itself at $\tau=0$.

We then turned to the four-dimensional case and its higher-dimensional generalization. In flat
space, the bulk local state can be constructed directly from the stabilizer condition in the induced
representation. On the AdS side, we developed a momentum-descendant basis built from the
rotationally invariant descendants $(P^2)^n|\Delta\rangle$, and showed that the corresponding
Euclidean bulk local state reproduces the AdS Green's function in a hypergeometric form. In the
flat limit $\Delta=\xi l+O(1)$, this discrete descendant expansion turns into the standard continuum
momentum representation of the flat bulk local state,
\begin{equation}
|\phi^{E}_{\mathrm{flat}_{d+1}}(\tau,\mathbf 0)\rangle
=
\int\frac{d^d p}{(2\pi)^d\,2E_{\vec p}}\,
e^{-\tau E_{\vec p}}\,|\vec p\rangle,
\end{equation}
and the corresponding Green's function becomes the massive Euclidean propagator written in terms
of the modified Bessel function $K_{\frac{d-1}{2}}$.

A key lesson of the AdS$_4$ and general AdS$_{d+1}$ analysis is that the flat limit is non-uniform
in the descendant level. A naive termwise limit at fixed level fails to reproduce the correct flat
answer. Instead, the dominant contribution comes from the scaling window $n\sim l$. In this regime,
the descendant sum reorganizes into a Riemann sum, which becomes a continuum spectral integral
over energy or momentum. This mechanism explains, at the level of both states and Green's
functions, how the AdS hypergeometric expression reduces to the flat-space Bessel propagator. In
this respect, the momentum basis provides the cleanest description of the flat limit and also makes
clear why the AdS$_3$ case is special, where the growth of the coefficient is level independent.

We also discussed an alternative description based on the tilde basis in AdS$_4$. This basis is
useful for analyzing the algebraic structure of the flat limit and for making contact with the
flat-space descendant construction. At the same time, the momentum-basis description appears to
be more robust conceptually, since it avoids the dual-basis subtlety that arises in the AdS$_3$
flat analysis and generalizes more naturally to arbitrary spacetime dimension.

\subsubsection*{Interacting case}
Finally, let us note that our method remains robust even when interactions are introduced.
We derived the scalar bulk local excitation in Flat$_{d+1}$ by solving two conditions.
One is the covariance condition under isometry transformations at $(t,0,0,0)$, given in eq.\,\eqref{eq:flat_stabilizer}.
The other is the Casimir associated with translations, which is nothing but the equation of motion for a free scalar field.

When interactions are introduced, as long as they preserve the structure of spacetime, the first condition, namely the covariance under isometry transformations at $(t,0,0,0)$, remains unchanged.
By contrast, the equation of motion is modified to the interacting one, and this corresponds to the bulk mass spectrum.
As a result, the bulk local state in the interacting theory can be expressed as a superposition of the free local states:
\begin{align}
    \ket*{\phi(t,\mathbf 0)}^{\text{int}}=\sum_{\xi}c_\xi \ket*{\phi(t,\mathbf 0)}^{\text{free}}_\xi,
\end{align}
where $c_\xi$ is theory-dependent coefficient.

The bulk local state of an interacting scalar in AdS$_{d+1}$ was discussed in Ref.\,\cite{Nakayama:2015mva}, where it is also concluded that it can be represented as a superposition of free local states.
Since we know the flat limit of each free excitation in AdS$_{d+1}$, we can naturally expect that the flat limit of the interacting bulk local excitation is given by a superposition of them.
As a result, it is also natural to expect that our calculation can be straightforwardly reformulated in real time, with the $i \epsilon$ prescription.\footnote{Strictly speaking, however, one must prove that the superposition and the flat limit commute.}

\subsubsection*{HKLL reconstruction}
There are several natural directions for future work. It would be important to extend the present
construction to spinning states and to representations beyond the scalar sector, to understand the
role of the full BMS$_4$ algebra beyond its global Poincar\'e subgroup, and to clarify how the present
state-based reconstruction is related to a genuinely boundary-local description in flat holography.
It would also be interesting 
to investigate whether the momentum-basis picture developed here can serve as a useful starting
point for a flat-space analogue of HKLL-type bulk reconstruction \cite{Hamilton:2005ju,Hamilton:2006az}.

We hope that the results of this paper help clarify the representation-theoretic foundation of bulk
local states in flat holography, and that they provide a concrete bridge between AdS bulk
reconstruction and its flat-space counterpart.

\section*{Acknowledgments}
We are grateful to Bin Chen, Song He, Wenxin Lai, Wei Song, Tadashi Takayanagi, and Jie-Qiang Wu for valuable discussions.

PH is supported by the NSFC special fund for theoretical physics No.\,12447108 and MEXT KAKENHI Grant-in-Aid for Transformative Research Areas (A) through the Extreme Universe collaboration: Grant Number 21H05182. 
KS is supported by Grant-in-Aid for JSPS Fellows No.25KJ1498. 
YS is supported by Grant-in-Aid for JSPS Fellows No.23KJ1337.
ST is supported by Division of Graduate Studies Donor Designated Scholarship by Fujitsu Limited and Grant for Overseas Research by the Division of Graduate Studies. 
We are grateful to the YITP workshop YITP-T-25-01 held at YITP, Kyoto University, where a part of this work was done. PH thanks RIKEN Center for Interdisciplinary Theoretical and Mathematical Sciences where discussions during "New computational methods in quantum field theory 2026" were useful in this work.

\appendix
\section{Explicit expressions for symmetry generators}
\label{app:explicit_generators}

In this appendix, we collect the explicit differential operator realizations of the symmetry generators that were omitted from the main text for brevity. These expressions are useful for explicit computations and for verifying the algebraic relations discussed in Sec.~\ref{sec:algebra_representation}.

\subsection{Poincar\'e generators in spherical coordinates}
\label{app:poincare_spherical}

In the spherical coordinates $(t,r,\theta,\phi)$ of 4d Minkowski space, defined by
\[
x^0=t,\quad x^1=r\sin\theta\cos\phi,\quad x^2=r\sin\theta\sin\phi,\quad x^3=r\cos\theta,
\]
the Poincar\'e generators take the following form:
\begin{align}
    P_0 &= i\partial_t, \\
    P_1 &= i\Big(\sin\theta\cos\phi\,\partial_r+\frac{\cos\theta\cos\phi}{r}\,\partial_\theta-\frac{\sin\phi}{r\sin\theta}\,\partial_\phi\Big), \\
    P_2 &= i\Big(\sin\theta\sin\phi\,\partial_r+\frac{\cos\theta\sin\phi}{r}\,\partial_\theta+\frac{\cos\phi}{r\sin\theta}\,\partial_\phi\Big), \\
    P_3 &= i\Big(\cos\theta\,\partial_r-\frac{\sin\theta}{r}\,\partial_\theta\Big), \\[4pt]
    J_{01} &= i\Big(-t\sin\theta\cos\phi\,\partial_r - t\frac{\cos\theta\cos\phi}{r}\,\partial_\theta + t\frac{\sin\phi}{r\sin\theta}\,\partial_\phi - r\sin\theta\cos\phi\,\partial_t\Big), \\
    J_{02} &= i\Big(-t\sin\theta\sin\phi\,\partial_r-\frac{t\cos\theta\sin\phi}{r}\,\partial_\theta-\frac{t\cos\phi}{r\sin\theta}\,\partial_\phi-r\sin\theta\sin\phi\,\partial_t\Big), \\
    J_{03} &= i\Big(-t\cos\theta\,\partial_r+\frac{t\sin\theta}{r}\,\partial_\theta-r\cos\theta\,\partial_t\Big), \\
    J_{12} &= i\partial_{\phi}, \\
    J_{13} &= i\Big(-\cos\phi\,\partial_\theta + \frac{\cos\theta\sin\phi}{\sin\theta}\,\partial_\phi\Big), \\
    J_{23} &= i\Big(-\sin\phi\,\partial_\theta - \frac{\cos\theta\cos\phi}{\sin\theta}\,\partial_\phi\Big).
\end{align}
These expressions satisfy the Poincar\'e algebra \eqref{eq:poincare_lorentz}-\eqref{eq:poincare_trans}.

\subsection{AdS$_4$ isometry generators in global coordinates}
\label{app:ads_generators}

For AdS$_4$ with radius $l$, embedded in $\mathbb{R}^{2,3}$ and parametrized by global coordinates $(\tau,\rho,\theta,\phi)$\footnote{The metric of global coordinates is given by $ds^2=-\cosh^2\rho \,\,d\tau^2+ d\rho^2+\sinh^2\rho(d\theta^2+\sin^2\theta d\phi^2)$.} , the ten isometry generators $\mathcal{J}_{\mu\nu} = i(x_\mu\partial_\nu - x_\nu\partial_\mu)$ ($\mu,\nu=-1,0,1,2,3$) are given by:
\begin{align}
    \mathcal{J}_{-10} &= -i\partial_\tau, \\
    \mathcal{J}_{12} &= i\partial_\phi, \\
    \mathcal{J}_{13} &= i\Big(-\cos\phi\partial_\theta + \frac{\cos\theta\sin\phi}{\sin\theta}\partial_\phi\Big), \\
    \mathcal{J}_{23} &= i\Big(-\sin\phi\partial_\theta - \frac{\cos\theta\cos\phi}{\sin\theta}\partial_\phi\Big), \\[4pt]
    \mathcal{J}_{01} &= i\Big(-\sin\tau\sin\theta\cos\phi\partial_\rho - \frac{\sin\tau\cos\theta\cos\phi}{\tanh\rho}\partial_\theta + \frac{\sin\tau\sin\phi}{\tanh\rho\sin\theta}\partial_\phi \notag\\
    &\qquad -\cos\tau\sin\theta\cos\phi\tanh\rho\partial_\tau\Big), \\[4pt]
    \mathcal{J}_{02} &= i\Big(-\sin\tau\sin\theta\sin\phi\partial_\rho - \frac{\sin\tau\cos\theta\sin\phi}{\tanh\rho}\partial_\theta - \frac{\sin\tau\cos\phi}{\tanh\rho\sin\theta}\partial_\phi \notag\\
    &\qquad -\cos\tau\sin\theta\sin\phi\tanh\rho\partial_\tau\Big), \\[4pt]
    \mathcal{J}_{03} &= i\Big(-\sin\tau\cos\theta\partial_\rho + \frac{\sin\tau\sin\theta}{\tanh\rho}\partial_\theta -\cos\tau\cos\theta\tanh\rho\partial_\tau\Big), \\[4pt]
    \mathcal{J}_{-11} &= i\Big(-\cos\tau\sin\theta\cos\phi\partial_\rho - \frac{\cos\tau\cos\theta\cos\phi}{\tanh\rho}\partial_\theta + \frac{\cos\tau\sin\phi}{\tanh\rho\sin\theta}\partial_\phi \notag\\
    &\qquad +\sin\tau\sin\theta\cos\phi\tanh\rho\partial_\tau\Big), \\[4pt]
    \mathcal{J}_{-12} &= i\Big(-\cos\tau\sin\theta\sin\phi\partial_\rho - \frac{\cos\tau\cos\theta\sin\phi}{\tanh\rho}\partial_\theta - \frac{\cos\tau\cos\phi}{\tanh\rho\sin\theta}\partial_\phi \notag\\
    &\qquad +\sin\tau\sin\theta\sin\phi\tanh\rho\partial_\tau\Big), \\[4pt]
    \mathcal{J}_{-13} &= i\Big(-\cos\tau\cos\theta\partial_\rho + \frac{\cos\tau\sin\theta}{\tanh\rho}\partial_\theta +\sin\tau\cos\theta\tanh\rho\partial_\tau\Big).
\end{align}
These satisfy the $\mathfrak{so}(2,3)$ commutation relations:
\[
[\mathcal{J}_{\mu\nu},\mathcal{J}_{\rho\sigma}] = i\big(\eta_{\nu\rho}\mathcal{J}_{\mu\sigma}-\eta_{\mu\rho}\mathcal{J}_{\nu\sigma}-\eta_{\nu\sigma}\mathcal{J}_{\mu\rho}+\eta_{\mu\sigma}\mathcal{J}_{\nu\rho}\big),
\]
with $\eta = \mathrm{diag}(-1,-1,+1,+1,+1)$. The basis for AdS/CFT, $\{\mathcal{H},\mathcal{P}_a,\mathcal{K}_a,\mathcal{M}_{ab}\}$, is obtained via the linear combinations:
\[
\mathcal{H} = \mathcal{J}_{-10},\quad \mathcal{M}_{ab} = -i\mathcal{J}_{ab},\quad \mathcal{P}_a = \mathcal{J}_{-1a}+i\mathcal{J}_{0a},\quad \mathcal{K}_a = \mathcal{J}_{-1a}-i\mathcal{J}_{0a}.
\]

\subsection{Poincar\'e generators in boundary coordinates}
\label{app:poincare_boundary}

In the boundary coordinates $(z,\bar z, y)$ introduced in Sec.~\ref{subsec:boundary_algebra}, the Poincar\'e generators take the following explicit form as differential operators:
\begin{align}
    P_0 &= \frac{1+z\bar z}{2}\partial_y, \\
    P_1 &= -\frac{z+\bar z}{2}\partial_y, \\
    P_2 &= -\frac{z-\bar z}{2i}\partial_y, \\
    P_3 &= \frac{z\bar z-1}{2}\partial_y, \\[4pt]
    J_1 &= \frac{1}{2}\big[(z^2-1)\partial_z - (\bar z^2-1)\partial_{\bar z} + (z-\bar z)y\partial_y\big], \\
    J_2 &= -\frac{i}{2}\big[(z^2+1)\partial_z + (\bar z^2+1)\partial_{\bar z} + (z+\bar z)y\partial_y\big], \\
    J_3 &= (1-i)z\partial_z - (1+i)\bar z\partial_{\bar z} - i y\partial_y, \\[4pt]
    K_1 &= \frac{i}{2}\big[(1-z^2)\partial_z + (1-\bar z^2)\partial_{\bar z} - (z+\bar z)y\partial_y\big], \\
    K_2 &= \frac{1}{2}\big[-(1+z^2)\partial_z + (1+\bar z^2)\partial_{\bar z} + (\bar z - z)y\partial_y\big], \\
    K_3 &= (1-i)(z\partial_z + \bar z\partial_{\bar z} + y\partial_y).
\end{align}
These expressions are obtained by inverting the linear relations between the Poincar\'e and BMS$_4$ generators given in Sec.~\ref{subsec:boundary_algebra}. One can verify that they satisfy the Poincar\'e algebra \eqref{eq:poincare_lorentz}-\eqref{eq:poincare_trans} when acting on functions of $(z,\bar z, y)$. The finite transformations generated by these operators are listed in Table~\ref{tab:poincare_finite_transformations}.

\subsection{Relation between Poincar\'e and BMS$_4$ generators}
\label{app:poincare_bms_map}

For completeness, we repeat here the linear map between the Poincar\'e generators $\{P_\mu, J_a, K_a\}$ and the global BMS$_4$ generators $\{L_n,\bar L_n, M_{r,s}\}$ used in the main text:
\begin{align}
    M_{0,0} &= -P_0 + P_3, \quad M_{1,1} = -P_0 - P_3, \quad M_{1,0} = P_1 + iP_2, \quad M_{0,1} = P_1 - iP_2, \\
    L_{-1} &= \tfrac{1}{2}(iK_1+K_2+J_1-iJ_2), \quad L_1 = \tfrac{1}{2}(-iK_1+K_2-J_1-iJ_2), \\
    L_0 &= \tfrac{1}{2}(-J_3-iK_3), \quad \bar L_0 = \tfrac{1}{2}(J_3-K_3), \\
    \bar L_{-1} &= \tfrac{1}{2}(iK_1-K_2-J_1-iJ_2), \quad \bar L_1 = \tfrac{1}{2}(-iK_1-K_2+J_1-iJ_2).
\end{align}
These relations are essential for translating bulk physical conditions into boundary conditions, as done in Sec.~\ref{subsec:boundary_conditions}.

    \section{Conventions and OS reflection}
    \label{app:OS}
     In the Lorentzian signature, the time evolution of bra and ket is described by
 \be
\ket*{\Psi(t)}=e^{-itH}\ket*{\Psi(0)},\quad \bra*{\Psi(t)}=e^{itH}\bra*{\Psi(0)}.
 \ee
However, in the Euclidean signature, the time evolution is given by
\be
\ket*{\Psi(\tau)}=e^{-\tau H}\ket*{\Psi(0)},\quad \bra*{\Psi(t)}=e^{-\tau H}\bra*{\Psi(0)}.
 \ee
Thus, to reproduce the correct behavior of the Green's function, we  need to take into account of time reflection or Osterwalder-Schrader reflection.

    We Wick-rotate $t=-i\tau$, $\tau\in\mathbb{R}$, and define the Osterwalder-Schrader reflected bra as
    \begin{equation}
        \bra*{\Phi(\tau)}_{\rm OS} := \big(\ket*{\Phi(-\tau)}\big)^\dagger .
    \end{equation}
    For $\tau_2>\tau_1$ the Euclidean two-point function along the time axis is
    \begin{equation}
        G^E(\tau_2-\tau_1) := {}_{\rm OS}\langle \Psi^E(\tau_1)|\,\Psi^E(\tau_2)\rangle .
    \end{equation}

\section{Norm of the descendant  basis}
\label{subsubsec:derive_Gn}

We work with the global conformal algebra $\mathfrak{so}(2,d)$, with Euclidean spatial indices
$a,b=1,\dots,d$ and commutators
\begin{equation}
[\mathcal{H},\mathcal{P}_a]=\mathcal{P}_a,\qquad [\mathcal{H},\mathcal{K}_a]=-\mathcal{K}_a,\qquad [\mathcal{K}_a,\mathcal{P}_b]=2(\delta_{ab}\mathcal{H}-\mathcal{M}_{ab}),
\end{equation}
together with
\begin{equation}
[\mathcal{M}_{ab},\mathcal{P}_c]=\delta_{bc}\mathcal{P}_a-\delta_{ac}\mathcal{P}_b,
\end{equation}
and the Hermitian rule $\mathcal{P}_a^\dagger=\mathcal{K}_a$, $\mathcal{H}^\dagger=\mathcal{H}$, $\mathcal{M}_{ab}^\dagger=\mathcal{M}_{ab}$.

\subsection{Primary and descendant}
Let $|\Delta\rangle$ be a scalar highest-weight state
\begin{equation}
\mathcal{H}|\Delta\rangle=\Delta|\Delta\rangle,\qquad \mathcal{K}_a|\Delta\rangle=0,\qquad \mathcal{M}_{ab}|\Delta\rangle=0,
\end{equation}
normalized by
\begin{equation}
\langle\Delta|\Delta\rangle \equiv G_0.
\end{equation}
We define
\begin{equation}
\mathcal{P}^2:=\mathcal{P}_a\mathcal{P}_a,\quad \mathcal{K}^2:=\mathcal{K}_a\mathcal{K}_a,
\end{equation}
and the rotationally invariant descendants
\begin{equation}
|e_n\rangle := (\mathcal{P}^2)^n|\Delta\rangle,\qquad n=0,1,2,\dots,
\quad
\langle e_n| := \langle\Delta|(\mathcal{K}^2)^n.
\end{equation}
Since $[\mathcal{H},\mathcal{P}^2]=2\mathcal{P}^2$, we have $\mathcal{H}|e_n\rangle=(\Delta+2n)|e_n\rangle$, hence
$\langle e_m|e_n\rangle=0$ for $m\neq n$ by Hermiticity of $\mathcal{H}$.

First let us  compute $[\mathcal{K}_a,\mathcal{P}^2]$.
Using $[\mathcal{K}_a,\mathcal{P}_b]=2(\delta_{ab}\mathcal{H}-\mathcal{M}_{ab})$, we obtain
\begin{align}
[\mathcal{K}_a,\mathcal{P}^2]
&=[\mathcal{K}_a,\mathcal{P}_b\mathcal{P}_b]=[\mathcal{K}_a,\mathcal{P}_b]\mathcal{P}_b+\mathcal{P}_b[\mathcal{K}_a,\mathcal{P}_b] \nonumber\\
&=2(\delta_{ab}\mathcal{H}-\mathcal{M}_{ab})\mathcal{P}_b+2\mathcal{P}_b(\delta_{ab}\mathcal{H}-\mathcal{M}_{ab}) \nonumber\\
&=2(\mathcal{H}\mathcal{P}_a+\mathcal{P}_a\mathcal{H})-2(\mathcal{M}_{ab}\mathcal{P}_b+\mathcal{P}_b\mathcal{M}_{ab}). \label{eq:KaP2_mid}
\end{align}
From $[\mathcal{H},\mathcal{P}_a]=\mathcal{P}_a$, we get $\mathcal{H}\mathcal{P}_a+\mathcal{P}_a\mathcal{H}=2\mathcal{P}_a\mathcal{H}+\mathcal{P}_a$.
Moreover, contracting $[\mathcal{M}_{ab},\mathcal{P}_c]=\delta_{bc}\mathcal{P}_a-\delta_{ac}\mathcal{P}_b$ with $c=b$ gives
\begin{equation}
[\mathcal{M}_{ab},\mathcal{P}_b]=(\delta_{bb}\mathcal{P}_a-\delta_{ab}\mathcal{P}_b)=(d-1)\mathcal{P}_a,
\end{equation}
hence $\mathcal{M}_{ab}\mathcal{P}_b=\mathcal{P}_b\mathcal{M}_{ab}+(d-1)\mathcal{P}_a$ and
$\mathcal{M}_{ab}\mathcal{P}_b+\mathcal{P}_b\mathcal{M}_{ab}=2\mathcal{P}_b\mathcal{M}_{ab}+(d-1)\mathcal{P}_a$.
Substituting into \eqref{eq:KaP2_mid}, we have
\begin{equation}
[\mathcal{K}_a,\mathcal{P}^2]=4\mathcal{P}_a\mathcal{H}-4\mathcal{P}_b\mathcal{M}_{ab}-2(d-2)\mathcal{P}_a.
\label{eq:KaP2_final}
\end{equation}

Next, let us compute the action of $\mathcal{K}_a$ on $|e_n\rangle$.
Using $\mathcal{K}_a|\Delta\rangle=0$, we see
\begin{equation}
\mathcal{K}_a|e_n\rangle = \mathcal{K}_a(\mathcal{P}^2)^n|\Delta\rangle=[\mathcal{K}_a,(\mathcal{P}^2)^n]|\Delta\rangle
=\sum_{m=0}^{n-1}(\mathcal{P}^2)^{n-1-m}[\mathcal{K}_a,\mathcal{P}^2](\mathcal{P}^2)^m|\Delta\rangle.
\end{equation}
Since $|e_m\rangle=(\mathcal{P}^2)^m|\Delta\rangle$ is a scalar, $\mathcal{M}_{ab}|e_m\rangle=0$, and
$\mathcal{H}|e_m\rangle=(\Delta+2m)|e_m\rangle$. Applying \eqref{eq:KaP2_final} on $|e_m\rangle$, we find
\begin{equation}
[\mathcal{K}_a,\mathcal{P}^2]|e_m\rangle
=\big(4(\Delta+2m)-2(d-2)\big)\mathcal{P}_a|e_m\rangle
=4\Big(\Delta+2m-\frac d2+1\Big)\mathcal{P}_a|e_m\rangle.
\end{equation}
Since $[\mathcal{P}^2,\mathcal{P}_a]=0$, all terms in the sum are proportional to $\mathcal{P}_a(\mathcal{P}^2)^{n-1}|\Delta\rangle$
\begin{align}
\mathcal{K}_a|e_n\rangle
&=4\left[\sum_{m=0}^{n-1}\Big(\Delta-\frac d2+1+2m\Big)\right]\mathcal{P}_a(\mathcal{P}^2)^{n-1}|\Delta\rangle \nonumber\\
&=4n\Big(\Delta-\frac d2+n\Big)\,\mathcal{P}_a|e_{n-1}\rangle.
\end{align}
Therefore, we obtain
\begin{equation}
\mathcal{K}_a|e_n\rangle
=
4n\Big(\Delta-\frac d2+n\Big)\,\mathcal{P}_a|e_{n-1}\rangle.
\label{eq:Ka_en}
\end{equation}

Next, let us also evaluate $(\mathcal{K}_a\mathcal{P}_a)|e_m\rangle$.
We write
\begin{equation}
(\mathcal{K}_a\mathcal{P}_a)|e_m\rangle = \big([\mathcal{K}_a,\mathcal{P}_a]+\mathcal{P}_a\mathcal{K}_a\big)|e_m\rangle.
\end{equation}
Contracting $[\mathcal{K}_a,\mathcal{P}_b]=2(\delta_{ab}\mathcal{H}-\mathcal{M}_{ab})$, we see
\begin{equation}
[\mathcal{K}_a,\mathcal{P}_a]=2(\delta_{aa}\mathcal{H}-\mathcal{M}_{aa})=2d\,\mathcal{H},
\end{equation}
since $\mathcal{M}_{aa}=0$. Hence $[\mathcal{K}_a,\mathcal{P}_a]|e_m\rangle=2d(\Delta+2m)|e_m\rangle$.
Next, using eq.\,\eqref{eq:Ka_en} with $n=m$ and contracting with $\mathcal{P}_a$, we obtain
\begin{equation}
\mathcal{P}_a\mathcal{K}_a|e_m\rangle
=
4m\Big(\Delta-\frac d2+m\Big)\,\mathcal{P}_a\mathcal{P}_a|e_{m-1}\rangle
=
4m\Big(\Delta-\frac d2+m\Big)\,|e_m\rangle.
\end{equation}
Combining these, we find the factorized form
\begin{equation}
(\mathcal{K}_a\mathcal{P}_a)|e_m\rangle
=
4(\Delta+m)\Big(\frac d2+m\Big)\,|e_m\rangle.
\label{eq:KaPa_em}
\end{equation}

To proceed, let us derive the action of $\mathcal{K}^2$ on $|e_n\rangle$.
From eq.\,\eqref{eq:Ka_en},
\begin{equation}
\mathcal{K}^2|e_n\rangle
=
\mathcal{K}_a\!\left(4n\Big(\Delta-\frac d2+n\Big)\mathcal{P}_a|e_{n-1}\rangle\right)
=
4n\Big(\Delta-\frac d2+n\Big)\,(\mathcal{K}_a\mathcal{P}_a)|e_{n-1}\rangle.
\end{equation}
Using eq.\,\eqref{eq:KaPa_em} at $m=n-1$ gives
\begin{equation}
\mathcal{K}^2|e_n\rangle
=
16\,n\Big(\Delta-\frac d2+n\Big)(\Delta+n-1)\Big(\frac d2+n-1\Big)\,|e_{n-1}\rangle.
\label{eq:K2_en}
\end{equation}

Finally, we can derive the   the norm $G_n$ using the recursion relation.
We define $G_n:=\langle e_n|e_n\rangle=\langle\Delta|(\mathcal{K}^2)^n(\mathcal{P}^2)^n|\Delta\rangle$.
Using eq.\,\eqref{eq:K2_en},
\begin{align}
G_n
&=\langle\Delta|(\mathcal{K}^2)^{n-1}\big(\mathcal{K}^2|e_n\rangle\big) \nonumber\\
&=16\,n\Big(\Delta-\frac d2+n\Big)(\Delta+n-1)\Big(\frac d2+n-1\Big)\,
\langle\Delta|(\mathcal{K}^2)^{n-1}(\mathcal{P}^2)^{n-1}|\Delta\rangle \nonumber\\
&=16\,n\Big(\frac d2+n-1\Big)(\Delta+n-1)\Big(\Delta-\frac d2+n\Big)\,G_{n-1}.
\end{align}
Thus, we have
\begin{equation}
G_n
=
16\,n\Big(\frac d2+n-1\Big)(\Delta+n-1)\Big(\Delta-\frac d2+n\Big)\,G_{n-1}.
\label{eq:Gn_recursion}
\end{equation}

Iterating \eqref{eq:Gn_recursion} with $G_0=\langle\Delta|\Delta\rangle$ gives
\begin{align}
G_n
&=G_0\prod_{k=1}^n
16k\Big(\frac d2+k-1\Big)(\Delta+k-1)\Big(\Delta-\frac d2+k\Big) \nonumber\\
&=
16^n\,n!\,\Big(\frac d2\Big)_n\,(\Delta)_n\,
\Big(\Delta-\frac d2+1\Big)_n\,G_0,
\end{align}
where $(x)_n$ denotes the Pochhammer symbol $(x)_n=\Gamma(x+n)/\Gamma(x)$.
Therefore, we obtain
\begin{equation}
G_n
=16^n\,n!\,\Big(\frac d2\Big)_n\,(\Delta)_n\,
\Big(\Delta-\frac d2+1\Big)_n\,G_0.
\label{eq:Gn_closed}
\end{equation}


\subsection{Derivation of the bulk-local state at the AdS center}
\label{subsubsec:derive_bulk_local_center_PminusK}

A bulk point at the AdS center is stabilized by an $SO(1,d)$ subgroup 
\begin{equation}
\mathcal{M}_{ab}\,|\Psi(0)\rangle=0,\qquad (\mathcal{P}_a-\mathcal{K}_a)\,|\Psi(0)\rangle=0.
\label{eq:center_Ishibashi_conditions_PminusK}
\end{equation}
Restricting to the $s$-wave (rotational singlet) sector, we expand the bulk local state as
\begin{equation}
|\Psi(0)\rangle=\sum_{n=0}^\infty c_n\,|e_n\rangle,
\quad |e_n\rangle=(P^2)^n|\Delta\rangle.
\label{eq:center_state_ansatz_PminusK}
\end{equation}
The condition $M_{ab}|\Psi(0)\rangle=0$ is automatic because each $|e_n\rangle$ is a scalar.

Let us derive the recursion relation for the coefficients $c_n$ (with $P_a-K_a$).
Imposing $(P_a-K_a)|\Psi(0)\rangle=0$ on \eqref{eq:center_state_ansatz_PminusK} gives
\begin{equation}
0=\sum_{n=0}^\infty c_n\,\mathcal{P}_a|e_n\rangle-\sum_{n=0}^\infty c_n\,\mathcal{K}_a|e_n\rangle.
\end{equation}
Using (\ref{eq:Ka_en}) in the second term and shifting the summation index, we have
\begin{align}
\sum_{n=0}^\infty c_n\,\mathcal{K}_a|e_n\rangle
&=\sum_{n=1}^\infty c_n\,4n\Big(\Delta-\frac d2+n\Big)\,\mathcal{P}_a|e_{n-1}\rangle\nonumber\\
&=\sum_{m=0}^\infty c_{m+1}\,4(m+1)\Big(\Delta-\frac d2+m+1\Big)\,\mathcal{P}_a|e_m\rangle.
\end{align}
Thus, we find
\begin{equation}
0=\sum_{m=0}^\infty
\Big[c_m-4(m+1)\Big(\Delta-\frac d2+m+1\Big)c_{m+1}\Big]\,\mathcal{P}_a|e_m\rangle,
\end{equation}
and independence of the $\mathcal{P}_a|e_m\rangle$ implies the recursion
\begin{equation}
c_{m+1}
=
\frac{c_m}{4(m+1)\left(\Delta-\frac d2+m+1\right)}.
\label{eq:cn_recursion_PminusK}
\end{equation}
Iterating \eqref{eq:cn_recursion_PminusK} yields
\begin{equation}
c_n=\frac{c_0}{4^n\,n!\,\left(\Delta-\frac d2+1\right)_n}.
\label{eq:cn_solution_PminusK}
\end{equation}
Therefore, the bulk local state at center is
\begin{equation}
|\Psi(0)\rangle
=
c_0\sum_{n=0}^\infty
\frac{1}{4^n\,n!\,\left(\Delta-\frac d2+1\right)_n}\,|e_n\rangle.
\label{eq:center_state_series_PminusK}
\end{equation}

We can obtain the state at $\tau$ via time evolution.
We define the Euclidean state at $(\tau,\vec 0)$ by
\begin{equation}
|\Psi^E(\tau)\rangle := e^{-(\tau/l)\mathcal{H}}\,|\Psi(0)\rangle,\qquad (\tau>0).
\end{equation}
Since $\mathcal{H}|e_n\rangle=(\Delta+2n)|e_n\rangle$, we have
\begin{equation}
e^{-(\tau/l)\mathcal{H}}|e_n\rangle=e^{-(\Delta+2n)\tau/l}|e_n\rangle
=e^{-\Delta\tau/l}\,q^n\,|e_n\rangle,
\qquad q:=e^{-2\tau/l}.
\end{equation}
Substituting into \eqref{eq:center_state_series_PminusK} gives
\begin{equation}
|\Psi^E(\tau)\rangle
=
c_0\,e^{-\Delta\tau/l}\sum_{n=0}^{\infty}
\frac{q^n}{4^n\,n!\,\left(\Delta-\frac d2+1\right)_n}\,|e_n\rangle,
\quad q=e^{-2\tau/l}.
\label{eq:bulk_local_ket_en_derived_PminusK}
\end{equation}

\subsection{Flat limit of Green's function in AdS$_4$ }\label{legendre}
We can rewrite the hypergeometric function using the identity as follows
\be
G_{\rm AdS_4}^E(\tau)
=G_0\cdot W^\Delta {}_2F_1\left[\Delta,\Delta-1,2\Delta-2,-4W \right],\quad W=\frac{1}{2\left(\cosh\frac{\tau}{l}-1\right)}, 
\ee
which matches the known result \cite{Berenstein:1998ij} with the normalization constant 
\be
G_0=\frac{\Gamma(\Delta)}{2\pi^{\frac{3}{2}}\Gamma\left(\Delta-\frac{1}{2}\right)l^{2}},
\ee
where the conformal weight is related with the mass of the bulk scalar field as 
\be
\Delta=\frac{3}{2}+\sqrt{m^2l^2+\frac{9}{4}}.
\ee
Notice that now the normalization of primary states are different from that in AdS$_3$ since we have
\be
\langle\Delta|\Delta\rangle=G_0=\frac{\Gamma(\Delta)}{2\pi^{\frac{3}{2}}\Gamma\left(\Delta-\frac{1}{2}\right)l^{2}}.
\ee
 Let us take the flat limit $l\to \infty$. We can convert the hypergeometric function into the associated Legendre function of the second kind
 \be
    Q^\mu_{\nu-\frac{1}{2}}(\xi)=\frac{e^{\mu\pi i}\sqrt{\pi}\Gamma\left(\mu+\nu+\frac{1}{2}\right)(\xi^2-1)^{\frac{\mu}{2}}}{2^{\nu+\frac{1}{2}}\Gamma(\nu+1)}\xi^{-\mu-\nu-\frac{1}{2}}\cdot{}_2F_1\left[\frac{\mu+\nu+\frac{1}{2}}{2},\frac{\mu+\nu+\frac{3}{2}}{2},\nu+1;\frac{1}{\xi^2}\right].
 \ee
 Using the transformation law of the hypergeometric functions
 \begin{align}
     F[a,b,c;z]&=(1-z)^{-a}{}_2F_1\left[a,c-b,c;\frac{z}{z-1}\right],\nn\\
     F[a,b,2b;z]&=(1-z)^{-\frac{a}{2}}{}_2F_1\left[\frac{a}{2},b-\frac{a}{2},b+\frac{1}{2};\frac{z^2}{4(z-1)}\right],
 \end{align}
 the Green's function can be expressed as
 \be
G_{\rm AdS_4}^E(\tau)=-\frac{Q^1_{\Delta-2}\left(\cosh\frac{\tau}{l}\right)}{4l^2\pi^2\sinh\frac{\tau}{l}}.
 \ee
 In the flat limit, we can confirm the following relations
 \begin{align}
     \cosh\frac{\tau}{l}&\sim1+\frac{\tau^2}{2l^2},\quad \Delta\sim ml=\xi l,\nn\\
     Q^\mu_\nu(\cosh\chi)&\sim e^{\mu\pi i}\nu^\mu\sqrt{\frac{\chi}{\sinh\chi}}K_{\mu}\left(\left(\nu+\frac{1}{2}\right)\chi\right),
 \end{align}
 where $K_\mu$ is the modified Bessel function. Substituting these forms, we obtain 
 \be
G_{\rm AdS_4}^E(\tau)\sim \frac{1}{4\pi^2}\frac{\xi}{\tau}K_1(\xi\tau)=G_{\rm flat}, \quad \quad (l\rightarrow\infty).
 \ee

\section{Existence of the tilde basis in AdS$_{d+1}$} \label{app:ads tilde}
The existence of a basis $|\tilde n\rangle_l$ satisfying eq.\,\eqref{eq:modified_tilde_AdS} can be proven by induction on $n$. Assume that for all $j < n$ we have constructed upper-triangular matrices $S_{jk}$ such that
\[
|\tilde j\rangle_l = |j\rangle_l + \sum_{k=1}^{j-1} S_{jk}\,|k\rangle_l,\qquad S_{jj}=1.
\]
We wish to find coefficients $S_{nk}$ for $k=1,\dots,n-1$ (with $S_{nn}=1$) so that eq.\,\eqref{eq:modified_tilde_AdS} holds. Expanding both sides in the basis $\{|m\rangle_l\}$ and equating coefficients of each level $2k-1$ (for $k=1,\dots,n$) yields a linear system for the unknowns $S_{nk}$. Because the operator $(P_a(l)+\frac{i}{l}K_a(l))$ is lower-triangular in the index $n$, the equations at level $2k-1$ involve only $S_{nj}$ with $j\ge k$. Moreover, the coefficient of $S_{nk}$ in the equation at level $2k-1$ is proportional to the same combination that appears in \eqref{eq:lambda_n_AdS_general} and is nonzero for generic parameters. The system therefore has a lower-triangular form with non-zero diagonal entries, guaranteeing a unique solution. This completes the induction and establishes the existence of the tilde basis for any $n$.

\section{The descendant states and the tilde basis}
\subsection{AdS$_4$ case} \label{subsec:relation_descendant}
 We can relate the descendant basis \eqref{eq:e_n_def} to the tilde basis $\{|\tilde n\rangle_l\}_n$ in AdS$_4$.
The bulk local state in the descendant basis reads (See eq.\,\eqref{eq:ads4_solution_des})
\begin{equation}
|\Psi^E_{\rm AdS_4}(\tau)\rangle
=
e^{-\Delta \tau/l}
\sum_{k=0}^\infty
\frac{q^k}{4^k\,k!\,(\Delta-\tfrac12)_k}\;
|e_k\rangle, \quad q:=e^{-2\tau/l}.\quad (|q|<1)
\label{eq:bulk_ek}
\end{equation}
On the other hand, solving the stabilizer condition \eqref{eq:AdS_bulk_condition_original} in the tilde basis yields
\begin{equation}
|\Psi^E_{\rm AdS_4}(\tau)\rangle
=
f_0^E(\tau)\sum_{n=0}^\infty f_n(q)\,|\tilde n\rangle_l,
\quad
f_n(q)=\frac{1}{n!\,(\Delta-\tfrac12)_n}\left(\frac{q}{1-q}\right)^n,
\label{eq:bulk_tilden}
\end{equation}
where the overall factor $f_0^E(\tau)$ is a convention.  This can be obtained by rewriting eq.\,\eqref{eq:AdS_f_n_solution}. We require these two expressions to match and derive the relation between the two bases.
For $n\ge1$, we use
\begin{equation}
\frac{q^n}{(1-q)^n}
=\sum_{k=n}^\infty \binom{k-1}{n-1}\,q^k,
\qquad (|q|<1),
\label{eq:q_over_1q_expand}
\end{equation}
and $q^0/(1-q)^0=1$ for $n=0$.
Inserting eq.\,\eqref{eq:q_over_1q_expand} into eq.\,\eqref{eq:bulk_tilden} yields a power series in $q$
\begin{align}
|\Psi^E_{\rm AdS_4}(\tau)\rangle
&=
f_0^E(\tau)\Bigg[
|\tilde0\rangle_l
+\sum_{k=1}^\infty q^k
\sum_{n=1}^{k}
\binom{k-1}{n-1}\;
\frac{1}{n!\,(\Delta-\tfrac12)_n}\;
|\tilde n\rangle_l
\Bigg].
\label{eq:bulk_tilden_qseries}
\end{align}
Since eqs.\,\eqref{eq:bulk_ek} and \eqref{eq:bulk_tilden_qseries} are the same vector for all $|q|<1$,
the coefficient of each power $q^k$ must coincide.
Choosing $f_0^E(\tau)=e^{-\Delta\tau/l}$ and comparing the $q^0$ term gives
\begin{equation}
|e_0\rangle=|\tilde0\rangle_l.
\end{equation}
For each $k\ge1$, matching the $q^k$ coefficient gives the lower-triangular relation
\begin{equation}
|e_k\rangle
=
\,4^k\,k!\,(\Delta-\tfrac12)_k
\sum_{n=1}^{k}
\binom{k-1}{n-1}\;
\frac{1}{n!\,(\Delta-\tfrac12)_n}\;
|\tilde n\rangle_l
\qquad (k\ge1).
\label{eq:ek_to_tilden}
\end{equation}
The matrix in eq.\,\eqref{eq:ek_to_tilden} is a lower-triangular matrix with non-vanishing diagonal part,
hence it is invertible.  The inverse follows from the binomial inversion, which claims that the condition
\begin{equation}
u_k=\sum_{n=1}^{k}\binom{k-1}{n-1}v_n,\quad\iff\quad v_n=\sum_{k=1}^{n}(-1)^{\,n-k}\binom{n-1}{k-1}u_k.
\end{equation}
Applying this with $u_k:=\frac{1}{4^k k!(\Delta-\tfrac12)_k}\,|e_k\rangle$, $v_n:=\frac{1}{n!(\Delta-\tfrac12)_n}\,|\tilde n\rangle_l$,
we obtain
\begin{equation}
|\tilde 0\rangle_l = |e_0\rangle,
\end{equation}
and for $n\ge1$,
\begin{equation}
|\tilde n\rangle_l
=
\,n!\,(\Delta-\tfrac12)_n
\sum_{k=1}^{n}(-1)^{\,n-k}
\binom{n-1}{k-1}\;
\frac{1}{4^k\,k!\,(\Delta-\tfrac12)_k}\;
|e_k\rangle
\qquad (n\ge1).
\label{eq:tilden_to_ek}
\end{equation}
The $|e_k\rangle$ basis is orthogonal
\begin{equation}
\langle e_m|e_n\rangle=\delta_{mn}G_n,
\qquad
G_n=16^n\,n!\,\Big(\tfrac32\Big)_n\,(\Delta)_n\,(\Delta-\tfrac12)_n\;G_0,
\label{eq:en_norm}
\end{equation}
where $G_0=\langle\Delta|\Delta\rangle$. Using eq.\,\eqref{eq:tilden_to_ek}, we can write
\begin{equation}
|\tilde n\rangle_l = \sum_{k=0}^{n}A_{n k}\,|e_k\rangle,
\end{equation}
where
\be
A_{00}=1,\quad
A_{n0}=0\ (n\ge1),\quad
A_{nk}=\,n!\,(\Delta-\tfrac12)_n\,
(-1)^{\,n-k}\binom{n-1}{k-1}\frac{1}{4^k k!(\Delta-\tfrac12)_k}\ (1\le k\le n).
\ee
Then the Gram matrix in the tilde basis is
\begin{equation}
\tilde G_{mn}(l):={}_l\langle\tilde m|\tilde n\rangle_l
=\sum_{k=0}^{\min(m,n)} A_{mk}A_{nk}\,G_k.
\label{eq:Gram_def_sum}
\end{equation}
In particular, we get $\tilde G_{00}=G_0$, $\tilde G_{0n}=0\,(n\ge1)$.
For $m,n\ge1$, substituting eq.\,\eqref{eq:en_norm}, we obtain
\begin{align}
\tilde G_{mn}(l)
& =
(-1)^{m+n}\,m!\,n!\,(\Delta-\tfrac12)_m(\Delta-\tfrac12)_n\,G_0
\sum_{k=1}^{\min(m,n)}
\binom{m-1}{k-1}\binom{n-1}{k-1}\;
\frac{(\tfrac32)_k(\Delta)_k}{k!\,(\Delta-\tfrac12)_k}
\quad (m,n\ge1).
\label{eq:Gram_tilde_finite_sum} \\
& =
(-1)^{m+n}\,m!\,n!\,(\Delta-\tfrac12)_m(\Delta-\tfrac12)_n\,G_0\;
\frac{3\Delta}{2\Delta-1}\;
{}_4F_3\!\left(
\begin{array}{c}
1-m,\ 1-n,\ \tfrac52,\ \Delta+1\\
1,\ \Delta+\tfrac12,\ 2
\end{array}
;1\right),
\label{eq:Gram_tilde_hypergeom}
\end{align}
where $m,n\geq1$.
Finally, the dual bra basis is defined by the inverse Gram matrix
\begin{equation}
{}^\vee\!\langle \tilde m|:=\sum_{n\ge0}(\tilde G^{-1})_{mn}\,\langle\tilde n|,
\qquad
{}^\vee\!\langle\tilde m|\tilde n\rangle=\delta_{mn}.
\end{equation}

\subsection{General AdS$_{d+1}$ case}
\label{subsec:relation_descendant_general}

Then we provide the explicit transformation between the tilde basis $\{|\tilde n\rangle_l\}$ constructed above and the descendant basis $\{|e_n\rangle\}$ defined in eq.\,\eqref{eq:arb:AdS_en_def}.  These relations generalize the four-dimensional formulas \eqref{eq:ek_to_tilden} and \eqref{eq:tilden_to_ek} to arbitrary dimension $d$ and are useful for computing overlaps and for understanding the flat limit.

Recall that the descendant basis is defined by
\begin{equation}
|e_n\rangle := (\mathcal{P}^2)^n |\Delta\rangle ,\qquad n=0,1,2,\dots ,
\end{equation}
and satisfies the orthogonality relation
\begin{equation}
\langle e_m | e_n \rangle = \delta_{mn} G_n ,\qquad 
G_n = 16^n\,n!\,\Bigl(\frac{d}{2}\Bigr)_n(\Delta)_n\Bigl(\Delta-\frac{d}{2}+1\Bigr)_n\,G_0 ,
\label{eq:arb:Gn_repeat}
\end{equation}
with $G_0 = \langle\Delta|\Delta\rangle$ given in eq.\,\eqref{eq:arb:AdS_G0}.  It is convenient to introduce the orthonormal descendant basis
\begin{equation}
|n\rangle := \frac{1}{\sqrt{G_n}} |e_n\rangle ,\qquad \langle m|n\rangle = \delta_{mn}.
\end{equation}

The tilde basis $\{|\tilde n\rangle_l\}$ is defined by a lower-triangular transformation
\begin{equation}
|\tilde n\rangle_l = \sum_{k=0}^{n} S_{nk}(l)\,|k\rangle ,\qquad S_{nn}=1,
\label{eq:arb:tilde_transform}
\end{equation}
with coefficients $S_{nk}(l)$ chosen so that the action of $P_a(l)+\frac{i}{l}K_a(l)$ takes the simple form \eqref{eq:modified_tilde_AdS}.  
Solving the recurrsive conditions order by order (or, equivalently, generalizing the four-dimensional derivation to arbitrary $d$) yields a closed-form expression for the transformation matrix. 
For $n\ge 1$ and $1\le k\le n$,
\begin{equation}
 \; S_{nk}(l) = n!\,\Bigl(\Delta-\frac{d}{2}+1\Bigr)_n\;
\frac{(-1)^{n-k}}{4^k k!\,\bigl(\Delta-\frac{d}{2}+1\bigr)_k}\binom{n-1}{k-1}\;  .
\label{eq:arb:Snk}
\end{equation}
We also have $S_{00}=1$ and $S_{n0}=0$ for $n\ge1$.

In terms of the original descendant states $|e_k\rangle$, the tilde basis is therefore
\begin{align}
|\tilde n\rangle_l &= n!\,\Bigl(\Delta-\frac{d}{2}+1\Bigr)_n
\sum_{k=1}^{n}\frac{(-1)^{n-k}}{4^k k!\,\bigl(\Delta-\frac{d}{2}+1\bigr)_k}\binom{n-1}{k-1}\,
\frac{1}{\sqrt{G_k}}\,|e_k\rangle \qquad (n\ge1), \label{eq:arb:tilde_to_en}\\
|\tilde0\rangle_l &= |e_0\rangle .
\end{align}
Comparing with eq.\,\eqref{eq:tilden_to_ek} we see that the four-dimensional case $d=3$ is recovered after setting $d=3$ and noting that $(\Delta-\frac{d}{2}+1)_k = (\Delta-\frac12)_k$.

Using the transformation \eqref{eq:arb:tilde_transform} together with the orthogonality of $|k\rangle$, the Gram matrix of the tilde basis is
\begin{equation}
\tilde G_{mn}(l) := {}_l\langle\tilde m|\tilde n\rangle_l = \sum_{k=0}^{\min(m,n)} S_{mk}(l)\,S_{nk}(l)\, .
\end{equation}
Substituting eq.\,\eqref{eq:arb:Snk} and returning to the $|e_k\rangle$ basis (which carries the norm $G_k$) we obtain, for $m,n\ge1$,
\begin{align}
\tilde G_{mn}(l) &= (-1)^{m+n} m!\,n!\,\Bigl(\Delta-\tfrac{d}{2}+1\Bigr)_m\Bigl(\Delta-\tfrac{d}{2}+1\Bigr)_n G_0 \nonumber\\
&\qquad\times \sum_{k=1}^{\min(m,n)}\binom{m-1}{k-1}\binom{n-1}{k-1}\,
\frac{\bigl(\frac{d}{2}\bigr)_k(\Delta)_k}{k!\,\bigl(\Delta-\frac{d}{2}+1\bigr)_k}. \label{eq:arb:Gram_sum}
\end{align}
The sum can be expressed as a generalized hypergeometric function
\begin{align}
\tilde G_{mn}(l) &= (-1)^{m+n} m!\,n!\,\Bigl(\Delta-\tfrac{d}{2}+1\Bigr)_m\Bigl(\Delta-\tfrac{d}{2}+1\Bigr)_n G_0\;
\frac{\frac{d}{2}\Delta}{\Delta-\frac{d}{2}+1} \nonumber\\
&\qquad\times {}_4F_3\!\left(\begin{array}{c}
1-m,\;1-n,\;\frac{d}{2}+1,\;\Delta+1\\[2pt]
1,\;\Delta-\frac{d}{2}+2,\;2
\end{array};1\right),\qquad m,n\ge1. \label{eq:arb:Gram_hyper}
\end{align}
For $m=0$ or $n=0$ we have $\tilde G_{0n}=0\ (n\ge1)$ and $\tilde G_{00}=G_0$.

The flat limit $l\to\infty$ (with $\Delta = \xi l + \mathcal{O}(1)$) of these expressions is controlled by the scaling window $n\sim l$, exactly as discussed in Subsubsection \ref{subsubsec:momentum_general}.
In that regime the sums turn into Riemann integrals and reproduce the inner products of the flat-space momentum basis. 
In particular, one can verify that the continuum limit of $\tilde G_{mn}(l)$ yields the expected orthonormality of the rescaled basis $|x\rangle = \sqrt{l}\,|n\rangle|_{n=\lfloor lx\rfloor}$.  Thus the tilde basis provides a convenient algebraic bridge between the discrete AdS descendant states and the continuous momentum basis of Minkowski space.

The relations presented here hold for any spacetime dimension $d+1$ and demonstrate that the algebraic structure underlying the tilde basis is universal.  They are essential for explicit calculations involving bulk local states and for understanding how the flat limit is implemented at the level of states.

    \bibliographystyle{JHEP}
    \bibliography{paper.bib}

@article{Barnich:2010eb,
    author = "Barnich, Glenn and Troessaert, Cedric",
    title = "{Aspects of the BMS/CFT correspondence}",
    eprint = "1001.1541",
    archivePrefix = "arXiv",
    primaryClass = "hep-th",
    reportNumber = "ULB-TH-09-28",
    doi = "10.1007/JHEP05(2010)062",
    journal = "JHEP",
    volume = "05",
    pages = "062",
    year = "2010"
}

@article{Bagchi:2010zz,
    author = "Bagchi, Arjun",
    title = "{Correspondence between Asymptotically Flat Spacetimes and Nonrelativistic Conformal Field Theories}",
    eprint = "1006.3354",
    archivePrefix = "arXiv",
    primaryClass = "hep-th",
    doi = "10.1103/PhysRevLett.105.171601",
    journal = "Phys. Rev. Lett.",
    volume = "105",
    pages = "171601",
    year = "2010"
}

@article{Fareghbal:2013ifa,
    author = "Fareghbal, Reza and Naseh, Ali",
    title = "{Flat-Space Energy-Momentum Tensor from BMS/GCA Correspondence}",
    eprint = "1312.2109",
    archivePrefix = "arXiv",
    primaryClass = "hep-th",
    doi = "10.1007/JHEP03(2014)005",
    journal = "JHEP",
    volume = "03",
    pages = "005",
    year = "2014"
}

@article{Bagchi:2025vri,
    author = "Bagchi, Arjun and Banerjee, Aritra and Dhivakar, Prateksh and Mondal, Saikat and Shukla, Ashish",
    title = "{The Carrollian Kaleidoscope}",
    eprint = "2506.16164",
    archivePrefix = "arXiv",
    primaryClass = "hep-th",
    month = "6",
    year = "2025"
}

@article{Barnich:2012xq,
    author = "Barnich, Glenn",
    title = "{Entropy of three-dimensional asymptotically flat cosmological solutions}",
    eprint = "1208.4371",
    archivePrefix = "arXiv",
    primaryClass = "hep-th",
    reportNumber = "ULB-TH-12-14",
    doi = "10.1007/JHEP10(2012)095",
    journal = "JHEP",
    volume = "10",
    pages = "095",
    year = "2012"
}

@article{Bagchi:2012xr,
    author = "Bagchi, Arjun and Detournay, St{\'e}phane and Fareghbal, Reza and Sim{\'o}n, Joan",
    title = "{Holography of 3D Flat Cosmological Horizons}",
    eprint = "1208.4372",
    archivePrefix = "arXiv",
    primaryClass = "hep-th",
    reportNumber = "EMPG-12-18",
    doi = "10.1103/PhysRevLett.110.141302",
    journal = "Phys. Rev. Lett.",
    volume = "110",
    number = "14",
    pages = "141302",
    year = "2013"
}

@article{Barnich:2012rz,
    author = "Barnich, Glenn and Gomberoff, Andr{\'e}s and Gonz{\'a}lez, Hern{\'a}n A.",
    title = "{Three-dimensional Bondi-Metzner-Sachs invariant two-dimensional field theories as the flat limit of Liouville theory}",
    eprint = "1210.0731",
    archivePrefix = "arXiv",
    primaryClass = "hep-th",
    doi = "10.1103/PhysRevD.87.124032",
    journal = "Phys. Rev. D",
    volume = "87",
    number = "12",
    pages = "124032",
    year = "2013"
}

@article{Chen:2023naw,
    author = "Chen, Bin and Hu, Zezhou",
    title = "{Bulk reconstruction in flat holography}",
    eprint = "2312.13574",
    archivePrefix = "arXiv",
    primaryClass = "hep-th",
    doi = "10.1007/JHEP03(2024)064",
    journal = "JHEP",
    volume = "03",
    pages = "064",
    year = "2024"
}

@article{Hao:2025btl,
    author = "Hao, Peng-Xiang and Shinmyo, Kotaro and Suzuki, Yu-ki and Takahashi, Shunta and Takayanagi, Tadashi",
    title = "{Bulk reconstruction of scalar excitations in Flat$_{3}$/CCFT$_{2}$ and the flat limit from (A)dS$_{3}$/CFT$_{2}$}",
    eprint = "2505.20084",
    archivePrefix = "arXiv",
    primaryClass = "hep-th",
    reportNumber = "YITP-25-83, KUNS-3052",
    doi = "10.1007/JHEP11(2025)054",
    journal = "JHEP",
    volume = "11",
    pages = "054",
    year = "2025"
}

@article{Bagchi:2014iea,
    author = "Bagchi, Arjun and Basu, Rudranil and Grumiller, Daniel and Riegler, Max",
    title = "{Entanglement entropy in Galilean conformal field theories and flat holography}",
    eprint = "1410.4089",
    archivePrefix = "arXiv",
    primaryClass = "hep-th",
    reportNumber = "TUW-14-14",
    doi = "10.1103/PhysRevLett.114.111602",
    journal = "Phys. Rev. Lett.",
    volume = "114",
    number = "11",
    pages = "111602",
    year = "2015"
}

@article{Hao:2025naz,
    author = "Hao, Peng-Xiang and Takahashi, Shunta",
    title = "{Conformal Blocks in 2d Carrollian/Galilean CFTs and Excited State Entanglement Entropy}",
    eprint = "2510.25688",
    archivePrefix = "arXiv",
    primaryClass = "hep-th",
    reportNumber = "KUNS-3076",
    month = "10",
    year = "2025"
}

@article{Jiang:2017ecm,
    author = "Jiang, Hongliang and Song, Wei and Wen, Qiang",
    title = "{Entanglement Entropy in Flat Holography}",
    eprint = "1706.07552",
    archivePrefix = "arXiv",
    primaryClass = "hep-th",
    doi = "10.1007/JHEP07(2017)142",
    journal = "JHEP",
    volume = "07",
    pages = "142",
    year = "2017"
}

@article{Bagchi:2009ca,
    author = "Bagchi, Arjun and Mandal, Ipsita",
    title = "{On Representations and Correlation Functions of Galilean Conformal Algebras}",
    eprint = "0903.4524",
    archivePrefix = "arXiv",
    primaryClass = "hep-th",
    reportNumber = "HRI-ST-0910",
    doi = "10.1016/j.physletb.2009.04.030",
    journal = "Phys. Lett. B",
    volume = "675",
    pages = "393--397",
    year = "2009"
}

@article{Bagchi:2016geg,
    author = "Bagchi, Arjun and Gary, Mirah and Zodinmawia",
    title = "{Bondi-Metzner-Sachs bootstrap}",
    eprint = "1612.01730",
    archivePrefix = "arXiv",
    primaryClass = "hep-th",
    doi = "10.1103/PhysRevD.96.025007",
    journal = "Phys. Rev. D",
    volume = "96",
    number = "2",
    pages = "025007",
    year = "2017"
}

@article{Bagchi:2017cpu,
    author = "Bagchi, Arjun and Gary, Mirah and Zodinmawia",
    title = "{The nuts and bolts of the BMS Bootstrap}",
    eprint = "1705.05890",
    archivePrefix = "arXiv",
    primaryClass = "hep-th",
    doi = "10.1088/1361-6382/aa8003",
    journal = "Class. Quant. Grav.",
    volume = "34",
    number = "17",
    pages = "174002",
    year = "2017"
}

@article{Chen:2020vvn,
    author = "Chen, Bin and Hao, Peng-Xiang and Liu, Reiko and Yu, Zhe-Fei",
    title = "{On Galilean conformal bootstrap}",
    eprint = "2011.11092",
    archivePrefix = "arXiv",
    primaryClass = "hep-th",
    doi = "10.1007/JHEP06(2021)112",
    journal = "JHEP",
    volume = "06",
    pages = "112",
    year = "2021"
}

@article{Chen:2022jhx,
    author = "Chen, Bin and Hao, Peng-xiang and Liu, Reiko and Yu, Zhe-fei",
    title = "{On Galilean conformal bootstrap. Part II. {\ensuremath{\xi}} = 0 sector}",
    eprint = "2207.01474",
    archivePrefix = "arXiv",
    primaryClass = "hep-th",
    doi = "10.1007/JHEP12(2022)019",
    journal = "JHEP",
    volume = "12",
    pages = "019",
    year = "2022"
}

@article{Hao:2021urq,
    author = "Hao, Peng-xiang and Song, Wei and Xie, Xianjin and Zhong, Yuan",
    title = "{BMS-invariant free scalar model}",
    eprint = "2111.04701",
    archivePrefix = "arXiv",
    primaryClass = "hep-th",
    doi = "10.1103/PhysRevD.105.125005",
    journal = "Phys. Rev. D",
    volume = "105",
    number = "12",
    pages = "125005",
    year = "2022"
}

@article{Chen:2021xkw,
    author = "Chen, Bin and Liu, Reiko and Zheng, Yu-fan",
    title = "{On higher-dimensional Carrollian and Galilean conformal field theories}",
    eprint = "2112.10514",
    archivePrefix = "arXiv",
    primaryClass = "hep-th",
    doi = "10.21468/SciPostPhys.14.5.088",
    journal = "SciPost Phys.",
    volume = "14",
    number = "5",
    pages = "088",
    year = "2023"
}

@article{Yu:2022bcp,
    author = "Yu, Zhe-fei and Chen, Bin",
    title = "{Free field realization of the BMS Ising model}",
    eprint = "2211.06926",
    archivePrefix = "arXiv",
    primaryClass = "hep-th",
    doi = "10.1007/JHEP08(2023)116",
    journal = "JHEP",
    volume = "08",
    pages = "116",
    year = "2023"
}

@article{Hao:2022xhq,
    author = "Hao, Peng-Xiang and Song, Wei and Xiao, Zehua and Xie, Xianjin",
    title = "{BMS-invariant free fermion models}",
    eprint = "2211.06927",
    archivePrefix = "arXiv",
    primaryClass = "hep-th",
    doi = "10.1103/PhysRevD.109.025002",
    journal = "Phys. Rev. D",
    volume = "109",
    number = "2",
    pages = "025002",
    year = "2024"
}

@article{Banerjee:2022ocj,
    author = "Banerjee, Aritra and Dutta, Sudipta and Mondal, Saikat",
    title = "{Carroll fermions in two dimensions}",
    eprint = "2211.11639",
    archivePrefix = "arXiv",
    primaryClass = "hep-th",
    doi = "10.1103/PhysRevD.107.125020",
    journal = "Phys. Rev. D",
    volume = "107",
    number = "12",
    pages = "125020",
    year = "2023"
}

@article{deBoer:2023fnj,
    author = "de Boer, Jan and Hartong, Jelle and Obers, Niels A. and Sybesma, Watse and Vandoren, Stefan",
    title = "{Carroll stories}",
    eprint = "2307.06827",
    archivePrefix = "arXiv",
    primaryClass = "hep-th",
    reportNumber = "NORDITA-2023-036",
    doi = "10.1007/JHEP09(2023)148",
    journal = "JHEP",
    volume = "09",
    pages = "148",
    year = "2023"
}

@article{Hao:2025hfa,
    author = "Hao, Peng-Xiang and Lai, Wen-Xin and Song, Wei and Xiao, Zehua",
    title = "{Modular Hamiltonian and entanglement entropy in the BMS free fermion theory}",
    eprint = "2507.10503",
    archivePrefix = "arXiv",
    primaryClass = "hep-th",
    doi = "10.1007/JHEP02(2026)099",
    journal = "JHEP",
    volume = "02",
    pages = "099",
    year = "2026"
}

@article{Pasterski:2016qvg,
    author = "Pasterski, Sabrina and Shao, Shu-Heng and Strominger, Andrew",
    title = "{Flat Space Amplitudes and Conformal Symmetry of the Celestial Sphere}",
    eprint = "1701.00049",
    archivePrefix = "arXiv",
    primaryClass = "hep-th",
    doi = "10.1103/PhysRevD.96.065026",
    journal = "Phys. Rev. D",
    volume = "96",
    number = "6",
    pages = "065026",
    year = "2017"
}

@article{Pasterski:2017kqt,
    author = "Pasterski, Sabrina and Shao, Shu-Heng",
    title = "{Conformal basis for flat space amplitudes}",
    eprint = "1705.01027",
    archivePrefix = "arXiv",
    primaryClass = "hep-th",
    doi = "10.1103/PhysRevD.96.065022",
    journal = "Phys. Rev. D",
    volume = "96",
    number = "6",
    pages = "065022",
    year = "2017"
}

@article{Donnay:2022aba,
    author = "Donnay, Laura and Fiorucci, Adrien and Herfray, Yannick and Ruzziconi, Romain",
    title = "{Carrollian Perspective on Celestial Holography}",
    eprint = "2202.04702",
    archivePrefix = "arXiv",
    primaryClass = "hep-th",
    doi = "10.1103/PhysRevLett.129.071602",
    journal = "Phys. Rev. Lett.",
    volume = "129",
    number = "7",
    pages = "071602",
    year = "2022"
}

@article{Donnay:2022wvx,
    author = "Donnay, Laura and Fiorucci, Adrien and Herfray, Yannick and Ruzziconi, Romain",
    title = "{Bridging Carrollian and celestial holography}",
    eprint = "2212.12553",
    archivePrefix = "arXiv",
    primaryClass = "hep-th",
    doi = "10.1103/PhysRevD.107.126027",
    journal = "Phys. Rev. D",
    volume = "107",
    number = "12",
    pages = "126027",
    year = "2023"
}

@article{Bagchi:2022emh,
    author = "Bagchi, Arjun and Banerjee, Shamik and Basu, Rudranil and Dutta, Sudipta",
    title = "{Scattering Amplitudes: Celestial and Carrollian}",
    eprint = "2202.08438",
    archivePrefix = "arXiv",
    primaryClass = "hep-th",
    doi = "10.1103/PhysRevLett.128.241601",
    journal = "Phys. Rev. Lett.",
    volume = "128",
    number = "24",
    pages = "241601",
    year = "2022"
}

@article{Bagchi:2023fbj,
    author = "Bagchi, Arjun and Dhivakar, Prateksh and Dutta, Sudipta",
    title = "{AdS Witten diagrams to Carrollian correlators}",
    eprint = "2303.07388",
    archivePrefix = "arXiv",
    primaryClass = "hep-th",
    doi = "10.1007/JHEP04(2023)135",
    journal = "JHEP",
    volume = "04",
    pages = "135",
    year = "2023"
}

@article{Kulp:2024scx,
    author = "Kulp, Justin and Pasterski, Sabrina",
    title = "{Multiparticle states for the flat hologram}",
    eprint = "2501.00462",
    archivePrefix = "arXiv",
    primaryClass = "hep-th",
    doi = "10.1007/JHEP08(2025)091",
    journal = "JHEP",
    volume = "08",
    pages = "091",
    year = "2025"
}

@article{Ruzziconi:2026isv,
    author = "Ruzziconi, Romain and West, Peter",
    title = "{Extended BMS representations and strings}",
    eprint = "2601.00662",
    archivePrefix = "arXiv",
    primaryClass = "hep-th",
    month = "1",
    year = "2026"
}

@article{Bekaert:2024uuy,
    author = "Bekaert, Xavier and Donnay, Laura and Herfray, Yannick",
    title = "{Bondi-Metzner-Sachs Particles}",
    eprint = "2412.06002",
    archivePrefix = "arXiv",
    primaryClass = "hep-th",
    doi = "10.1103/8376-fync",
    journal = "Phys. Rev. Lett.",
    volume = "135",
    number = "13",
    pages = "131602",
    year = "2025"
}

@article{Bekaert:2025kjb,
    author = "Bekaert, Xavier and Herfray, Yannick",
    title = "{BMS Representations for Generic Supermomentum}",
    eprint = "2505.05368",
    archivePrefix = "arXiv",
    primaryClass = "hep-th",
    doi = "10.1007/s00220-025-05513-0",
    journal = "Commun. Math. Phys.",
    volume = "407",
    number = "2",
    pages = "35",
    year = "2026"
}

@article{Nakayama:2015mva,
    author = "Nakayama, Yu and Ooguri, Hirosi",
    title = "{Bulk Locality and Boundary Creating Operators}",
    eprint = "1507.04130",
    archivePrefix = "arXiv",
    primaryClass = "hep-th",
    reportNumber = "CALT-TH-2015-037, IPMU-15-0105",
    doi = "10.1007/JHEP10(2015)114",
    journal = "JHEP",
    volume = "10",
    pages = "114",
    year = "2015"
}

@article{Barnich:2014kra,
    author = "Barnich, Glenn and Oblak, Blagoje",
    title = "{Notes on the BMS group in three dimensions: I. Induced representations}",
    eprint = "1403.5803",
    archivePrefix = "arXiv",
    primaryClass = "hep-th",
    doi = "10.1007/JHEP06(2014)129",
    journal = "JHEP",
    volume = "06",
    pages = "129",
    year = "2014"
}

@article{Barnich:2015uva,
    author = "Barnich, Glenn and Oblak, Blagoje",
    title = "{Notes on the BMS group in three dimensions: II. Coadjoint representation}",
    eprint = "1502.00010",
    archivePrefix = "arXiv",
    primaryClass = "hep-th",
    doi = "10.1007/JHEP03(2015)033",
    journal = "JHEP",
    volume = "03",
    pages = "033",
    year = "2015"
}

@article{Miyaji:2015fia,
    author = "Miyaji, Masamichi and Numasawa, Tokiro and Shiba, Noburo and Takayanagi, Tadashi and Watanabe, Kento",
    title = "{Continuous Multiscale Entanglement Renormalization Ansatz as Holographic Surface-State Correspondence}",
    eprint = "1506.01353",
    archivePrefix = "arXiv",
    primaryClass = "hep-th",
    reportNumber = "YITP-15-46, IPMU15-0077, YITP-15-46, IPMU15-0077",
    doi = "10.1103/PhysRevLett.115.171602",
    journal = "Phys. Rev. Lett.",
    volume = "115",
    number = "17",
    pages = "171602",
    year = "2015"
}

@article{Berenstein:1998ij,
    author = "Berenstein, David Eliecer and Corrado, Richard and Fischler, Willy and Maldacena, Juan Martin",
    title = "{The Operator product expansion for Wilson loops and surfaces in the large N limit}",
    eprint = "hep-th/9809188",
    archivePrefix = "arXiv",
    reportNumber = "UTTG-05-98, HUTP-98-A066",
    doi = "10.1103/PhysRevD.59.105023",
    journal = "Phys. Rev. D",
    volume = "59",
    pages = "105023",
    year = "1999"
}

@article{barnich2015one,
  title={One loop partition function of three-dimensional flat gravity},
  author={Barnich, Glenn and Gonzalez, Hernan A and Maloney, Alexander and Oblak, Blagoje},
  journal={Journal of High Energy Physics},
  volume={2015},
  number={4},
  pages={1--8},
  year={2015},
  publisher={Springer}
}

@article{Lipstein:2025jfj,
    author = "Lipstein, Arthur and Ruzziconi, Romain and Yelleshpur Srikant, Akshay",
    title = "{Towards a flat space Carrollian hologram from AdS$_{4}$/CFT$_{3}$}",
    eprint = "2504.10291",
    archivePrefix = "arXiv",
    primaryClass = "hep-th",
    doi = "10.1007/JHEP06(2025)073",
    journal = "JHEP",
    volume = "06",
    pages = "073",
    year = "2025"
}

@article{Bagchi:2016bcd,
    author = "Bagchi, Arjun and Basu, Rudranil and Kakkar, Ashish and Mehra, Aditya",
    title = "{Flat Holography: Aspects of the dual field theory}",
    eprint = "1609.06203",
    archivePrefix = "arXiv",
    primaryClass = "hep-th",
    doi = "10.1007/JHEP12(2016)147",
    journal = "JHEP",
    volume = "12",
    pages = "147",
    year = "2016"
}

@article{Nguyen:2025zhg,
    author = "Nguyen, Kevin",
    title = "{Lectures on Carrollian Holography}",
    eprint = "2511.10162",
    archivePrefix = "arXiv",
    primaryClass = "hep-th",
    month = "11",
    year = "2025"
}

@article{Hamilton:2005ju,
    author = "Hamilton, Alex and Kabat, Daniel N. and Lifschytz, Gilad and Lowe, David A.",
    title = "{Local bulk operators in AdS/CFT: A Boundary view of horizons and locality}",
    eprint = "hep-th/0506118",
    archivePrefix = "arXiv",
    reportNumber = "BROWN-HET-1448, CU-TP-1130",
    doi = "10.1103/PhysRevD.73.086003",
    journal = "Phys. Rev. D",
    volume = "73",
    pages = "086003",
    year = "2006"
}

@article{Hamilton:2006az,
    author = "Hamilton, Alex and Kabat, Daniel N. and Lifschytz, Gilad and Lowe, David A.",
    title = "{Holographic representation of local bulk operators}",
    eprint = "hep-th/0606141",
    archivePrefix = "arXiv",
    reportNumber = "CU-TP-1149",
    doi = "10.1103/PhysRevD.74.066009",
    journal = "Phys. Rev. D",
    volume = "74",
    pages = "066009",
    year = "2006"
}

@article{Doi:2024nty,
    author = "Doi, Kazuki and Ogawa, Naoki and Shinmyo, Kotaro and Suzuki, Yu-ki and Takayanagi, Tadashi",
    title = "{{Probing de Sitter Space Using CFT States}}",
    eprint = "2405.14237",
    archivePrefix = "arXiv",
    primaryClass = "hep-th",
    reportNumber = "YITP-24-66",
    doi = "10.1007/JHEP02(2025)093",
    journal = "JHEP",
    volume = "02",
    pages = "093",
    year = "2025"
}

@article{Fitzpatrick:2011hu,
    author = "Fitzpatrick, A. Liam and Kaplan, Jared",
    title = "{Analyticity and the Holographic S-Matrix}",
    eprint = "1111.6972",
    archivePrefix = "arXiv",
    primaryClass = "hep-th",
    reportNumber = "SLAC-PUB-14841",
    doi = "10.1007/JHEP10(2012)127",
    journal = "JHEP",
    volume = "10",
    pages = "127",
    year = "2012"
}

@article{Sachs:1962wk,
    author = "Sachs, R. K.",
    title = "{Gravitational waves in general relativity. 8. Waves in asymptotically flat space-times}",
    doi = "10.1098/rspa.1962.0206",
    journal = "Proc. Roy. Soc. Lond. A",
    volume = "270",
    pages = "103--126",
    year = "1962"
}

@article{Bondi:1962px,
    author = "Bondi, H. and van der Burg, M. G. J. and Metzner, A. W. K.",
    title = "{Gravitational waves in general relativity. 7. Waves from axisymmetric isolated systems}",
    doi = "10.1098/rspa.1962.0161",
    journal = "Proc. Roy. Soc. Lond. A",
    volume = "269",
    pages = "21--52",
    year = "1962"
}

@article{tHooft:1993dmi,
    author = "'t Hooft, Gerard",
    title = "{Dimensional reduction in quantum gravity}",
    eprint = "gr-qc/9310026",
    archivePrefix = "arXiv",
    reportNumber = "THU-93-26",
    journal = "Conf. Proc. C",
    volume = "930308",
    pages = "284--296",
    year = "1993"
}

@article{Susskind:1994vu,
    author = "Susskind, Leonard",
    title = "{The World as a hologram}",
    eprint = "hep-th/9409089",
    archivePrefix = "arXiv",
    reportNumber = "SU-ITP-94-33",
    doi = "10.1063/1.531249",
    journal = "J. Math. Phys.",
    volume = "36",
    pages = "6377--6396",
    year = "1995"
}

@article{Maldacena:1997re,
    author = "Maldacena, Juan Martin",
    title = "{The Large $N$ limit of superconformal field theories and supergravity}",
    eprint = "hep-th/9711200",
    archivePrefix = "arXiv",
    reportNumber = "HUTP-97-A097, HUTP-98-A097",
    doi = "10.4310/ATMP.1998.v2.n2.a1",
    journal = "Adv. Theor. Math. Phys.",
    volume = "2",
    pages = "231--252",
    year = "1998"
}

@article{Gubser:1998bc,
    author = "Gubser, S. S. and Klebanov, Igor R. and Polyakov, Alexander M.",
    title = "{Gauge theory correlators from noncritical string theory}",
    eprint = "hep-th/9802109",
    archivePrefix = "arXiv",
    reportNumber = "PUPT-1767",
    doi = "10.1016/S0370-2693(98)00377-3",
    journal = "Phys. Lett. B",
    volume = "428",
    pages = "105--114",
    year = "1998"
}

@article{Witten:1998qj,
    author = "Witten, Edward",
    title = "{Anti de Sitter space and holography}",
    eprint = "hep-th/9802150",
    archivePrefix = "arXiv",
    reportNumber = "IASSNS-HEP-98-15",
    doi = "10.4310/ATMP.1998.v2.n2.a2",
    journal = "Adv. Theor. Math. Phys.",
    volume = "2",
    pages = "253--291",
    year = "1998"
}

@article{Strominger:2001pn,
    author = "Strominger, Andrew",
    title = "{The dS / CFT correspondence}",
    eprint = "hep-th/0106113",
    archivePrefix = "arXiv",
    doi = "10.1088/1126-6708/2001/10/034",
    journal = "JHEP",
    volume = "10",
    pages = "034",
    year = "2001"
}

@article{Maldacena:2002vr,
    author = "Maldacena, Juan Martin",
    title = "{Non-Gaussian features of primordial fluctuations in single field inflationary models}",
    eprint = "astro-ph/0210603",
    archivePrefix = "arXiv",
    doi = "10.1088/1126-6708/2003/05/013",
    journal = "JHEP",
    volume = "05",
    pages = "013",
    year = "2003"
}

@article{Hikida:2021ese,
    author = "Hikida, Yasuaki and Nishioka, Tatsuma and Takayanagi, Tadashi and Taki, Yusuke",
    title = "{Holography in de Sitter Space via Chern-Simons Gauge Theory}",
    eprint = "2110.03197",
    archivePrefix = "arXiv",
    primaryClass = "hep-th",
    reportNumber = "YITP-21-105; IPMU21-0059",
    doi = "10.1103/PhysRevLett.129.041601",
    journal = "Phys. Rev. Lett.",
    volume = "129",
    number = "4",
    pages = "041601",
    year = "2022"
}

\end{document}